\documentclass[twocolumn]{aastex631}

\usepackage{threeparttable}
\usepackage{amsmath} 
\newcommand{\angstrom}{\text{\normalfont\AA}}
\newcommand{\change}[1]{#1}           
\usepackage{hyperref}

\begin{document}

\title{MINDS. Anatomy of a water-rich, inclined, brown dwarf disk: lack of abundant hydrocarbons}

\author[0000-0002-8545-6175]{Giulia Perotti}
\affil{Max-Planck-Institut f\"{u}r Astronomie, K\"{o}nigstuhl 17, 69117 Heidelberg, Germany}
\affil{Niels Bohr Institute, University of Copenhagen, NBB BA2, Jagtvej 155A, 2200 Copenhagen, Denmark}

\author[0000-0002-2358-4796]{Nicol\'as T. Kurtovic}
\affil{Max-Planck Institut f\"{u}r Extraterrestrische Physik (MPE), Giessenbachstr. 1, 85748, Garching, Germany}

\author[0000-0002-1493-300X]{Thomas Henning}
\affil{Max-Planck-Institut f\"{u}r Astronomie, K\"{o}nigstuhl 17, 69117 Heidelberg, Germany}

\author[0000-0003-3747-7120]{G\"oran Olofsson}
\affil{Department of Astronomy, Stockholm University, AlbaNova University Center, 10691 Stockholm, Sweden}

\author[0000-0001-8407-4020]{Aditya M. Arabhavi}
\affil{Kapteyn Astronomical Institute, Rijksuniversiteit Groningen, Postbus 800, 9700AV Groningen, The Netherlands}

\author[0000-0002-6429-9457]{Kamber Schwarz}
\affil{Max-Planck-Institut f\"{u}r Astronomie, K\"{o}nigstuhl 17, 69117 Heidelberg, Germany}

\author[0000-0003-0386-2178]{Jayatee Kanwar}
\affil{Kapteyn Astronomical Institute, Rijksuniversiteit Groningen, Postbus 800, 9700AV Groningen, The Netherlands}
\affil{Space Research Institute, Austrian Academy of Sciences, Schmiedlstr. 6, A-8042, Graz, Austria}
\affil{TU Graz, Fakultät für Mathematik, Physik und Geodäsie, Petersgasse 16 8010 Graz, Austria}

\author[0000-0002-2190-3108]{Roy van Boekel}
\affil{Max-Planck-Institut f\"{u}r Astronomie, K\"{o}nigstuhl 17, 69117 Heidelberg, Germany}

\author[0000-0001-7455-5349]{Inga Kamp}
\affil{Kapteyn Astronomical Institute, Rijksuniversiteit Groningen, Postbus 800, 9700AV Groningen, The Netherlands}

\author[0000-0003-0041-4599]{Ilaria Pascucci}
\affil{Lunar and Planetary Laboratory, The University of Arizona, Tucson, AZ 85721, USA}


\author[0000-0001-7591-1907]{Ewine F. van Dishoeck}
\affil{Leiden Observatory, Leiden University, PO Box 9513, 2300 RA Leiden, the Netherlands}
\affil{Max-Planck Institut f\"{u}r Extraterrestrische Physik (MPE), Giessenbachstr. 1, 85748, Garching, Germany}

\author[0000-0001-9818-0588]{Manuel G\"udel}
\affil{Dept. of Astrophysics, University of Vienna, T\"urkenschanzstr. 17, A-1180 Vienna, Austria}
\affil{ETH Z\"urich, Institute for Particle Physics and Astrophysics, Wolfgang-Pauli-Str. 27, 8093 Z\"urich, Switzerland}

\author{Pierre-Olivier Lagage}
\affil{Universit\'e Paris-Saclay, Universit\'e Paris Cit\'e, CEA, CNRS, AIM, F-91191 Gif-sur-Yvette, France}


\author[0000-0002-5971-9242]{David Barrado}
\affil{Centro de Astrobiolog\'ia, CSIC-INTA, ESAC Campus, Camino Bajo del Castillo s/n, 28692 Villanueva de la Ca\~nada, Madrid, Spain}

\author[0000-0001-8876-6614]{Alessio Caratti o Garatti}
\affil{INAF – Osservatorio Astronomico di Capodimonte, Salita Moiariello 16, 80131 Napoli, Italy}
\affil{Dublin Institute for Advanced Studies, 31 Fitzwilliam Place, D02 XF86 Dublin, Ireland}

\author[0000-0001-9250-1547]{Adrian M. Glauser}
\affil{ETH Z\"urich, Institute for Particle Physics and Astrophysics, Wolfgang-Pauli-Str. 27, 8093 Z\"urich, Switzerland}

\author[0000-0002-8443-9488]{F. Lahuis}
\affil{SRON Netherlands Institute for Space Research, PO Box 800, 9700 AV, Groningen, The Netherlands}

\author[0000-0002-0101-8814]{Valentin Christiaens}
\affil{Institute of Astronomy, KU Leuven, Celestijnenlaan 200D, 3001 Leuven, Belgium}
\affil{STAR Institute, Universit\'e de Li\`ege, All\'ee du Six Ao\^ut 19c, 4000 Li\`ege, Belgium}

\author[0000-0002-8889-2992]{Riccardo Franceschi}
\affil{LESIA, Observatoire de Paris, Universit\'e PSL, CNRS, Sorbonne Universit\'e, Universit\'e de Paris, 5 place Jules Janssen, 92195 Meudon, France}

\author[0000-0002-1257-7742]{Danny Gasman}
\affil{Institute of Astronomy, KU Leuven, Celestijnenlaan 200D, 3001 Leuven, Belgium}

\author[0000-0002-4022-4899]{Sierra L. Grant}
\affil{Max-Planck Institut f\"{u}r Extraterrestrische Physik (MPE), Giessenbachstr. 1, 85748, Garching, Germany}

\author[0000-0002-6592-690X]{Hyerin Jang}
\affil{Department of Astrophysics/IMAPP, Radboud University, PO Box 9010, 6500 GL Nijmegen, The Netherlands}

\author[0000-0001-8240-978X]{Till Kaeufer}
\affil{Department of Physics and Astronomy, University of Exeter, Exeter EX4 4QL, UK}

\author[0000-0001-9526-9499]{Maria Morales-Calder\'on}
\affil{Centro de Astrobiolog\'ia, CSIC-INTA, ESAC Campus, Camino Bajo del Castillo s/n, 28692 Villanueva de la Ca\~nada,
Madrid, Spain}

\author[0000-0002-7935-7445]{Milou Temmink}
\affil{Leiden Observatory, Leiden University, PO Box 9513, 2300 RA Leiden, the Netherlands}

\author[0000-0002-3135-2477]{Marissa Vlasblom}
\affil{Leiden Observatory, Leiden University, PO Box 9513, 2300 RA Leiden, the Netherlands}


\begin{abstract}
\change{2MASS~J04381486+2611399 (or J0438) is one of the few young brown dwarfs (BD) with a highly inclined ($i\!\sim\!70^\circ$) disk. Here we report results from JWST-MIRI MRS, HST-ACS and ALMA Band~7 observations. Despite its late spectral type (M7.25), the spectrum of J0438 resembles those of inner disks around earlier-type stars (K1–M5, T Tauri stars), with a volatile reservoir lacking hydrocarbons (except for acetylene, C$_2$H$_2$) and dominated by water. Other identified species are H$_2$, CO$_2$, HCN, [Ar$^{+}$], and [Ne$^{+}$]. The dominance of water over hydrocarbons is driven by multiple factors such as disk dynamics, young disk age, low accretion rate and possible inner disk clearing. J0438 appears highly dynamic, showing a seesaw-like variability and extended emission in H$_2 \,\,\, S$(1), $S$(3), $S$(5), [Ne$^{+}$] and CO ($J=3-2$). Interestingly, the CO  emission reaches up to 400~au from the brown dwarf, suggesting ongoing infalling/outflowing activity impacting the disk chemistry. These observations underscore the combined power of MIRI, HST and ALMA in characterizing the chemical diversity and dynamics of brown dwarf disks.}
\end{abstract}


\section{Introduction} \label{sec:intro}
Brown dwarfs (BDs) are sub-stellar objects covering the mass range between the heaviest gas giant planets and the lightest stars. Observations of young brown dwarfs have revealed the presence of disks around them in both the near and mid-infrared (e.g., \citealt{Comeron1998,Natta2001,Natta2002,Apai2005}) as well as in the far-infrared and (sub-)millimeter spectral domains (e.g., \citealt{Klein2003,Harvey2012a,Daemgen2016,testi2016,sanchis2020,Rilinger2021}). Interestingly, both rocky and giant planet companions have been detected around brown dwarfs (e.g., 
\citealt{chauvin2005,Todorov2010,Han2013}), leading to considerable interest in disentangling what are the chemical and physical properties of the disks surrounding such very low-mass ($\leq0.075~M_\odot$) objects.  

Compared to disks around young solar analogs (i.e., T~Tauri disks), brown dwarf disks are typically smaller and less dense (e.g., \citealt{Hendler2017}). They are characterized by lower accretion rates ($\leq 10^{-10}-10^{-12} M_\odot \mathrm{yr}^{-1}$; \citealt{Herczeg2009}) and plausibly longer lifetimes (e.g., \citealt{Carpenter2006,Harvey2012b}). 
From a chemical point of view, surveys using the \textit{Spitzer} Space Telescope InfraRed Spectrograph (SST-IRS; $R=\lambda/\Delta \lambda \sim 700$) reported differences in the volatile content of the inner disks around very low-mass stars (VLMS) and BD (i.e., M5$-$M9) with respect to T~Tauri (i.e., K1$-$M5) disks (e.g., \citealt{Pascucci2009a}). The mid-infrared spectra of the former are typically dominated by the bright emission of C- and H-bearing compounds \citep{Pascucci2013}, whereas the latter are characterized by transitions of O-rich compounds, most notably H$_2$O and CO$_2$ (e.g., \citealt{Pontoppidan2010}). 

Recent mid-infrared observations with the \textit{James Webb} Space Telescope (JWST; $R \sim 3000$) are now characterizing a larger sample of inner disks to draw firm conclusions on the variations of inner disk chemistry across stellar mass (e.g., \citealt{Henning2024}). Indeed, early JWST spectra of VLMS have shown orders of magnitude higher column densities of \change{acetylene (C$_2$H$_2$)} and a much richer hydrocarbon chemistry, including isotopologs and species like C$_2$H$_4$, C$_4$H$_2$, C$_3$H$_4$ and C$_6$H$_6$ (e.g., \citealt{tabone2023,Arabhavi2024,Kanwar2024,Morales-Calderon2025}), than found in the pioneering SST-IRS data. Interestingly, among all VLMS and BD disks targeted in JWST-MIRI Cycles 1 and 2 so far ($\sim$ 25) all of them appear to have a close to face-on configuration except for the target of this work.   
 
Here, we present the JWST-MIRI MRS spectrum of the Class~II disk of 2MASS~J04381486+2611399 (or ITG-3, hereafter J0438), a M7.25 brown dwarf ($M_* \sim 0.05\,M_\odot$; \citealt{Luhman2004,Manara2023}) located at $140.26~\pm~10.38~$pc in the Taurus star-forming region \citep{gaia2021}. J0438 has been classified as a burster from \change{\textit{Kepler}-K2} observations \citep{Cody2022} with an estimated mass accretion rate of \change{log$_{10}\,$}$\dot{M}$ [$M_\odot$ yr$^{-1}$] = $-10.8$ \citep{Muzerolle2005}. Interestingly, it does not show variability in the TESS light curve \citep{Kumbhakar2023}. 
The small disk of J0438 was imaged with the \textit{Hubble} Space Telescope (HST), and modeling of the HST scattered light images and spectral energy distribution (SED) suggested that the dusty disk extends up to $\sim 20-40~$au\footnote{SED modelling including data from \textit{Herschel}-PACS suggest a $R_\mathrm{out}\sim$10~au \citep{Hendler2017}. This is in agreement with scattered light images estimating on average larger disk outer radii compared to millimeter observations probing larger, more settled, dust.} and presents an \textit{inner hole} at $\sim 0.28~$au. These data also confirmed that the disk is close-to-edge-on ($67-71^{\circ}$; \citealt{Luhman2004,Scholz2006}), making it \change{one of the very few} highly-inclined brown dwarf disk known to date \citep{Luhman2010}. 

J0438 has the most complete SED for a brown dwarf disk, encompassing a near-infrared IRTF spectrum \citep{Luhman2007}, a SST-IRS spectrum \citep{Luhman2007,Pascucci2013}, SST-IRAC, MIPS \citep{Monin2010} and $Herschel$-PACS photometry \citep{Bulger2014,Hendler2017} in addition to ALMA \citep{Ward-Duong2018,Pinilla2017}, CARMA \citep{Phan-Bao2014} and IRAM~30\,m measurements \citep{Scholz2006}. Compared to most very low-mass star and brown dwarf disks, the SST-IRS spectrum of J0438 ($R\!\sim\!700$) exhibits an anomalously red SED, two tentative detections of water vapour at 17.22 and $18.17~\mu$m and does not present emission from other organic species \change{\citep{Pascucci2013}}. 

Thanks to the high spectral resolving power of JWST-MIRI MRS ($R\!\sim\!3000$; \citealt{Labiano2021}) we confirm the presence of faint water emission throughout the JWST mid-infrared (mid-IR) spectrum of J0438 as well as of a set of atomic, molecular and ionic emission lines (Sec.~\ref{sec:observations} and \ref{sec:results}). 
We compare molecular flux ratios relative to water for J0438 with those obtained for other VLMS and T~Tauri disks in Sec.~\ref{sec:discussion}. 
We conclude by summarizing the most relevant findings of our study and by listing avenues for future work (Sec.~\ref{sec:conclusions}). 

\begin{table*}
    \centering
    \renewcommand{\arraystretch}{1.1}
    \caption{Stellar and Disk Properties of J04381486+2611399.}
    \begin{tabular}{lll}
    \hline
    \hline
    Property [unit] & Value & Ref. \\
     \hline
    Distance, $d$ [pc] & $ 140.26 \pm 10.38 $ & 1, 2\\
    Spectral type & M7.25 & 3\\
    Mass, $M_*$ [$M_\odot$]& 0.05& 2 \\
    Luminosity, \change{log$_{10}\!$} $L_*$ [$L_\odot$] & -2.7 & 2\\
    Effective temperature [K] & 2838 & 3 \\
    Mass accretion rate, \change{log$_{10}\!$} $\dot{M}$ [$M_\odot$ yr$^{-1}$] & -10.8 & 4\\
    Accretion luminosity,  \change{log$_{10}\!$} $L_\mathrm{acc}$ [$L_{\odot}$ yr$^{-1}$] & -4.1  &  5\\
    70$~\mu$m Flux [mJy] & $95 \pm 2$ & 6\\ 
    0.89 mm Flux [mJy] & 1.36 $\pm$ 0.11 & 7\\
    1.3 mm Flux [mJy] & $2.29 \pm 0.75$ & 8\\
    Disk inclination, $i$ [$^{\circ}$] & $67-71$ & 8, 9\\
    Inner disk radius, $R_\mathrm{in}$ [au] & $0.28$ & 9\\
    Outer disk radius, $R_\mathrm{out}$ [au] & $20-40$ & 9\\
    \hline
    \textbf{References.} 1. \citet{gaia2021}; & 2. \citet{Manara2023}; & 3. \citet{Luhman2004}; \\
     4. \citet{Muzerolle2005}; & 5. \citet{Xie2023}; & 6. \citet{Bulger2014};\\
     7. \citet{Ward-Duong2018} & 8. \citet{Scholz2006}; & 9. \citet{Luhman2007}; \\ 
    \end{tabular}
    \label{tab:properties}    
\end{table*}

\section{Observations and data reduction}
\label{sec:observations}
J0438 was observed with the JWST \citep{Rigby2023} Mid-Infrared Instrument (MIRI; \citealt{rieke2015,Wright2015,Wright2023}) in Medium Resolution Spectroscopy (MRS; \citealt{Wells2015,Argyriou2023}) mode on 1 October, 2023 with visit number 43 as part of the Cycle~1 GTO program “MIRI mid-INfrared Disk Survey” MINDS (PID: 1282; \citealt{Henning2024}). The source was initially acquired by adopting the target acquisition procedure in FAST readout mode with a neutral density filter (FND) and 10 groups per integration. The disk was then observed in FASTR1 readout mode employing a 4-point dither pattern optimized for point source in the positive direction and all three MRS grating settings (SHORT, MEDIUM, LONG) to obtain full spectral coverage. To ensure a signal-to-noise of at least 100 longward of $17.66~\mu$m, 27 groups per integration were selected for an exposure time per module of 1232~s. The total number of integrations per module is 16. 

\begin{figure*}[ht!]
\centering
\includegraphics[width=\hsize]{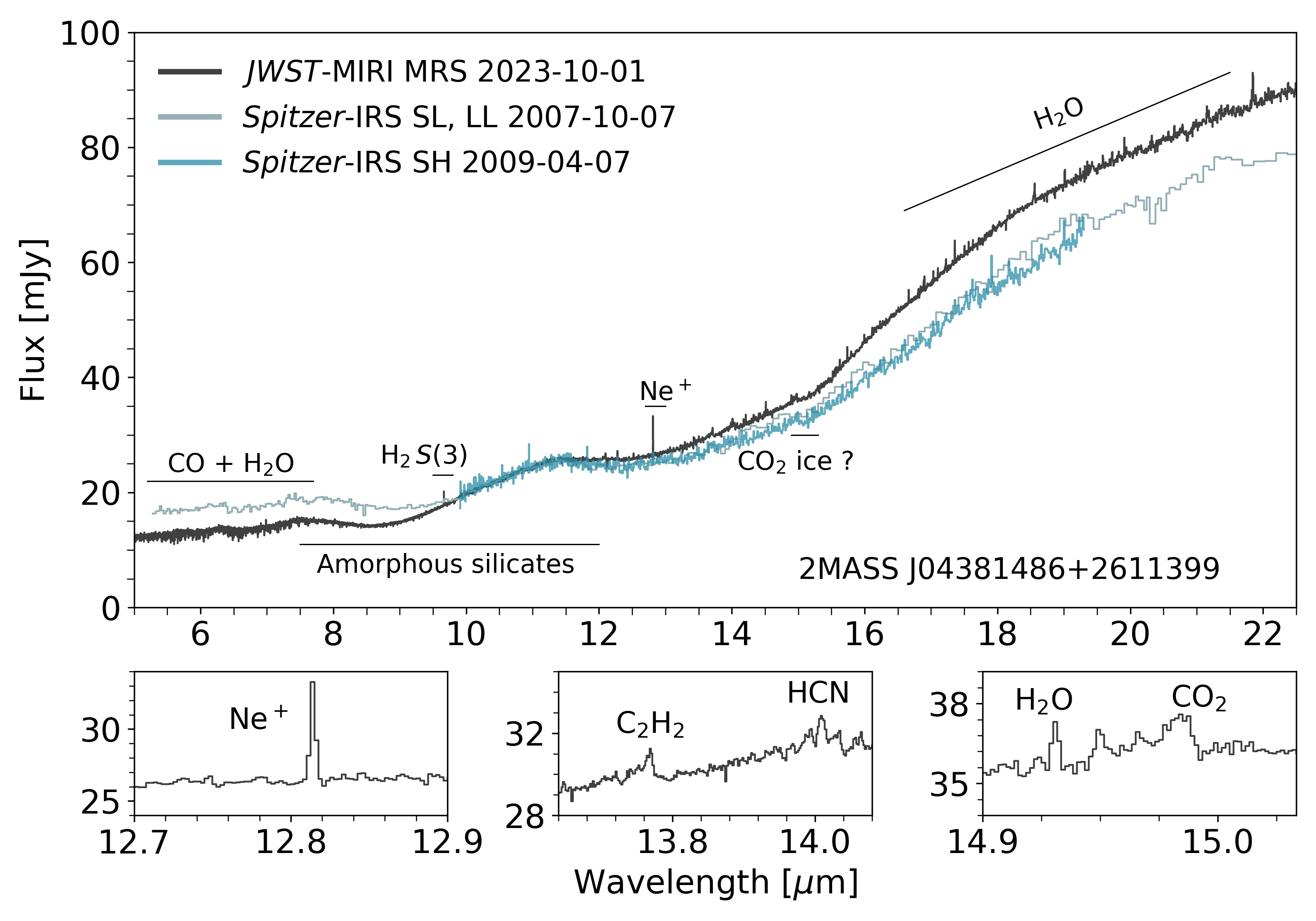}
\caption{\textbf{JWST-MIRI MRS spectrum of the young J0438 disk (black).} The spectrum is dominated by amorphous silicate absorption and emission bands typical of a highly inclined disk configuration. A very weak CO$_2$ ice absorption feature is tentatively detected (see Appendix~\ref{app:cont_sub}, Fig.~\ref{fig:co2_ice}) along with inner disk gas reservoir (Table~\ref{tab:fluxes}) composed by molecular hydrogen (H$_2$), water (H$_2$O), carbon dioxide (CO$_2$), hydrogen cyanide (HCN), acetylene (C$_2$H$_2$) and the singly ionized cations of argon and neon ([Ar$^{+}$], [Ne$^{+}]$; for zoom-ins see Appendices~\ref{app:stellar_contamination} and \ref{app:variability}). The comparison between the JWST-MIRI and the \textit{Spitzer}-IRS low- (SL, LL) and high- (SH) resolution spectra reveal continuum and line variability (grey and teal). The \textit{Spitzer}-IRS spectra were taken from CASSIS \citep{Lebouteiller2011,Lebouteiller2015} and \citet{Pascucci2013}.}
\label{fig:full_spectra}
\end{figure*}

The data were reduced using a publicly available hybrid pipeline which relies on the standard JWST pipeline \citep{Christiaens2024}. In short, the data were processed using version 1.14.1 of the JWST Calibration Pipeline \citep{Bushouse2024} and Calibration Reference Data System (CRDS) context \texttt{jwst\_1253.pmap} in combination with dedicated routines based on the Vortex Image Processing (\texttt{VIP}) package for background subtraction, bad pixel correction and spectral extraction \citep{GomezGonzalez2017,Christiaens2023}. To subtract the background a direct pair-wise dither subtraction was carried out. This method is particularly suited for faint sources
and guarantees the reduction of the noise level in the spectrum. We refer the reader to \change{\citet{Arabhavi2025b}} for further details on the data reduction. As the source is compact and unresolved, 1-D spectra were then extracted 
from the reduced datacubes with aperture photometry in 1.5-FWHM apertures centered on the centroid location, corrected for aperture size using correction factors from \citet{Argyriou2023}. After spectral extraction, an additional residual fringe correction was carried out. The final reduced spectrum of J0438 is presented in black in Figure~\ref{fig:full_spectra}.

\section{The mid-infrared spectrum of J0438} \label{sec:results}

\subsection{Dust continuum}
The JWST-MIRI MRS spectrum of the disk around J0438 shows a remarkably steep slope and the presence of both silicate absorption at $9~\mu$m and silicate emission at $11~\mu$m (Fig.~\ref{fig:full_spectra}), typical of edge-on disks (e.g., \citealt{Luhman2007}). Amorphous silicates dominate the spectrum, as no significant crystalline silicate features are identified. The spectrum lacks significant ice absorption features, and only a very weak CO$_2$ absorption band at $\sim\!15.2\, \mu$m is tentatively detected (see  Fig.~\ref{fig:co2_ice} of Appendix~\ref{app:cont_sub}). The absence of prominent ice bands is not surprising given the moderate disk inclination of $\leq 70^\circ$ (e.g., \citealt{Pontoppidan2005,Ballering2021,Arabhavi2022}) and the nature of the system (i.e., a small, low-mass dusty disk without remnant envelope material; \citealt{Luhman2007,Luhman2010}). 

The MIRI spectrum differs from those obtained by SST-IRS in either low- (SL, LL) and high- (SH) resolution spectroscopy modes (Fig.~\ref{fig:full_spectra} and Fig.~\ref{fig:nir_mir} of Appendix~\ref{app:stellar_contamination}). In particular, the MIRI spectrum shows weaker emission shortwards of $10~\mu$m with respect to the low-resolution observations and stronger emission longwards of $12\,\mu$m compared to both low- and high-resolution datasets (PID: 40302; PI: J.R.~Houck and PID: 50799; PI: G.~Herczeg). A good agreement is seen among the SST-IRS observations taken less than 2~years apart, excluding variability on $1-2$ years timescale. Interestingly, the three mid-infrared spectra overlap in the range between 10 and $\sim12\,\mu$m, at the location of the silicate emission band. The largest offsets are found between the MIRI (2023-10-01) and the low-resolution IRS observations (2007-10-07) taken 16 years apart with the MIRI spectrum roughly $30\%$ weaker at $6~ \mu$m and $15\%$ stronger at $20~\mu$m. 

\change{The reported spectrophotometric accuracy is 2–10\% for SST-IRS \citep{Furlan2006, Watson2009} and $5.6 \pm 0.7\%$ for JWST-MIRI MRS \citep{Argyriou2023},} which confidently rules out calibration uncertainties, especially for the shorter wavelengths. In addition, one should note that the continuum level of the JWST spectrum is consistent when performing different background subtraction methods for the MINDS pipeline (see \citealt{Christiaens2024} for more details). Finally, due to the larger aperture of SST-IRS $-$ and therefore the higher chances to recover the emission of neighboring sources $-$  it is reasonable to expect a higher SST-IRS flux relative to JWST-MIRI. However, this pattern is not observed for the full spectral coverage and instead, the MIRI spectrum is consistently stronger longwards of $12\, \mu$m, supporting the presence of both continuum and line variability (see also Appendix~\ref{app:variability}). 

Mid-IR variability in Class~II disks has been observed routinely when comparing SST-IRS and JWST-MIRI datasets (e.g., \citealt{Kospal2023,Perotti2023,Gaidos2024,Schwarz2024,Jang2024}). Most of the mid-IR variability observed prior to the launch of JWST (i.e., when comparing different epochs of \textit{Spitzer}-IRS observations) fell into two classes \change{\citep{Espaillat2011}}. The first, often referred to as \textit{seesaw} behaviour, is when the emission at shorter and longer wavelengths varies inversely due to dynamical changes in the inner disk (e.g., triggered by the presence of planet warps; \change{\citealt{Muzerolle2009,Flaherty2010}}). The second is when the intensity shift is consistently stronger (or weaker) throughout the mid-infrared range (e.g., due to a change in incident flux from the central object). 

The mid-IR variability of J0438 can be ascribed to a seesaw behaviour where the pivot point is at $\lambda\simeq11.5\,\mu$m. \change{At this wavelength, an anti-correlation between the flux of J0438 of the shortest and longest wavelengths is observed.} Interestingly, the variability profile of J0438 (which possibly has an inner hole; \citealt{Luhman2007}) is similar to the one reported for the (pre-) transition disk LkCa~15 which hosts a large cavity (Fig.1 of \citealt{Espaillat2011}). Possible explanations for the variability of J0438 are therefore the presence of a planet warp in the inner hole or simply changes of the inner rim height due to disk clearing. A detailed understanding of the dominant cause requires accurate radiative transfer modeling, which is outside the scope of this work. 

\subsection{Line detections}
At the shortest wavelengths ($<7.3 \,\mu$m), absorption features of H$_2$O and CO are found (Fig.~\ref{fig:BD_absorption}). These bands are attributed to the photosphere of the brown dwarf and not to the disk. The same behaviour was previously observed for MIRI MRS spectra of other disks around very low-mass stars (VLMS) such as ISO-ChaI~147 (Fig.~S6 of Supplementary Text of \citealt{Arabhavi2024}) and Sz~28 (Fig.~A.1 of \citealt{Kanwar2024}). In the case of J0438, near-infrared $0.65-2.5~\mu$m NASA InfraRed Telescope Facility (IRTF) data presented in \citet{Luhman2007} already revealed several broad H$_2$O and CO absorption features of photospheric origin (Figure~\ref{fig:nir_mir}). A comparison between the MIRI spectrum and a PHOENIX \citep{Husser2013} stellar model is presented in Fig.~\ref{fig:BD_absorption} in Appendix~\ref{app:stellar_contamination} to check for photospheric contamination at the shortest wavelengths. This analysis shows that the central object dominates the MIRI spectrum up until $\sim\!7.3~\mu$m.  

\begin{figure*}
\centering
\includegraphics[width=\hsize]{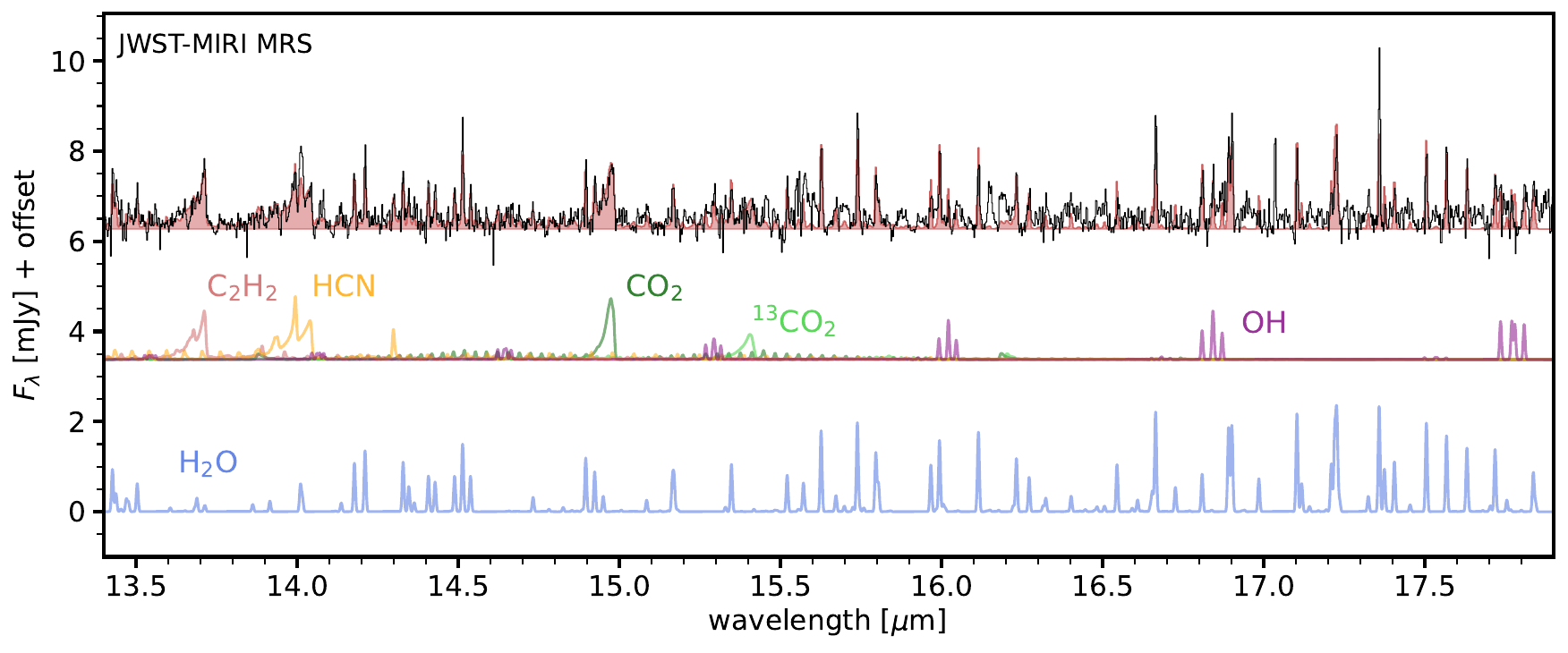}
\caption{\textbf{Comparison between the total synthetic (red shaded region) and the continuum-subtracted JWST-MIRI spectrum (black).} The total synthetic spectrum is a composite of water, C$_2$H$_2$, HCN, CO$_2$, $^{13}$CO$_2$ and OH slab models plotted in several colours (see Section~\ref{sec:slab_modelling} for more details).}
\label{fig:slabs}
\end{figure*}

Moving to longer wavelengths, beyond the [Ne$^{+}$] and the H$_2$~$S$(1) line securely identified with SST-IRS \citep{Pascucci2013}, we now provide ten other detections, plus three tentative assignments (Table~\ref{fig:full_spectra} and \ref{tab:fluxes}). Water lines are observed throughout the spectrum (Figures~\ref{fig:full_spectra} and \ref{fig:slabs}): these are pure rotational transitions and are found from $\sim 12~\mu$m up to $26~\mu$m (see also \citealt{Arabhavi2025a}). We note that J0438 is one of the two systems $-$ out of eight VMLS/BD disks in the sample of \citealt{Pascucci2013} $-$  for which two water lines ($17.22~\mu$m and $18.17~\mu$m) were already tentatively identified with SST-IRS. 

Apart from water, the molecular inventory of J0438 includes species routinely observed in the inner disks of T~Tauri stars (Fig.~\ref{fig:full_spectra} and Table~\ref{tab:fluxes}) such as five pure rotational bands of molecular hydrogen (H$_2$) and the ro-vibrational bands of carbon dioxide (CO$_2$), acetylene (C$_2$H$_2$) and hydrogen cyanide (HCN). The emissions of latter two molecules have been recently detected, for the first time, in the atmosphere of a Gyr~year-old brown dwarf \change{\citep{Matthews2025}}. Upper limits for the weaker species hydroxide (OH) and the isotopologue of carbon dioxide ($^{13}$CO$_2$) are also reported. 

\change{Forbidden emission lines are detected in the mid-IR spectrum of J0438. These are [Ar$^{+}$], [Ne$^{+}$] and tentatively [Ne$^{2+}$]. We note that [Ne$^{+}$] (Figure~\ref{fig:neon_zoomin}) is \textit{rarely} observed in VLMS/BD disks \citep{Pascucci2013} but often found in T~Tauri disks with SST-IRS (e.g., \citealt{Espaillat2007,Lahuis2007,Pascucci2007,
Guedel2010}) and JWST-MIRI (e.g., \citealt{Espaillat2023,Arulanantham2024,Bajaj2024,Schwarz2024}).} 
The presence of [Ne$^{+}$], together with the detection of other forbidden lines (i.e., O and S$^+$; \citealt{Luhman2004}) and of asymmetric extended emission in the HST images hinted at a weak outflowing activity in this system \citep{Luhman2007}.    
To disentangle whether the observed emission is probing the disk surface or jets/outflows, a detailed analysis of the MIRI datacubes and of archival HST and ALMA observations is carried out. These results are presented in Section~\ref{sec:discussion} and Appendix~\ref{app:extendedemission}. 
In the remainder of the paper we focus on the analysis of the water lines and of the ro-vibrational bands of CO$_2$, HCN and C$_2$H$_2$ to compare with other disks observed with SST-IRS and JWST-MIRI.

\subsection{Non-detections} \label{sec:nondetections}
Interestingly, several hydrocarbons commonly identified in disks around VLMS \change{(\citealt{Arabhavi2025b})} are not reported towards J0438. These species include: CH$_4$, \change{C$_3$HN}, C$_2$H$_4$, C$_2$H$_6$, C$_3$H$_4$, $^{13}$CCH$_2$ and C$_6$H$_6$. The same applies to hydrogen recombination lines which are often identified in VLMS disks (e.g., \citealt{Franceschi2024}) but not towards J0438. We note that hydrogen recombination lines were also not reported in the MIRI spectrum of another VLMS disk, Sz~28 \citep{Kanwar2024}, for which a comparably low mass accretion rate was estimated \change{(log$_{10} M_{\mathrm{acc}}$\,-10.27 $\pm$ 0.35 $M_\odot$ yr$^{-1}$; \citealt{Manara2017})}. 

In addition, although the spectrum of J0438 shows the emission of [Ar$^{+}$] and [Ne$^{+}$], only a tentative detection of [Ne$^{2+}$] is reported and other ions such as HCO$^+$ and CH$_3^+$, which have been found towards the disk around TW~Hya \citep{Henning2024} and towards the proplyd d203-506 (only CH$_3^+$; \citealt{Berne2023}), are not clearly identified. It remains unclear whether the non-detections $-$ especially of the latter two molecular cations $-$ are due to weak UV irradiation or simply due to the presence of water emission dominating at those wavelengths, and thus hampering the identification of these weaker species. Similarly, other typical jet tracers such as [Ni$^{+}$], [Fe$^{+}$] and [S$^{2+}$] which were previously reported in the T~Tauri edge-on disk Tau~042021 \citep{Arulanantham2024} and in young disks \citep{Tychoniec2024} are not found in J0438, possibly due to weak line emission and/or water masking. 

We also report the absence of emission from polycyclic aromatic hydrocarbons (PAHs), contrarily to what has been found for some other T~Tauri edge-on systems \citep{Arulanantham2024,Sturm2024} \change{as well as of rare isotopologues (C$^{18}$O$^{16}$O and H$^{13}$CN), recently detected in the similarly inclined T-Tauri disk MY Tau \citep{Salyk2025}}.

\begin{table*}
    \centering
    \renewcommand{\arraystretch}{1.1}
 \caption{Summary of the best-fit model parameters and $3\sigma$-confidence uncertainties.}
    \begin{tabular}{ccccc}
    \hline
    \hline
    Species & Temperature & log$_{10}$($N$) &  $R$ & $\mathcal{N}_\mathrm{tot}$\\
            & [K] & [cm$^{-2}$] & [au] & [molec.] \\
    \hline
    H$_2$O & 469 $^{+18}_{-17}$ & 18.7 $^{+0.17}_{-0.19}$ & 0.07  $^{+0.01}_{-0.01}$ & $ 1.7 \times 10^{43}$\\
    CO$_2$ & 534 $^{+138}_{-105}$ & $>$16.9 & 0.03 $^{+0.04}_{-0.01}$ & $>5.0 \times 10^{40}$ \\
    C$_2$H$_2$ & 684 $^{+276}_{-179}$ & 13.3 $^{+0.83}_{-0.29}$ & 0.92 $^{+0.41}_{-0.57}$ & $1.2 \times 10^{40}$\\
    HCN & 694 $^{+110}_{-91}$ & 15.6 $^{+0.79}_{-0.40}$ & 0.13 $^{+0.07}_{-0.08}$ & $4.7 \times 10^{40}$\\
    \hline
    \end{tabular}
    \label{tab:best_fit}
\end{table*}

\subsection{Slab modelling}
\label{sec:slab_modelling}
To perform a quantitative analysis of the molecular, ionic, and atomic gas in the disk around J0438 we initially subtracted the dust continuum emission from the spectrum. We followed the continuum subtraction procedure described in \citet{Temmink2024}, instead of identifying line-free channels tracing the dust emission by eye. A visual selection of dust continuum points would inevitably result in an inaccurate continuum determination due to the line richness of the MIRI spectrum. In short, the adopted method estimates the dust continuum emission using the \texttt{pybaselines} package \citep{Erb_2022}. As a first step, a dust continuum level is guessed using a Savitzky-Golay filter with a third order polynomial. Then, all emission lines extending above $2\sigma$ of the standard deviation are masked. Once no emission lines are left, the dust continuum is determined and subtracted. The last step of this procedure consists of masking the remaining $3\sigma$ downward spikes. For a more complete explanation of the continuum subtraction procedure we refer the reader to Sec.~2.2 of \citet{Temmink2024}. The resulting continuum-subtracted spectrum of J0438 is displayed in Figure~\ref{fig:cont_sub} of Appendix~\ref{app:cont_sub}. 

Next, we generate synthetic spectra to reproduce the continuum-subtracted spectrum assuming plane-parallel slabs of gas in local thermodynamic equilibrium (LTE) using spectroscopic data from the LAMDA and HITRAN databases \citep{Schoier2005,Gordon2022}. The same procedure has been largely used to analyse the emission observed in several other JWST-MIRI MRS spectra (e.g., \citealt{Grant2023,Gasman2023,Perotti2023,Ramirez2023,tabone2023,Temmink2024}), and therefore it will not be extensively described here again. Briefly, the synthetic spectra are produced following the prescription outlined in \citet{Salyk2011b}, and therefore assuming a Gaussian line profile with a full width at half maximum (FWHM) of $\Delta V = 4.7~$km~s$^{-1}$ ($\sigma=2~$km~s$^{-1}$). The slab models include three parameters namely, the column density of the molecular species ($N$), the excitation temperature of the gas ($T$), and the emitting area $\pi R^2$. Note that $R$ should not be ascribed to a radius within the disk as the adopted models do not take into account the multi-dimensional physical structure of the disk. As such, $R$ corresponds to the radius of a circle of area equivalent to the physical emitting area on the disk, which may be an annulus. 

We selected the major molecular emitters in the spectral region spanning from 13.4 to 18~$\mu$m: H$_2$O, CO$_2$, $^{13}$CO$_2$, C$_2$H$_2$, HCN, and OH. We did not produce synthetic spectra for the region below $13.4~\mu$m as it is largely dominated by the absorption lines from the central object up until $7.3~\mu$m (e.g., \citealt{Li2024}; Fig.~\ref{fig:BD_absorption}) and by the silicate band in absorption and emission (Fig.~\ref{fig:nir_mir}). \change{Beyond 18~$\mu$m,} the noise level increases considerably, therefore we restrict the slab modelling analysis to wavelengths shorter than this. However, the selected spectral range enables the comparison of J0438 with other disks observed with SST-IRS and JWST-MIRI which is the ultimate scope of this work. The resulting synthetic spectra are convolved with the resolution of the selected region of the MIRI spectrum ($R\approx2500$, \citealt{Labiano2021,Argyriou2023}) using the averaged values from \citet{Jones2023}, then resampled to the same wavelength grid as the observed data with \texttt{SpectRes} \citep{carnall2017}.

The slab modelling procedure consists of performing an Markov Chain Monte Carlo (MCMC; \citealt{Foreman-Mackey2013}) fit of all the six selected species simultaneously, instead of fitting iteratively one species at the time (i.e., subtracting the synthetic model of water first, then moving to the next species, and so on). This procedure has the advantage of reducing the bias introduced by the sequential subtraction of different synthetic spectra (e.g., \citealt{Kaufer2024}) and it has been successful in reproducing some of the most water-rich JWST-MIRI spectra obtained so far (e.g., \citealt{Grant2024}, Kurtovic et al. prep.). In particular, the fit is carried out using \texttt{emcee} \citep{emcee2013} and assuming a uniform prior for each of the free parameters (i.e., three per molecular species), as well as eight times the number of walkers with respect to the number of free parameters. The fit convergence is checked by comparing the distribution of walkers as a function of steps, and making sure it had become constant for over $10^3$ steps; for J0438 it is reached after $\sim10^5$ steps. A summary of the best-fit parameters and the $3\sigma$-uncertainties is provided in Table~\ref{tab:best_fit}. Additionally, the total number of molecules ($\mathcal{N}_\mathrm{tot}$) is listed for species characterized by optically thin emission, which has been computed by multiplying the column density by the emitting area (i.e., $\mathcal{N}_\mathrm{tot}=N \pi R^2$). The total synthetic spectrum is compared to the MIRI observations in Figure~\ref{fig:slabs}. Finally, the posterior distributions are shown in Appendix~\ref{app:correlations}, Figure~\ref{fig:correlations}. 

For most species, the emission is found to be coming from temperatures below 700~K, column densities $<10^{19}~$cm$^{-2}$ and very small radii ($\leq0.10~$au). Deviations from this behavior are seen for C$_2$H$_2$ ($N\!\sim\!10^{13}~$cm$^{-2}$ and $R\!\sim~0.9~$au), reporting a degeneracy between the column density and the emitting radius. Most emission is in the optically thin regime as indicated in Figure~\ref{fig:correlations}. We note that OH and $^{13}$CO$_2$ are tentatively identified and therefore poorly constrained. As such, we refrain from analysing further these two species. 
Due to the limited S/N, the column density of CO$_2$ remained challenging to constrain, and the current best slab is limited to a maximum column density of
$10^{17}$cm$^{-2}$. Thus, our results represents a lower limit for the column density of CO$_2$, and any increase on this number would lead to even smaller emitting radii. Finally, a careful analysis of the residuals \change{(Fig.~\ref{fig:residuals})} reveals a few regions where the observed data are not particularly well fitted by the total model. We do not attribute those to unidentified lines but to residual fringes.  

\begin{table*}
    \centering 
    \caption{Summary of line fluxes and $3\sigma$ upper limits.}
    \begin{tabular}{lcc}
    \hline
    \hline
    Species & Wavelength range & Integrated Flux \\
            &  [$\mu$m]        &  [$10^{-16}$erg s$^{-1}$cm$^{-2}$]\\
     \hline
     &  Atomic and Ionic & \\
     \hline
     \textnormal{[Ar$^{+}$]} & \change{6.98$-$6.99} & $1.02\pm 0.10$ \\
     \textnormal{[Ne$^{+}$]} & \change{12.80$-$12.82} & $5.48 \pm 0.07$ \\
     \textnormal{[Ne$^{2+}$]} & \change{15.55$-$15.56} & $<3$ \\
     \hline
     &  Molecular & \\
     \hline
     H$_2$ (0,0) $S$(1) & \change{17.02$-$17.04} & $1.35 \pm 0.03$ \\
     H$_2$ (0,0) $S$(2) & \change{12.27$-$12.28} & $2.30 \pm 0.06$ \\
     H$_2$ (0,0) $S$(3) & \change{9.66$-$9.67} & $3.45 \pm 0.03$ \\
     H$_2$ (0,0) $S$(4) & \change{8.02$-$8.03} & $1.24 \pm 0.08$ \\
     H$_2$ (0,0) $S$(5) & \change{6.90$-$6.91} & $1.75 \pm 0.07$\\
     H$_2$O$^a$ & \change{17.19$-$17.25} & $3.26 \pm 0.05$ \\
     H$_2$O & \change{18.13$-$18.23} & $3.24 \pm 0.10$ \\
     C$_2$H$_2$$^b$ & \change{13.55$-$13.76} & $8.40 \pm 0.22$\\
     HCN$^b$ & \change{13.83$-$14.07} & $12.76 \pm 0.14$\\
     CO$_2$$^b$ & \change{14.83$-$15.01} & $6.72 \pm 0.19$\\
     $^{13}$CO$_2$$^{b}$ &\change{15.39$-$15.43} & $<6$ \\
     OH$^{b}$ & \change{17.70$-$17.81} & $<9$ \\
    \hline
    \textbf{Notes.} & $^a$ Flux of the water complex emission at $17.22~\mu$m adopted in Fig.~\ref{fig:fluxes}. & \\ 
    & $^b$ Flux is estimated from the best-fit model flux due to line blending. & \\
    & \change{$''\left[ \,\,\right]''$ denote forbidden emission lines of ionized noble gases.} & \\
    \end{tabular}
    \label{tab:fluxes}
\end{table*}

\change{\section{Discussion}}
\label{sec:discussion}
\subsection{Why is the emission of hydrocarbons absent in J0438?} \label{sec:nohydrocarbons}
\change{\subsubsection{Line fluxes ratios}}
\label{Line fluxes ratios}
The slab modelling procedure described in Section~\ref{sec:slab_modelling} is employed primarily to distinguish between the blended contribution from multiple molecular emitters in the $13.4-18~\mu$m region and subsequently to calculate line fluxes of the detected species to compare with previous SST-IRS detections. To ensure a meaningful comparison, the line fluxes are determined by integrating the MIRI spectrum over the wavelength ranges used in previous SST-IRS studies. We note that since the emission of C$_2$H$_2$, HCN and CO$_2$ overlap with H$_2$O, the molecular fluxes of these three species are directly calculated from the synthetic spectrum and not from the MIRI spectrum as in \citet{Xie2023}. A summary of the line fluxes is reported in Table~\ref{tab:fluxes}.

\begin{figure*}
\centering
\includegraphics[width=5in,trim={0 0 1cm 2cm},clip]{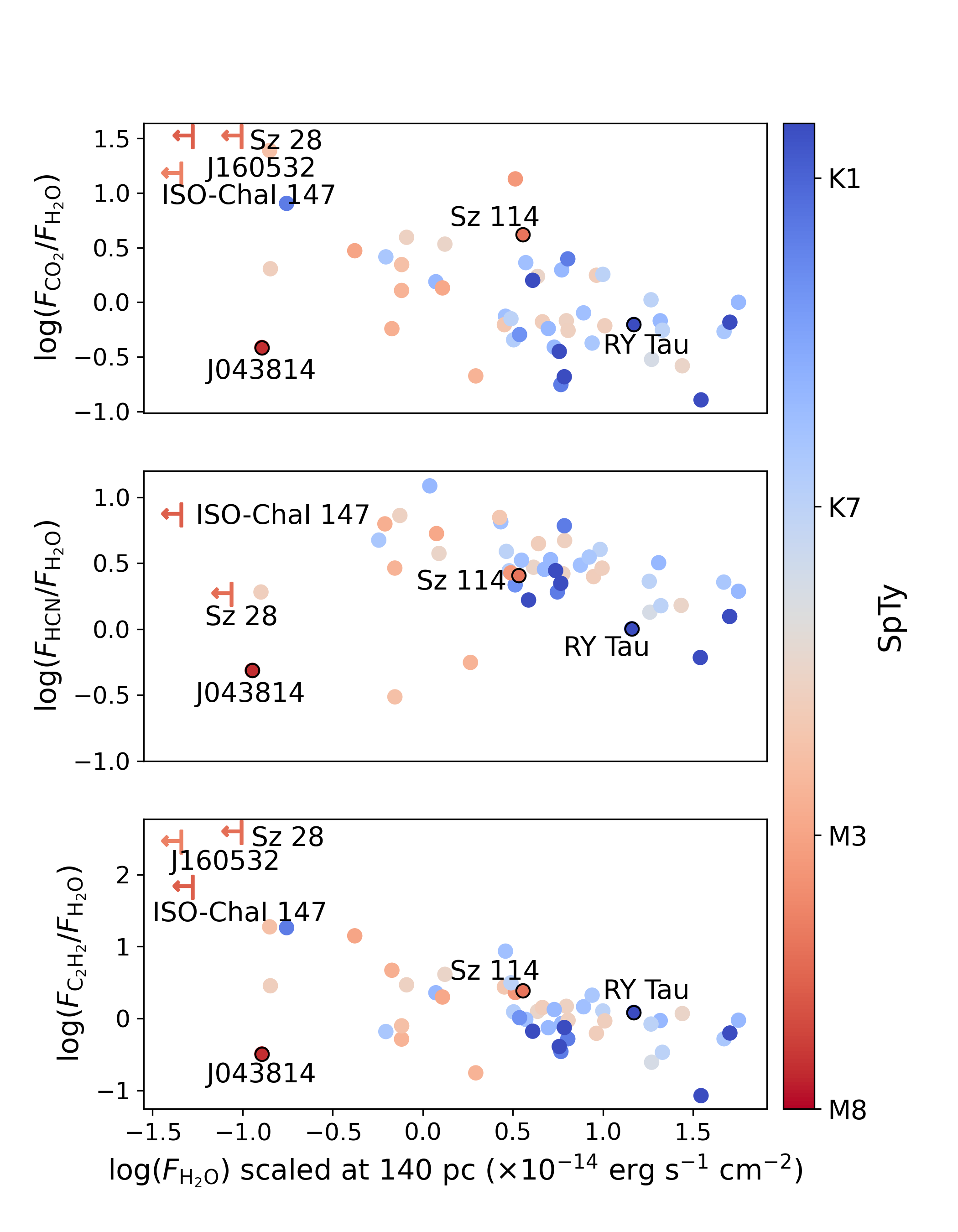}
\caption{\textbf{Comparison between CO$_2$, HCN and C$_2$H$_2$ flux ratios and H$_2$O flux for a sample of G, K, M-star disks.} The values for G, K and early M-star disks are \change{scaled from} \citet{banzatti2020,Xie2023} and for mid-to-late M-star disks from \citet{Pascucci2013} and \citet{Arabhavi2025b} and color-coded according to the spectral type (SpTy): from G (blue) to M (red). J160532 (M4.5), Sz~114 (M5), Sz~28 (M5.25), ISO-ChaI~147 (M.5.75) and J0438 (M7.25) are labelled. Arrows are upper limits. J1605 is not included in the middle panel because HCN was not detected in this disk \citep{tabone2023}. We also highlighted RY~Tau (K1), a T~Tauri disk with a disk inclination comparable to J0438 ($i\sim65^\circ$). For the comparisons, the water complex at $17.22~\mu$m is used and the fluxes are scaled to 140~pc using updated Gaia~DR3 distances collected in \citet{Manara2023}. Adapted from \citet{Xie2023}.}
\label{fig:fluxes}
\end{figure*}

\begin{figure*}
\centering
\includegraphics[width=5.5in]{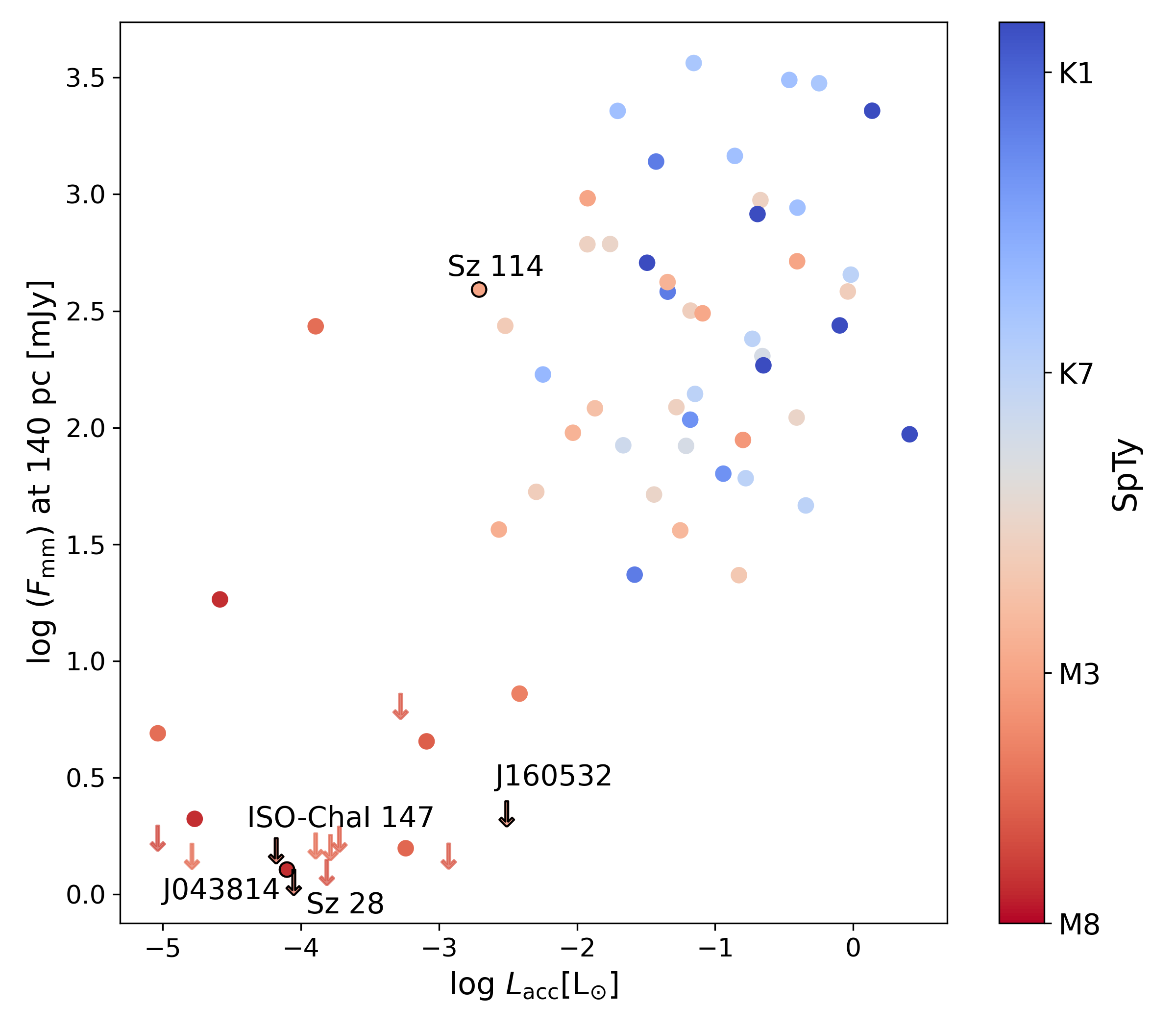}
\caption{\textbf{Comparison between flux at 0.89~mm scaled to 140~pc and accretion luminosity.} The value are color-coded according to the spectral type (SpTy) from G (blue) to M (red). The symbols corresponding to J160532 (M4.5), Sz~114 (M5), Sz~28 (M5.25), ISO-ChaI~147 (M5.75) and J0438 (M7.25) are surrounded by black edges. The values are from \citet{Pascucci2009a,Pascucci2016,banzatti2020,Xie2023,Franceschi2024}. Arrows represent upper limits. J0438 shows weak accretion luminosity and millimeter flux in agreement with late M-star disks. Adapted from \citet{Xie2023}.}
\label{fig:L_mmflux}
\end{figure*}

To place J0438 into context, we follow the same approach as in \citet{Xie2023} and plot in Fig.~\ref{fig:fluxes} the line flux ratios vs. the flux of the water complex at $17.22~\mu$m scaled at the distance of J0438 (i.e., 140~pc), and in Fig.~\ref{fig:L_mmflux} the accretion luminosity ($L_\mathrm{acc}$) vs the millimeter flux ($F_\mathrm{mm}$). As in \citet{Xie2023}, we include the sample of late M-type (M$7.25-$M4.5) stars from \citet{Pascucci2009a,Pascucci2013} and earlier-type (M$3-$G0) stars from \citet{banzatti2020}. Additionally, we plot available lower limits from recent JWST-MIRI observations of late M-type stars \citep{Arabhavi2025b}.

Figure~\ref{fig:fluxes} demonstrates that J0438 has a $17.22~\mu$m water flux that is fainter compared to disks around early M and K-type and to Sz~114, a large water-rich disk surrounding a late M-star (M5) \citep{Xie2023}. Additionally, J0438 appears to have consistently lower CO$_2$/H$_2$O, HCN/H$_2$O and C$_2$H$_2$/H$_2$O flux ratios with respect to disks around other late M-stars disks (e.g., J160532, ISO-ChaI~147, Sz~28; \citealt{tabone2023,Arabhavi2024,Kanwar2024}). Higher C$_2$H$_2$/H$_2$O ratios have been reconciled with high (i.e, $\geq0.8$) C/O ratios in the inner disks of late M-star disks \citep{Pascucci2013,tabone2023,Arabhavi2024,Kanwar2024} due to the production of hydrocarbons from carbon that is not locked up into CO or CO$_2$ (see Sect.~\ref{sec:nohydrocarbons} for more details). However, in the disk around J0438 neither CO (i.e., the identified CO absorption bands are of photospheric origin; Fig.~\ref{fig:BD_absorption}), nor hydrocarbons are detected except for C$_2$H$_2$, which suggests a lower elemental C/O, resembling that of early M-star disks.

\change{\subsubsection{Disk inclination}}
\label{inclination}
To check whether the disk configuration could prevent the detection of hydrocarbons in J0438 we ran a set of thermochemical models following the framework of \change{\citep{Kanwar2025}} and enabled the disk inclination to vary from 45$^{\circ}$ to 70$^{\circ}$. We found that the peak flux density of C$_2$H$_2$ decreases of approximately $33\%$ when the disk inclination changes from \textit{i}=45$^{\circ}$ to \textit{i}=70$^{\circ}$, which is well within our detection limit. This result made us confident that the inclination of our BD disk ($67^{\circ}-71^{\circ}$) is not responsible for the non-detection of hydrocarbons weaker than C$_2$H$_2$. Instead, in the case of J0438 (M7.25), MIRI appears to be probing a disk slowly evolving \textit{out} of a water-rich phase, an intermediate stage between the water-rich Sz~114 (M5) and the hydrocarbon-dominated J160532 (M4.5), Sz~28 (5.25) and ISO-ChaI~147 (M5.75) disks. The high spectral resolving power of MIRI MRS ($R\sim 3000$; \citealt{Labiano2021}) is revealing a diversity not only between disks around early and late spectral type stars but also within these two sub-categories due to factors such as different initial disk masses, mass accretion rates and presence/absence of sub-structures. 

\change{\subsubsection{Accretion luminosity and millimeter flux}}
\label{Line Fluxes}
To compare J0438 to other Class~II disks observed in the mid-infrared we have plotted in Figure~\ref{fig:L_mmflux} the accretion luminosity ($L_\mathrm{acc}$) and the millimeter flux ($F_\mathrm{mm}$) -- adopted as a diagnostic of the dust disk size \citep{Tripathi2017,Hendler2020} -- for the same sample shown in Figure~\ref{fig:fluxes}. The comparison between these two parameters was selected because \textit{Spitzer}-IRS surveys showed that the infrared water emission is positively correlated with $L_\mathrm{acc}$ whereas it is anti-correlated with $F_\mathrm{mm}$ (e.g., \citealt{banzatti2020}). 
The values for $L_\mathrm{acc}$ and $F_\mathrm{mm}$ are taken from the data table\footnote{\url{http://ppvii.org/chapter/15/}} of \citet{Manara2023} which collects measurements from several surveys (e.g., \citealt{Apai2005,Andrews2013,Pascucci2016,alcala2017,Ansdell2018,Ward-Duong2018}). As seen in Figure~\ref{fig:L_mmflux}, J0438 has a weak accretion luminosity, commensurate with the majority of late M-star disks targeted so far, in particular Sz~28 and ISO-ChaI~147. J0438 has also a comparatively weak $F_\mathrm{mm}$, advocating for a similar radial drift efficiency.   

\change{\subsubsection{Age}}
\label{Age}
Interestingly, the two late M-star disks showing water emission (Sz~114 and J0438) appear to be the youngest as J0438 belongs to the \change{$\sim0.6$~Myr} L1527 group in Taurus \citep{Luhman2023}, and Sz~114 is located in the Lupus~III sub-group ($\sim2.5~$Myr; \citealt{Galli2021}). On the other hand, the three hydrocarbon-dominated late M-star disks (J160532, ISO-ChaI~147 and Sz~28) could be older, as J160532 belongs to the Beta~Sco cluster ($\sim\!7.6~$Myr; \citealt{Ratzenbock2023}), ISO-ChaI~147 and Sz~28 are located in Chamaeleon~I with an estimated age older than the L1527 group in Taurus ($\sim\!3.8~$Myr; \citealt{Zucker2023,Ratzenbock2023}). Recent MIRI observations of a 30~Myr disk by \citet{Long2024} reveal a late-stage carbon-rich phase supporting this scenario. Although the age argument is valid for our comparison, one should note that at this stage of MIRI's exploration of VLMS/BD disks care should be taken when correlating the presence/absence of water emission with age, due to the large uncertainties in the age determination (see e.g., reviews by \citealt{Soderblom2014,Manara2023}).

\change{\subsubsection{Mass accretion rate}}
In the case of J160532, ISO-ChaI~147 and Sz~28, the supply of water-rich pebbles has plausibly stopped and the ongoing production of hydrocarbons from C-rich vapour resulted in an enhancement of the C/O ratio \citep{Kanwar2024}. On the other hand, for Sz~114 the pebble flux has not been consumed yet preserving an inner disk rich in water vapour \citep{Xie2023}. J0438 seems to fall between these two endpoints, with significantly fainter (i.e., approximately 2 orders of magnitude) water emission and lower mass accretion rate with respect to Sz~114 and no enhancement of hydrocarbons as in J160532, ISO-ChaI~147 and Sz~28, despite the late spectral type (M7.25). This may suggest that the hydrocarbon-rich phase for J0438 has yet to come, once the pebble flux will be exhausted.

Interestingly, the emission of several hydrocarbons was recently reported in the disk around a T~Tauri star ($1.1\,M_\odot$), DoAr~33 \citep{Colmenares2024}. The authors suggest that the presence of carbon-rich species in this disk is due to an unusually low mass accretion rate (i.e., $2.52\times10^{-10} M_\odot$ yr$^{-1}$; similar to that of late M-type stars) which prolongs the radial mixing timescales in the inner disk and enables a chemistry powered by carbon grain destruction to linger. A similar explanation (a very low mass accretion rate) could simply hold back the water reservoir in J0438 and slow down the development of a carbon-rich phase for this BD disk. 

\change{\subsubsection{Sub-structures}}
\citet{Mah2023} performed a series of 
simulations to explain the general trend in inner disk chemistry observed with SST and JWST so far using \texttt{chemcomp} \citep{Schneider2021}, a 1D disk evolution code that includes pebble drift, dust growth and ice sublimation at the major icelines (i.e., C, H$_2$O, CO$_2$, CH$_4$, CO). 
\citet{Mah2023} showed that, for VLMS disks presumably without deep gaps, the gaseous C/O ratio within the water iceline is initially sub-stellar due to the inward drift of water-rich pebbles and their sublimation. After the first 2~Myr, the C/O ratio increases to super-stellar values due to the loss of water vapour accreted onto the star and diffusion of C-rich vapour as a result of the sublimation of CH$_4$ ice. 

\begin{figure*}
\centering
\includegraphics[width=6in]{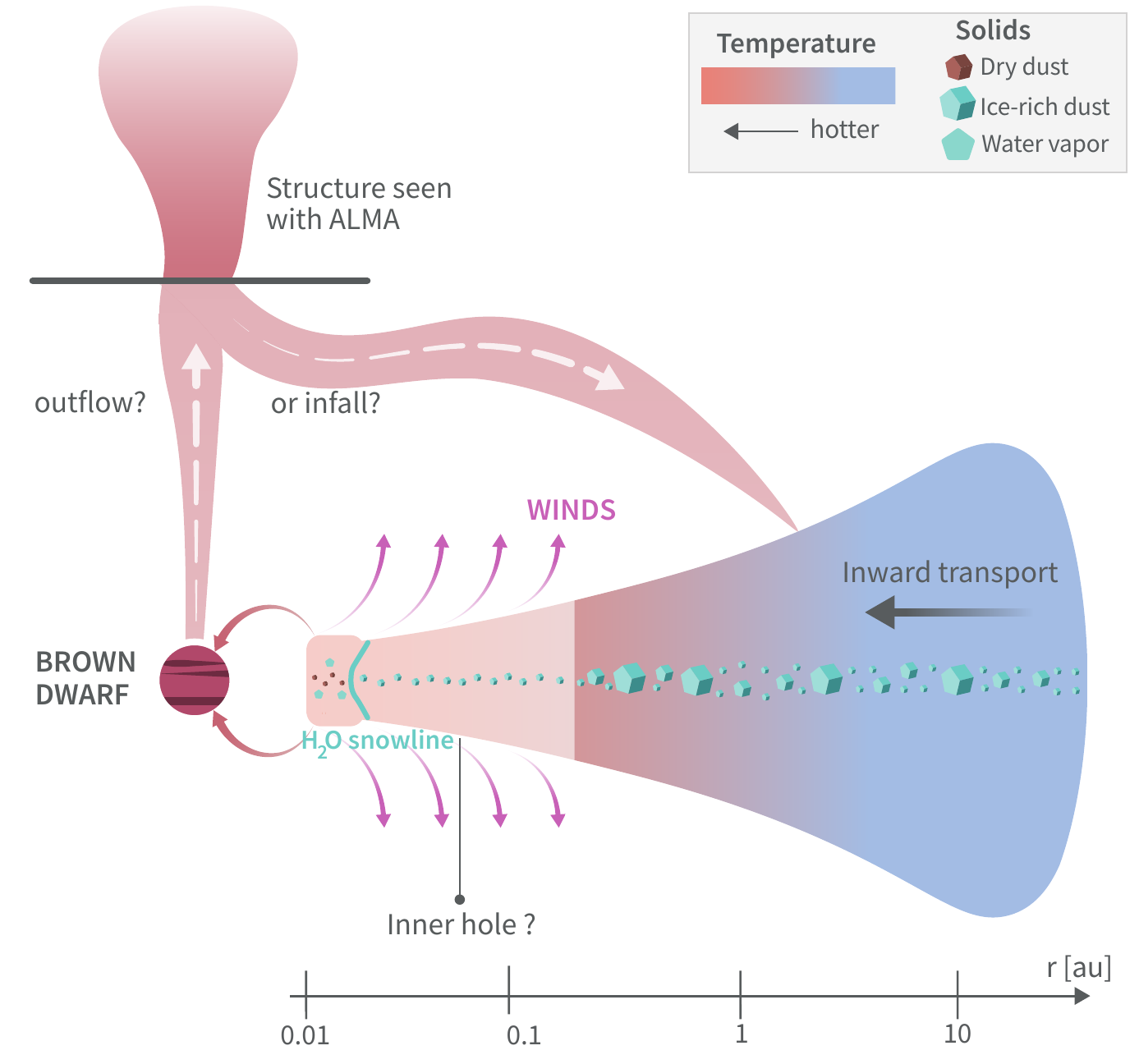}
\caption{\textbf{Schematic illustration of the inner disk around the young brown dwarf J0438}. The inner disk of J0438 is characterized by faint water vapour and \textit{not} by hydrocarbon-rich gas despite of the late spectral type (M7.25). The possible presence of an inner hole at $\sim0.28~$au reported by \citet{Luhman2007} and the low mass accretion rate ($\sim 10^{-11} M_\odot$~yr$^{-1}$; \citealt{Muzerolle2005}) are likely responsible for slowing down the inward drift of icy dust grains/pebbles and hence for prolonging a water-rich phase for this disk. \change{ALMA observations resolved extended emission from the BD disk suggesting ongoing outflowing/infalling activity}. Image credit: A.~Houge.}
\label{fig:cartoon}
\end{figure*}

We note that VLMS disks are characterized by shorter viscous timescales and faster radial drift with respect to T~Tauri disks \citep{pinilla2013,vanderMarel2023}, given smaller disks radii and the presumable lack of significant sub-structures. Additionally, icelines in VLMS are located further in, closer to the central star (e.g., \citealt{Mulders2015,Liu_Beibei_2019,Liu_Beibei_2020}). According to the formalism presented in \citet{Liu_Beibei_2019} (Eq.~19), we estimate the water iceline to lie at approximately $0.02~$au in the J0438 system (Fig.~\ref{fig:cartoon}). 

What made J0438 retain a water vapour reservoir? In contrast to Sz~114, J0438 does not have an extended mm-dust disk ($M_\mathrm{dust}$ equal to 1.28 vs 30~$M_\Earth$; \citealt{Manara2023,Ansdell2018}) explaining the fainter water emission, however akin to Sz~114, J0438 may host sub-structures \citep{Luhman2007,Huang2018,Jennings2022}. Several recent studies \citep{Kalyaan2023,Mah2024,Easterwood2024,Sellek2024} investigated what is the impact of gaps onto the water vapour abundance in the innermost disk regions. \citet{Mah2024} attributed the observed variations \textit{not} to the overall disk dust size (i.e., compact vs extended disk) but to the presence of gaps hindering the pebble flux. Importantly, they find that the effectiveness of gaps in acting as pebble barriers decreases if the gap forms late (i.e., $t_\mathrm{gap}\geq 0.1~$Myr) as by then most of the pebbles would \textit{already} populate the inner disk regions. 

Apart from the temporal component, the depth and the radial location of the gaps can also significantly alter the water enrichment. For disks around T~Tauri stars \citet{Kalyaan2023} found that the gap that is closest to the water iceline is the one that impacts the most the abundance of water vapour. Additionally, it has been shown that the inward transport of dust grains and pebbles results in an increased dust continuum optical depth in the inner disk which obscures the delivered water vapour \citep{Sellek2024,Houge2025}. Finally, although deep gaps can halt dust filtration, they do not completely block it \citep{Haugbolle2019}. The water reservoir can still be sustained by small, micron-sized dust particles well coupled to the gas leaking through planet-carved gaps as in the case of PDS~70 \citep{Perotti2023,Pinilla2024}. Interestingly, according to the SED and HST images modelling by \citet{Luhman2007} the disk around J0438 possibly hosts an inner hole at about 0.28~au $-$ and the water iceline is estimated to reside well inside it (Fig.~\ref{fig:cartoon}). This cavity may reduce the inward drift of pebbles which is already slowed down by the low mass accretion rate, and therefore, prolong the survival of a water reservoir. Dedicated high-resolution (sub-)millimeter observations of J0438 are needed to constrain the size of the inner hole and verify the presence of other potential sub-structures beyond the water iceline. Given that J0438 appears to be brighter and slightly larger than other VLMS/BD disks \citep{Hendler2017}, it is a prime candidate to explore the capability of (sub-)millimeter interferometers in studying the physical structure of disks around very low-mass star objects.    

\vspace{0.5cm}
\subsection{Is J0438 characterized by \change{outflowing/infalling} activity?} \label{sec:outflow}

\begin{figure*}
\centering
\includegraphics[width=6in]{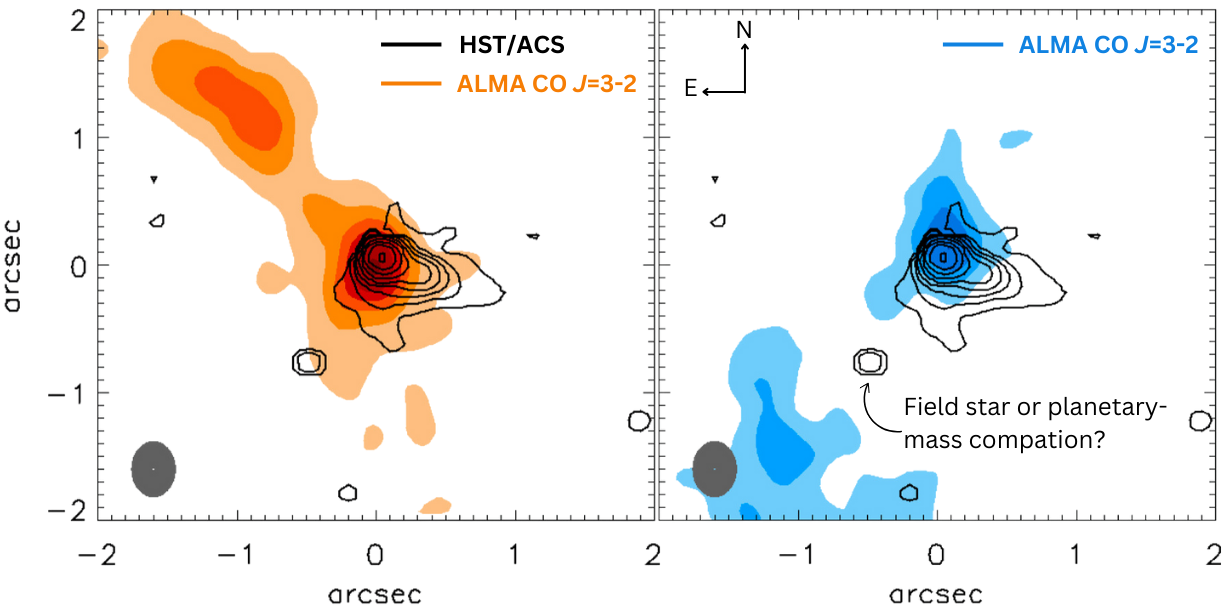}
\caption{\textbf{ALMA Band~7 observations of the J0438 disk}. CO ($J=3-2$) moment 1 maps highlighting the red-(\textit{left}) and blue-(\textit{right}) shifted emission with respect to the velocity of the central object. Boxes are $4\farcs0$ in size. The red and blue color scales represent the velocities with respect to the velocity of the central object between +0.4 to +1.7~km~s$^{-1}$ and $-0.4$ to $-1.7~$km~s$^{-1}$, respectively. The synthesized ALMA beam is marked as a grey ellipse. Black contours represent \change{archival HST/ACS scattered light observations (PID: 14212; PI: K. Stapelfeldt)} confirming the extended emission in the west direction previously reported by \citet{Luhman2007}.}
\label{fig:outflow}
\end{figure*}

\change{\subsubsection{Optical}}
Forbidden emission lines (i.e., O and S$^+$) have been identified in optical spectra of J0438 suggesting the presence of outflow activity in this system \citep{Luhman2004}. Follow-up HST/WFPC2 narrow- and broad-band imaging (F631N, F675W, F791W and F850LP) was performed in an attempt to detect resolved line emission from a jet or an outflow. However, no extended forbidden emission of atomic oxygen at $6300~\angstrom$ was observed in the narrow-band image using F631N \citep{Luhman2007}. Instead, broad-band imaging revealed the presence of elongated resolved emission in all three filters (F675W, F791W, and F850LP) reaching $0\farcs4 - 0\farcs5$ on one side and $\sim0\farcs1$ on the opposite side of the point source. The authors concluded that asymmetric bipolar extended emission, suggesting ongoing outflow activity, is detected in J0438. \change{More recent and deeper archival HST \change{ACS/WFC} observations (PID: 14212; PI: K. Stapelfeldt) were obtained using the F814W filter. The reduced dataset was downloaded from the Mikulski Archive for Space Telescopes (MAST) and the HST image was then sharpened by applying six iterations of the Lucy-Richardson algorithm \citep{Richardson1972} and using the nearby field star as the Point Spread Function (PSF). The resulting image (Figure~\ref{fig:outflow}) confirmed the elongated scattered light pattern previously observed by \citet{Luhman2007}.} 

\change{\subsubsection{Infrared}}
Moving to the infrared spectral regime, interestingly, among the BD disks observed with JWST-MIRI MRS in Cycle~1 so far \change{\citep{Arabhavi2025b}}, J0438 is the only one with [Ne$^{+}$] detection at $12.81~\mu$m. [Ne$^{2+}$] is tentatively detected (Fig.~\ref{fig:neon_zoomin}). [Ne$^{+}$] emission can act as a tracer of jets or probe the source of ionization (i.e., stellar EUV and/or X-ray photons) in the disk (e.g., \citealt{Herczeg2007,najita2009,vanBoekel2009}). One should note that [Ne$^{+}$] was previously observed in this system in 2009 with SST-IRS (\citealt{Pascucci2013}; Fig.~\ref{fig:neon_zoomin} of Appendix~\ref{app:variability}). Back then, a disk origin for the [Ne$^{+}$] emission in J0438 was suggested, with X-ray ionization as the main contributor of neon cations in the disk surface. 

Interestingly, when comparing the [Ne$^{+}$] peak line fluxes for J0438 obtained with both SST and JWST we observe line variability: the MIRI peak flux is approximately a factor of 2.5 brighter compared to the 2009 measurement obtained with IRS (Fig.~\ref{fig:neon_zoomin}). Beside J0438, the [Ne$^{+}$] line at $12.81~\mu$m was also detected for the BD 2MASS~J04442713+2512164 (M7.25), and tentatively, 2MASS~J04390163+2336029 (M6) \citep{Pascucci2013}. An outflow origin was attributed to 2MASS~J04442713+2512164 as its Ne$^{+}$ luminosity was similar to that of outflowing T~Tauri disks (Fig.~\ref{fig:L_Ne_L_acc}). The neon luminosity for J0438 measured with MIRI is still significantly weaker compared to 2MASS~J04442713+2512164 despite of the comparable accretion luminosity (Fig.~\ref{fig:L_Ne_L_acc}). 

To check for resolved extended emission indicative of winds or jets, we analysed further the MIRI MRS datacubes (see Appendix~\ref{app:extendedemission} for details on the adopted procedures). No clear patterns attributable to jets are detected such as the ones observed with JWST towards more massive disks \citep{Pascucci2024,Schwarz2024b}. However, the emission of two ortho-H$_2$ lines, the $S$(5) at $6.91~\mu$m and $S$(3) at $9.66~\mu$m appears to be \textit{marginally} extended. The same applies to [Ne$^{+}$] at $12.81~\mu$m which looks slightly more collimated than H$_2$.  
The $S$(1) line is extended but unresolved. All other main atomic, ionic and molecular species emitting in the MIRI bandwidth are also unresolved, including water. 

The results of this analysis are presented in Figures~\ref{fig:H2_extended} and \ref{fig:Ne_extended} of Appendix~\ref{app:extendedemission} and hint at most MIR emission tracing the disk, except for two H$_2$ lines and [Ne$^{+}$]. Interestingly, the emission of H$_2$ and [Ne$^{+}$] have a similar shape, suggesting that both species could be probing a wind. Alternatively, the emission of [Ne$^{+}$] could be tracing the disk surface and the H$_2$ lines both the disk surface and an uncollimated wind as they appear slightly more extended. \change{[Ne$^{+}$] does not seem} to be tracing a jet, as is often the case for more massive sources (e.g., \citealt{Arulanantham2024,Schwarz2024b}). We note that higher S/N MIRI exposures, and complementary NIRSpec observations targeting a plethora of jet and wind tracers (e.g., [Fe$^+$]), would enable a more in depth exploration of mass-loss events in this young brown dwarf system.   

\change{\subsubsection{Radio}}
We then moved to longer wavelengths to look for additional tracers and/or evidence for extended emission resembling the one observed with MIRI MRS. The forbidden atomic oxygen line at 63$~\mu$m probed with \textit{Herschel}-PACS remains undetected \citep{Hendler2017}. An initial search for water maser emission ($J=6_{16}–5_{25}$) with the 100 \!m Effelsberg radio telescope and for outflowing CO ($J=2-1$) emission with CARMA resulted in non-detections, pointing to negligible ongoing mass-loss in the system or simply to insufficient angular resolution \citep{Gomez2017,Phan-Bao2014}. 

The analysis of higher angular resolution ($\sim$0\farcs3) archival ALMA Band~7 ($340~$GHz) observations (PID: 2012.100743.S; PI: G. van der Plas; \change{see \citealt{Ward-Duong2018,Pinilla2017,kurtovic2021} for more details on the observational setup and the adopted data reduction parameters}) revealed spatially extended CO ($J=3-2$) emission with a velocity dispersion of approximately $\pm 2~$km~s$^{-1}$ from the brown dwarf. 
The CO emission is perpendicular to the line of sight and it does not follow a typical symmetric bipolar structure of the outflow. The blue- and red-shifted lobes are overlapping at the brown dwarf position; for clarity we present them separately in Figure~\ref{fig:outflow}. The red-shifted NE lobe extends up to approximately 280~au whereas the blue-shifted emission spans up to $420~$au. 
The SE lobe appears to be disconnected from the central object possibly due to the presence of an interfering planetary-mass companion robustly detected with a S/N $>5$ in the HST/ACS images but not observed in the ALMA dust continuum maps (Fig.~\ref{fig:outflow}). The point source is located 0\farcs49 east and 0\farcs86 south from J0438 and it is $6.6\pm 0.2~$mag fainter than the central object ($25.6~$mag vs $19~$mag at $0.8\,\mu$m). Further inspection to unveil its nature (i.e., a planetary-mass companion versus a field star) must be carried out. Additionally, future detailed modelling is needed to understand to what extent the emission traced by ALMA is possibly due to binary interactions \change{(e.g., \citealt{Kuruwita2020})} or a late-infall event (e.g., \citealt{Kueffmeier2023,Gupta2023}) given the lack of a bipolar pattern and the low gas velocities ($\pm 2~$km~s$^{-1}$). Follow-up observations combining a set of compact and very extended ALMA configurations in tandem with longer integration times would provide firmer constraints on the extended  emission that characterizes this BD system. 

\vspace{0.5cm}
\section{Summary and outlook} \label{sec:conclusions}
We present new JWST-MIRI MRS observations of the highly inclined ($i\!\sim\!70^\circ$) disk around the brown dwarf (M7.25) J0438. We pair the new data with complementary HST/ACS and ALMA Band~7 data to obtain new insights into this intriguing system. 

\begin{enumerate}
    \item The JWST MIRI-MRS spectrum of J0438 resembles that of a disk around a T~Tauri star with a C/O ratio below unity. The volatile reservoir is O-dominated, composed of H$_2$O, CO$_2$ and, contrarily to most VLMS and BD disks investigated so far, of only one hydrocarbon (i.e., C$_2$H$_2$). H$_2$ and HCN lines are also clearly identified as well as atomic and ionic species such as [Ne$^{+}$], and [Ar$^{+}$]. Amorphous silicates are seen in both absorption and emission due to the $\sim70^\circ$ disk inclination while prominent ice absorption bands are not observed, except for a weak CO$_2$ ice band which is tentatively detected.
    \item The comparison between the JWST spectrum taken in 2023 and two SST-IRS epochs from 2007 and 2009 reveals dust continuum and line variability. The continuum variability can be ascribed to a \textit{seesaw} behaviour, as the emission at shorter and longer wavelengths varies inversely due to dynamical changes of the inner disk. The line variability is clearly observed for [Ne$^{+}$] and the two H$_2$ lines which were previously detected with \textit{Spitzer}-IRS \citep{Pascucci2013} as well as for the H$_2$O complex tentatively detected with SST-IRS at $18.17~\mu$m and now confirmed with MIRI.
    \item The inspection of the JWST data cubes shows marginally extended emission for the H$_2 \, \, S$(1), $S$(3) and $S$(5) lines, as well as of [Ne$^{+}$], for the first time in a BD disk. [Ne$^{2+}$] emission is tentatively detected and unresolved. Interestingly, the emission patterns of H$_2$ and [Ne$^{+}$] are similar and are not attributable to a jet, but plausibly to a disk wind and/or to the disk surface. 
    \item  HST/ACS images report extended scattered light emission spanning in the western direction and reveal the robust detection of a point source at 0\farcs49 east and 0\farcs86 south from the brown dwarf. With our current data is not possible to disentangle whether this is a field star or a planetary-mass companion and further work is required to confirm its nature. 
    \item ALMA Band~7 observations show extended CO ($J=3-2$) emission spanning up to $420~$au from the brown dwarf with a velocity dispersion of $\pm 2$~km~s$^{-1}$, perpendicular to the line of sight. The low CO velocities and the lack of a bipolar pattern makes it challenging to attribute the extended emission to an outflow. A late-infall event cannot be ruled out at this stage without dedicated modelling.  
    \item Overall, these set of observations reveal a dynamic Class~II brown dwarf disk evolving out of a O-dominated phase $-$ an intermediate stage between the extremely water-rich Sz~114 (M5) disk and the C-dominated late-M star disks targeted with JWST so far. This is related to multiple factors, and most notably to a relatively young age ($\sim0.6~$Myr), to a low mass accretion rate ($\sim\!10^{-11}M_\odot$ yr$^{-1}$) and to the possible presence of an inner hole which is depleting the icy pebble reservoir at a slower pace compared to similar disks lacking significant inner disk clearing. 
\end{enumerate} 

The high spectral resolving power of MIRI MRS is unveiling a \textit{diversity} not only between disks around early and late spectral type stars but also within these two sub-categories due to variables such as different ages, initial disk masses, mass accretion rates, presence of deep sub-structures and the time of the last late-infall/replenishment event. As these JWST observations progress, it will be crucial to observe and analyze an unbiased sample of very low-mass stars and brown dwarf systems to determine what of the aforementioned factors impact planet formation in their inner disks the most. Lastly, we emphasize the value of pairing observations from multiple facilities to build a comprehensive understanding of protoplanetary disks (e.g., \citealt{Perotti2024}), especially of the poorly studied very-low stellar/brown dwarf regimes akin J0438.  

\section{Acknowledgments}
The authors thank the reviewer for the constructive comments. They also thank A. Houge for drawing the illustration in Figure~\ref{fig:cartoon}, 
P.~Hauschildt for providing the model atmosphere spectrum, C.~Xie for sharing the data to produce Figures~3 and 4, G. van der Plaas and K. Stapelfeld for enabling us to present results from their ALMA and HST programs and P.~Pinilla, A.~Scholz, M.~Benisty, H.~Jiang and A.~Johansen for constructive discussions on J0438. 
This work is based on observations made with the NASA/ESA/CSA \textit{James Webb} Space Telescope. The data were obtained from the Mikulski Archive for Space Telescopes at the Space Telescope Science Institute, which is operated by the Association of Universities for Research in Astronomy, Inc., under NASA contract NAS 5-03127 for JWST. In particular, the JWST-MIRI MRS observations are associated with the European MIRI GTO program MINDS (PID: 1282, PI: Th. Henning) with visit number 43. \change{The specific observations analyzed can be accessed via \dataset[doi:10.17909/789s-qb11]{https://doi.org/10.17909/789s-qb11}.} The following National and International Funding Agencies funded and supported the MIRI development: NASA; ESA; Belgian Science Policy Office (BELSPO); Centre Nationale d’Etudes Spatiales (CNES); Danish National Space Centre; Deutsches Zentrum fur Luft- und Raumfahrt (DLR); Enterprise Ireland; Ministerio De Econom\'ia y Competividad; Netherlands Research School for Astronomy (NOVA); Netherlands Organisation for Scientific Research (NWO); Science and Technology Facilities Council; Swiss Space Office; Swedish National Space Agency; and UK Space Agency. \change{This work is based on archival data obtained with the NASA Infrared Telescope Facility, which is operated by the University of Hawaii under a contract with the National Aeronautics and Space Administration.}
This research is based on observations made with the NASA/ESA Hubble Space Telescope obtained from the Space Telescope Science Institute, which is operated by the Association of Universities for Research in Astronomy, Inc., under NASA contract NAS 5–26555. These observations are associated with program 14212 (PI: K. Stapelfeldt).
The \textit{Spitzer}-IRS spectra are part of program PID: 50799 (SH, PI: G. Herczeg), and PID: 40302 (SL, LL; PI: J.R. Houck). This paper makes use of the following ALMA data: ADS/JAO.ALMA\#2012.1.00743.S (PI: G. van der Plas). ALMA is a partnership of ESO (representing its member states), NSF (USA), and NINS (Japan), together with NRC (Canada), NSC and ASIAA (Taiwan), and KASI (Republic of Korea), in cooperation with the Republic of Chile. The Joint ALMA Observatory is operated by ESO, AUI/NRAO, and NAOJ. G.P. gratefully acknowledges support from the Max Planck Society and from the Carlsberg Foundation, grant CF23-0481. A.C.G. acknowledges support from PRIN-MUR 2022 20228JPA3A “The path to star and planet formation in the JWST era (PATH)” funded by NextGeneration EU and by INAF-GoG 2022 “NIR-dark Accretion Outbursts in Massive Young stellar objects (NAOMY)” and Large Grant INAF 2022 “YSOs Outflows, Disks and Accretion: towards a global framework for the evolution of planet forming systems (YODA)”. E.v.D. acknowledges support from the ERC grant 101019751 MOLDISK and the Danish National Research Foundation through the Center of Excellence ``InterCat'' (DNRF150). T.H. and K.S. acknowledge support from the European Research Council under the Horizon 2020 Framework Program via the ERC Advanced Grant Origins 83 24 28. I.K., A.M.A., and E.v.D. acknowledge support from grant TOP-1 614.001.751 from the Dutch Research Council (NWO). I.K., J.K., and T.K. acknowledge funding from H2020-MSCA-ITN-2019, grant no. 860470 (CHAMELEON). B.T. is a Laureate of the Paris Region fellowship program, which is supported by the Ile-de-France Region and has received funding under the Horizon 2020 innovation framework program and Marie Sklodowska-Curie grant agreement No. 945298. V.C. acknowledges funding from the Belgian F.R.S.-FNRS. D.G. thanks the Belgian Federal Science Policy Office (BELSPO) for the provision of financial support in the framework of the PRODEX Programme of the European Space Agency (ESA). D.B. and M.M.C. has been funded by Spanish MCIN/AEI/10.13039/501100011033 grants PID2019-107061GB-C61 and No. MDM-2017-0737. M.T. and M.V. acknowledge support from the ERC grant 101019751 MOLDISK. 

\vspace{5mm}
\facilities{JWST, SST, IRTF}

\software{\change{{Astropy \citep{astropy:2013, astropy:2018, astropy:2022}}, SciPy \citep{2020SciPy-NMeth}, NUMPY \citep{2020NumPy-Array}}, MATPLOTLIB \citep{Hunter2007}, SpectRes \citep{carnall2017}, VIP \citep{GomezGonzalez2017,Christiaens2023}, emcee \citep{emcee2013}.
          }


\clearpage

\appendix

\section{Dust continuum subtraction}
\label{app:cont_sub}

\begin{figure}[ht!]
\centering
\includegraphics[width=\hsize]{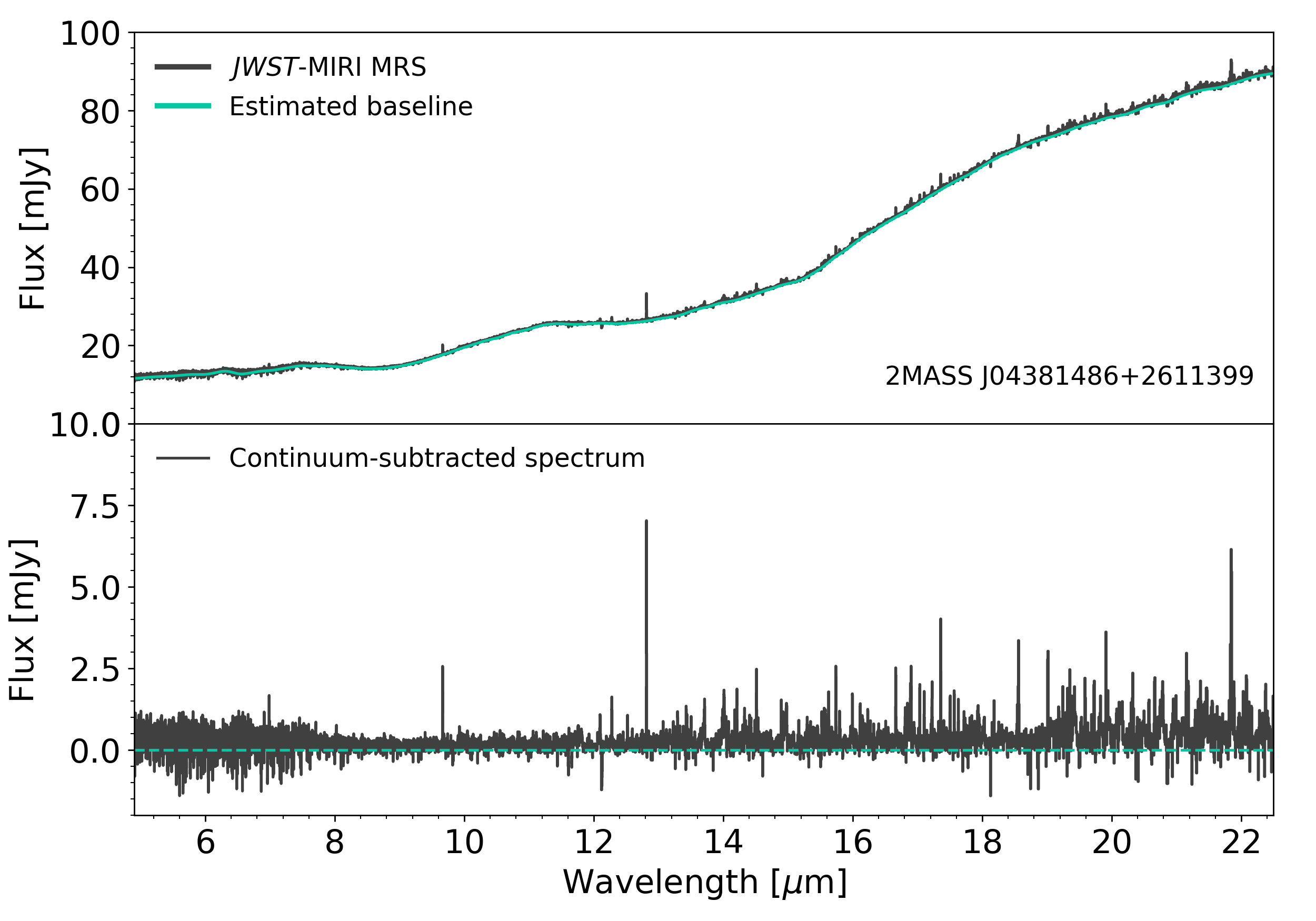}
\caption{\textbf{Dust continuum subtraction.} \textit{Top:} JWST-MIRI MRS spectrum of the disk around J0438 (black). Continuum baseline \change{(green)} estimated following the procedure in Sec.~\ref{sec:slab_modelling}. \textit{Bottom:} The continuum-subtracted spectrum (black). The dashed \change{green} line is used as a reference for the zero flux level.
\label{fig:cont_sub}}
\end{figure}

\begin{figure}
\centering
\includegraphics[width=\hsize]{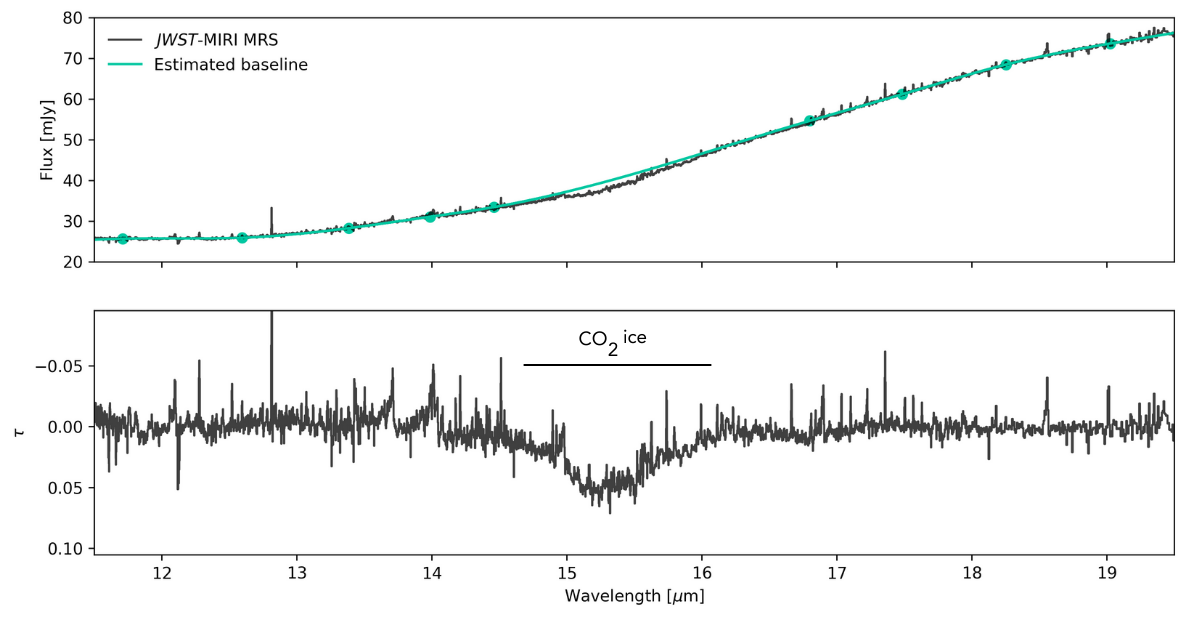}
\caption{\textbf{Tentative CO$_2$ ice absorption band}. \textit{Top:} JWST-MIRI MRS spectrum of the disk around J0438 (black). \change{Estimated baseline (green) obtained using the continuum points highlighted as green dots.} \textit{Bottom:} The continuum-subtracted spectrum (black) in an optical depth ($\tau$) scale revealing a weak CO$_2$ ice absorption band.}
\label{fig:co2_ice}
\end{figure}

The selected continuum points are displayed as green dots and the interpolated continuum is shown as a green line.

\clearpage

\section{Absorption bands from the brown dwarf} 
\label{app:stellar_contamination}
\begin{figure}[hb!]
\plotone{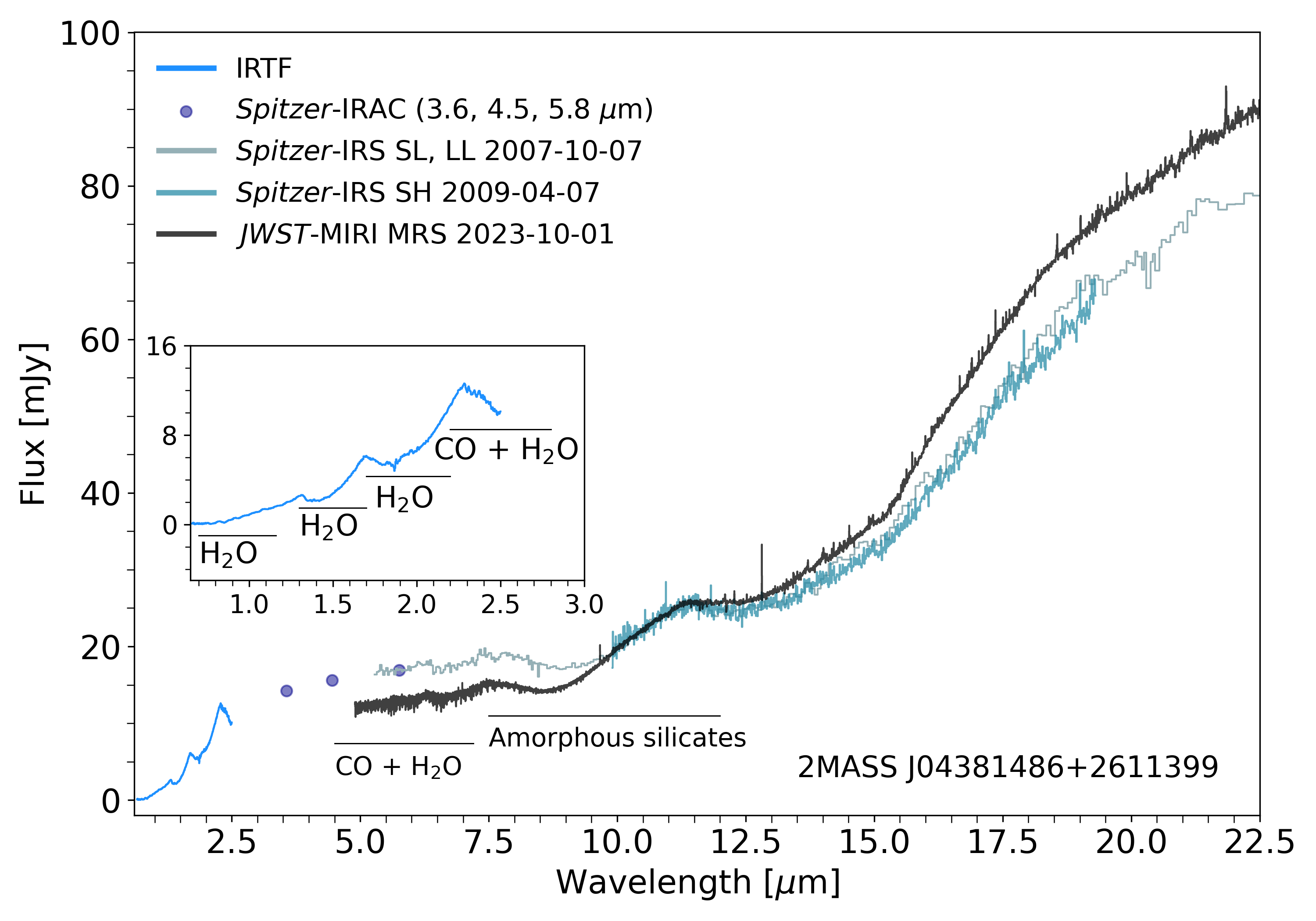}
\caption{\textbf{Near- and mid-IR spectra of the young brown dwarf J0438.} Near-infrared IRTF (light blue; \citealt{Luhman2007}), \textit{Spitzer}-IRAC photometry (blue points), \textit{Spitzer}-IRS SL, LL (grey; PID: 40302; PI: J.~R. Houck), \textit{Spitzer}-IRS SH (teal; \citealt{Pascucci2013}) and JWST-MIRI MRS (dark blue; this work). The inset shows a zoom-in of the IRTF data. The prominent H$_2$O, CO and amorphous silicate absorption bands are labelled.}
\label{fig:nir_mir}
\end{figure}

\begin{figure}
\centering
\includegraphics[width=\hsize]{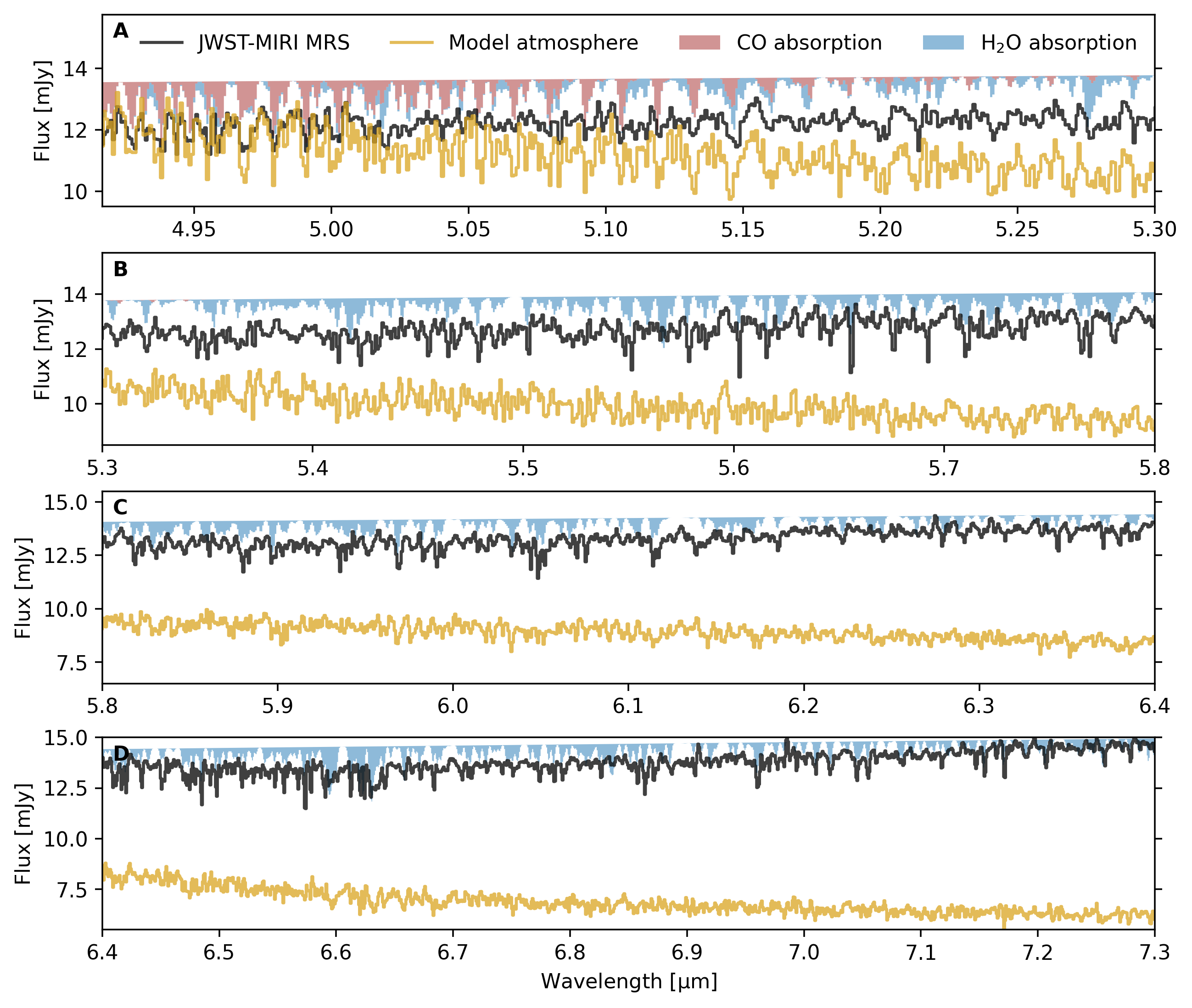}
\caption{\textbf{Absorption bands from the photosphere of the brown dwarf}. Comparison between the JWST-MIRI MRS spectrum of J0438 (black) and the PHOENIX model atmosphere spectrum (gold; \citealt{Husser2013}) described in Section~\ref{app:stellar_contamination}. For illustration purposes, slab models highlighting the absorption bands of H$_2$O (light blue) and CO (light red) are also shown.}
\label{fig:BD_absorption}
\end{figure}

The near-infrared IRTF spectrum of J0438 presents broad absorption bands of H$_2$O and CO originating from the brown dwarf (Figure~\ref{fig:nir_mir}). Similarly, the MIRI spectrum is dominated by absorption features of gas H$_2$O and CO below $<7.5~\mu$m. Figure~\ref{fig:BD_absorption} shows a comparison between the MIRI spectrum, a PHOENIX \citep{Husser2013} synthetic spectrum ($R=1-3 \times 10^6$) provided by P.~Hauschildt (personal communication, 2023) and slab absorption spectra of H$_2$O and CO. The parameters adopted to produce the model atmosphere spectrum are an effective temperature $T_\mathrm{eff} = 3000~$K, surface gravity log($g$) = 4.5 and solar metallicity. One should note that the PHOENIX model has not been tailored for J034814. However, the selected effective temperature well matches the value (i.e., 3100~K) determined by \citet{Scholz2006}. At wavelengths $<7.5~\mu$m the central object contributes significantly to the total flux.  Only longward of $\sim 6.5~\mu$m the brown dwarf contribution decreases to less than $50\%$ and the disk contribution starts to dominate. CO absorption bands are identified below $5~\mu$m in the PHOENIX synthetic spectrum as well as in the MIRI data. Any potential emission belonging to the CO fundamental band originating from the disk is unfortunately too weak to be detected. JWST-NIRSpec high-resolution grating (G395H) observations of J0438 probing most of the CO fundamental should be pursued to understand to what extent the disk contributes to the overall flux in that spectral region. Moving to slightly longer wavelengths, Figure~\ref{fig:nir_mir} shows that several H$_2$O absorption features from the brown dwarf are also detected. 

\begin{figure}[ht!]
\centering
\includegraphics[width=5in]{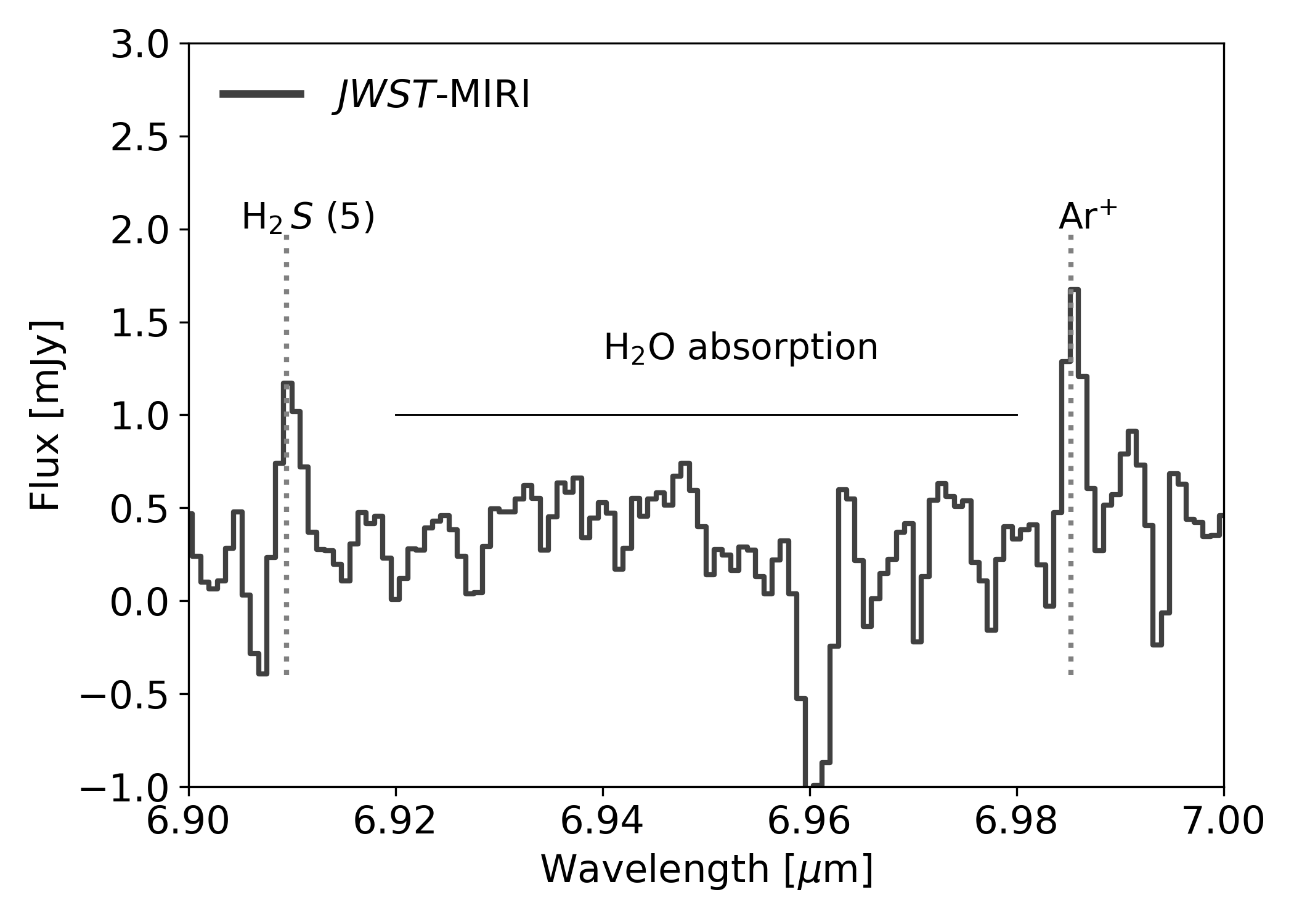}
 \caption{\textbf{H$_2\,S$(5) and [Ar$^{+}$] emission in J0438.} JWST-MIRI MRS continuum subtracted spectrum showing the emission of molecular hydrogen and ionized argon among H$_2$O absorption bands from the photosphere of the BD (see Fig.~\ref{fig:BD_absorption}).}
\label{fig:argon}
\end{figure}

\clearpage

\section{Line variability} 
\label{app:variability}

\begin{figure}[ht!]
\centering
\includegraphics[width=4.7in]{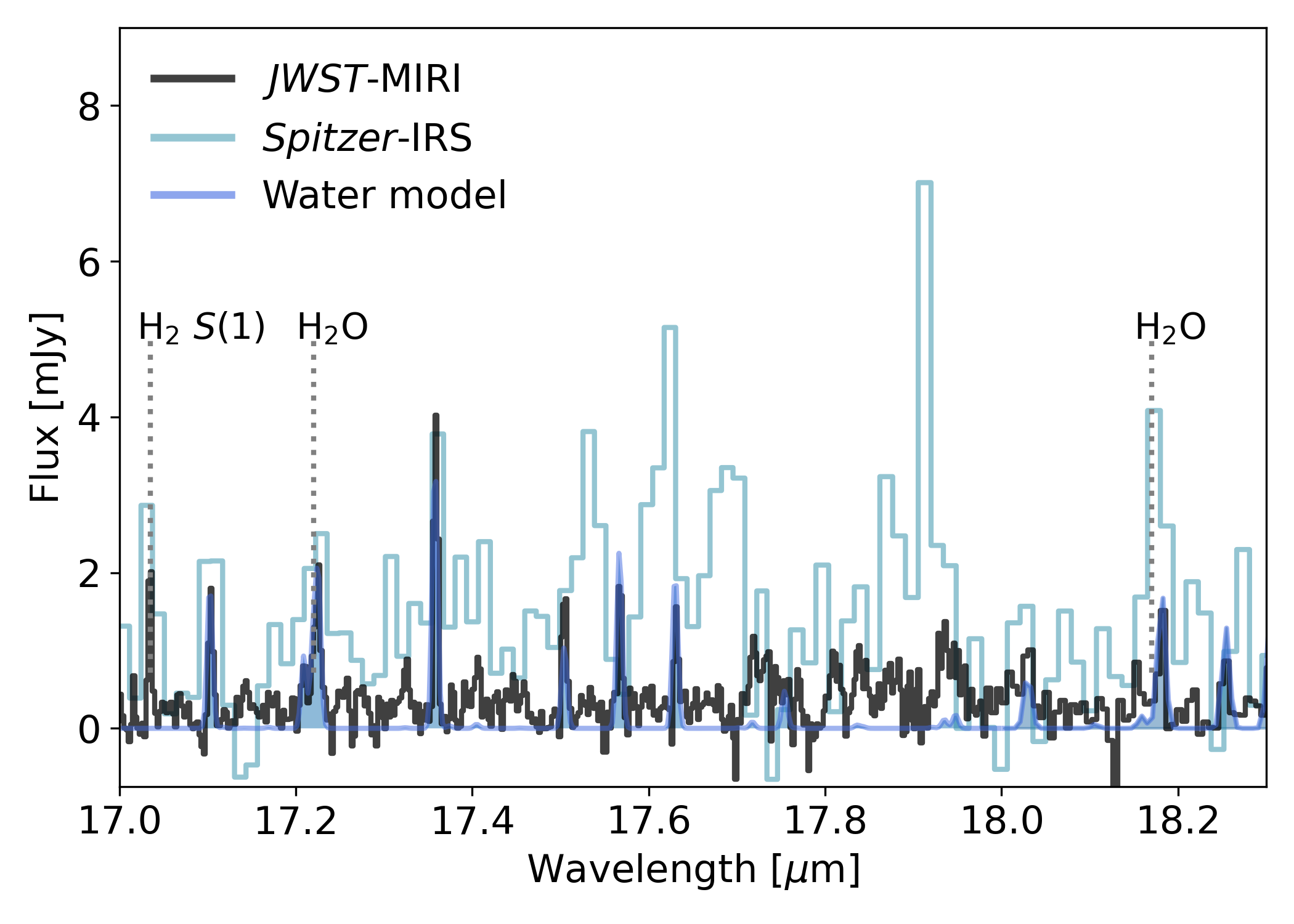}
 \caption{\textbf{Water and H$_2\,S$(1) emission in J0438.} Comparison between the H$_2$O lines at $17.22~\mu$m and $18.17~\mu$m tentatively identified in \textit{Spitzer}-IRS SH (teal) by \citet{Pascucci2013} and in the JWST-MIRI MRS (black) continuum-subtracted spectra. For clarity, a water slab model is also plotted (light blue). Line variability is observed for the H$_2\,S$(1) line as well as for the water complexes at $17.6~\mu$m and $18.17~\mu$m.}
\label{fig:water_H2_zoomin}
\end{figure}

\begin{figure}[hb!]
\centering
\includegraphics[width=\hsize]{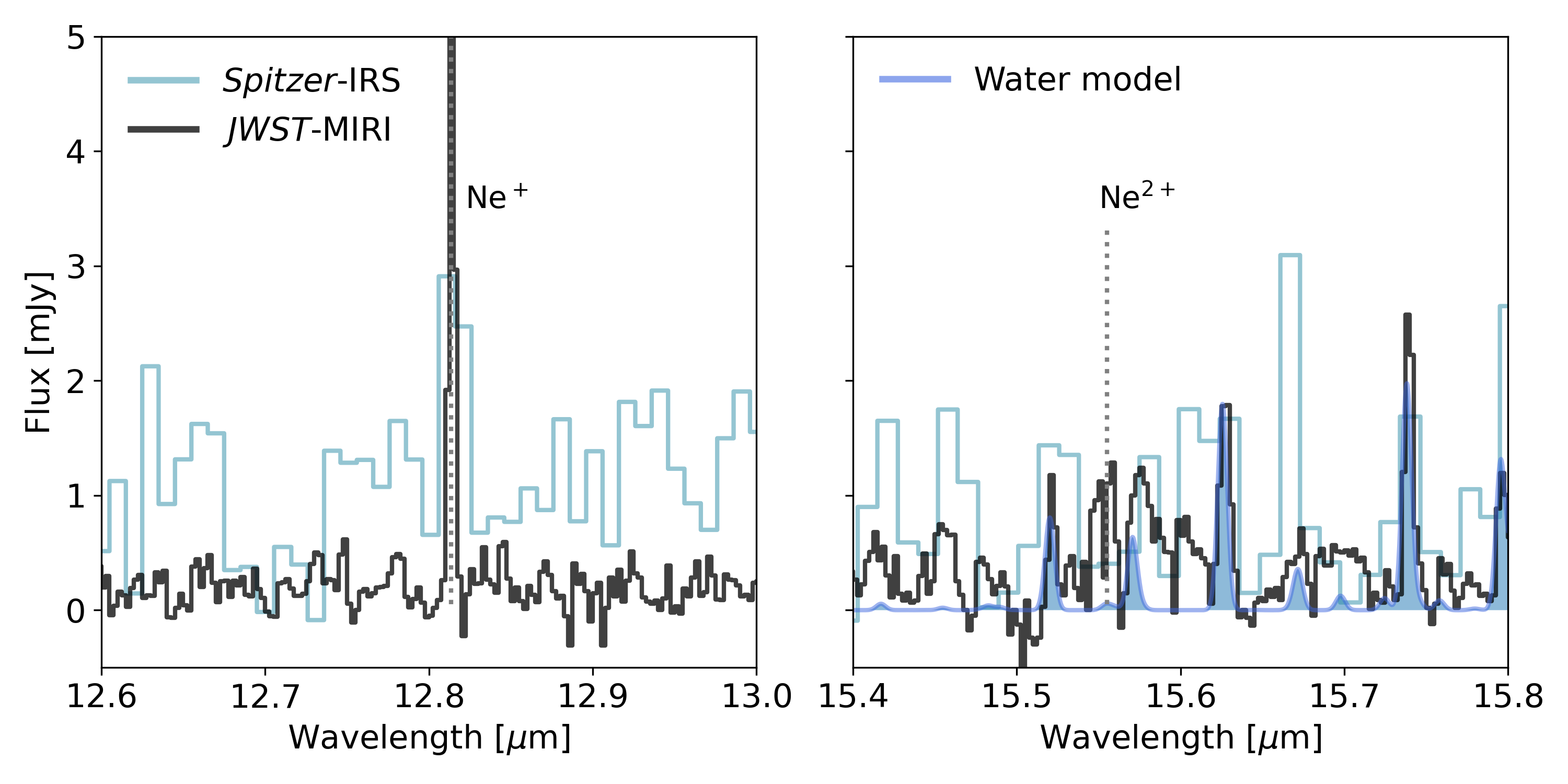}
 \caption{\textbf{Neon emission in J0438.} Comparison between the [Ne$^{+}$] (\textit{left}) and [Ne$^{2+}$] (\textit{right}) lines at $12.81~\mu$m and $15.56~\mu$m in the SST-IRS SH (teal) and JWST-MIRI MRS (black) continuum-subtracted spectra. For clarity, a water slab model is also plotted (light blue). Line variability is observed for [Ne$^{+}$], whereas the emission of [Ne$^{2+}$] is only tentatively detected.}
\label{fig:neon_zoomin}
\end{figure}

\clearpage

\section{Extended emission}
\label{app:extendedemission}
\begin{figure*}[hb!]
\centering
\includegraphics[width=\hsize]{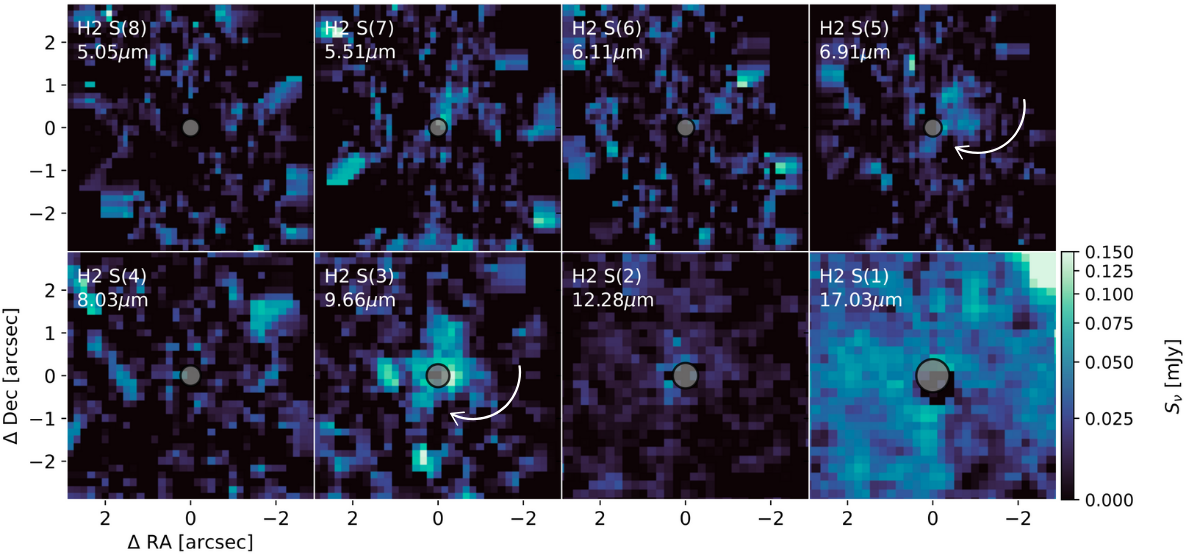}
\caption{\textbf{Moment zero maps of the rotational H$_2$ lines in the MIRI MRS wavelength range.} Marginal extended emission (white arrows) is resolved for $S$(3) and $S$(5). The $S$(1) line is spatially extended but unresolved. The \change{star} represents the position where the emission originated by a point source was subtracted.} 
\label{fig:H2_extended}
\end{figure*}

\begin{figure}
\centering
\includegraphics[width=5in]{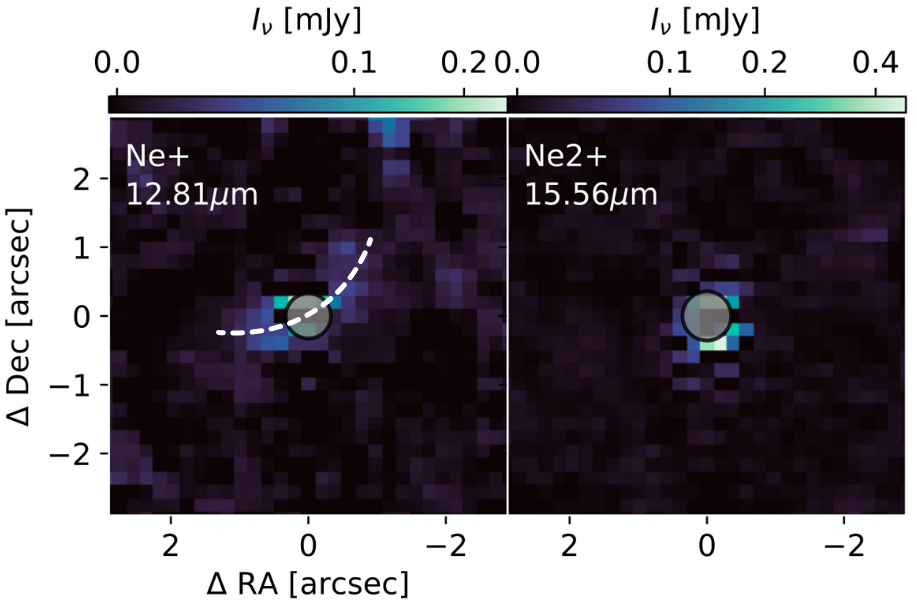}
\caption{\textbf{Moment zero maps of the neon lines in the MIRI MRS wavelength range.} The emission of [Ne$^{+}$] (\textit{left}) appears to be marginally extended (white dotted line). The [Ne$^{2+}$] line (\textit{right}) is unresolved. The grey circle represents the size of the full width at half maximum (FWHM) of the PSF at the position where the emission originated by a point source was subtracted.} 
\label{fig:Ne_extended}
\end{figure}

To check whether the molecular and ionic emission detected towards J0438 was extended, we first extracted spectra at the wavelength of the lines of interest using different aperture sizes (i.e., circular apertures with radii equal to 0\farcs5, 1\farcs0 and 2\farcs5). This preliminary approach is useful to infer the presence of extended emission towards faint objects like J0438. These initial results revealed signs of extended emission of three ortho-H$_2$ lines S(1), S(3) and S(5) and of [Ne$^{+}$]. We then continued our analysis following the prescription described in Kurtovic et al. (in prep.). Briefly, we normalize each image in a cube to the peak flux. Then, for each channel image, we approximate an empirical point spread function (PSF) by taking the median image of the ten surrounding channel images, but excluding the four that are closest in wavelength. This approximated PSF is then scaled to match the same peak flux as the channel image of interest, and it is subtracted to remove the emission originated by a point source. The remaining emission in the channel corresponds to all the flux which does not emit as the subtracted point source, therefore being extended in origin. These channels with a subtracted point source are then added together to produce a moment~0 map, as shown in Fig.~\ref{fig:H2_extended} for the rotational H$_2$ lines, and Fig.~\ref{fig:Ne_extended} for the ionized neon. These moment~0 maps have a flux of 0 at the position of the source, by construction. 

\begin{figure}
\centering
\includegraphics[width=4.5in, trim={0 0 0 0},clip]{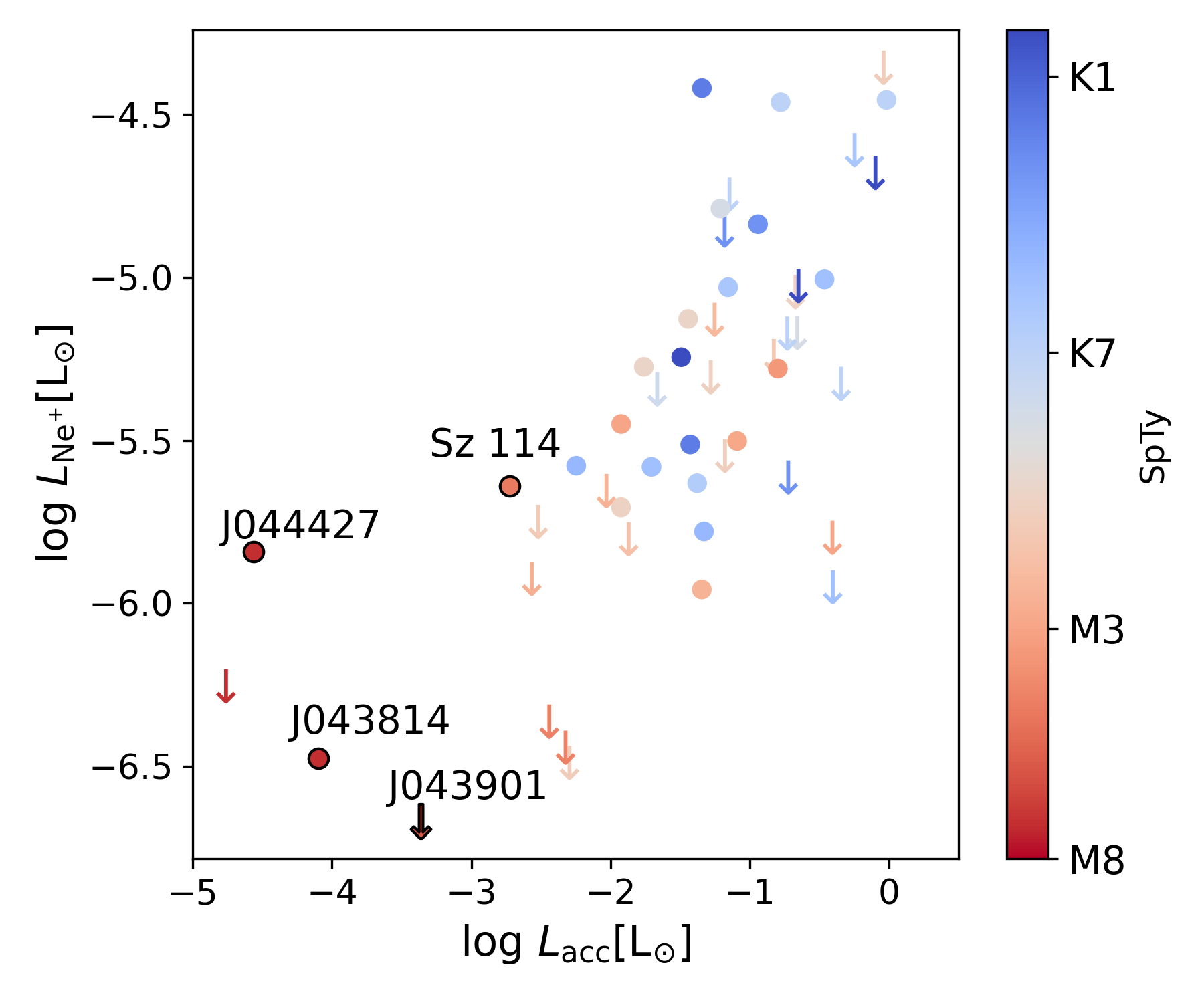}
\caption{\textbf{[Ne$^{+}$] luminosity versus accretion luminosity for a sample of late M-type stars and earlier type stars.} [Ne$^{+}$] luminosities are computed from SST-IRS and recent JWST-MIRI observations \citep{Guedel2010,Pascucci2013,Rigliaco2015,Xie2023}. All downward arrows represent  $3\sigma$ upper limits except for J043901, which is reported as a tentative ($2\sigma$) detection in \citet{Pascucci2013}. For the late M-type stars, only three show clear [Ne$^{+}$] emission: Sz~114 (M5), J044427 (M6) and J0438 (M7.25).} 
\label{fig:L_Ne_L_acc}
\end{figure}

\clearpage

\section{Slab modelling results}
\label{app:correlations}

\begin{figure}[ht!]
\centering
\includegraphics[width=\hsize]{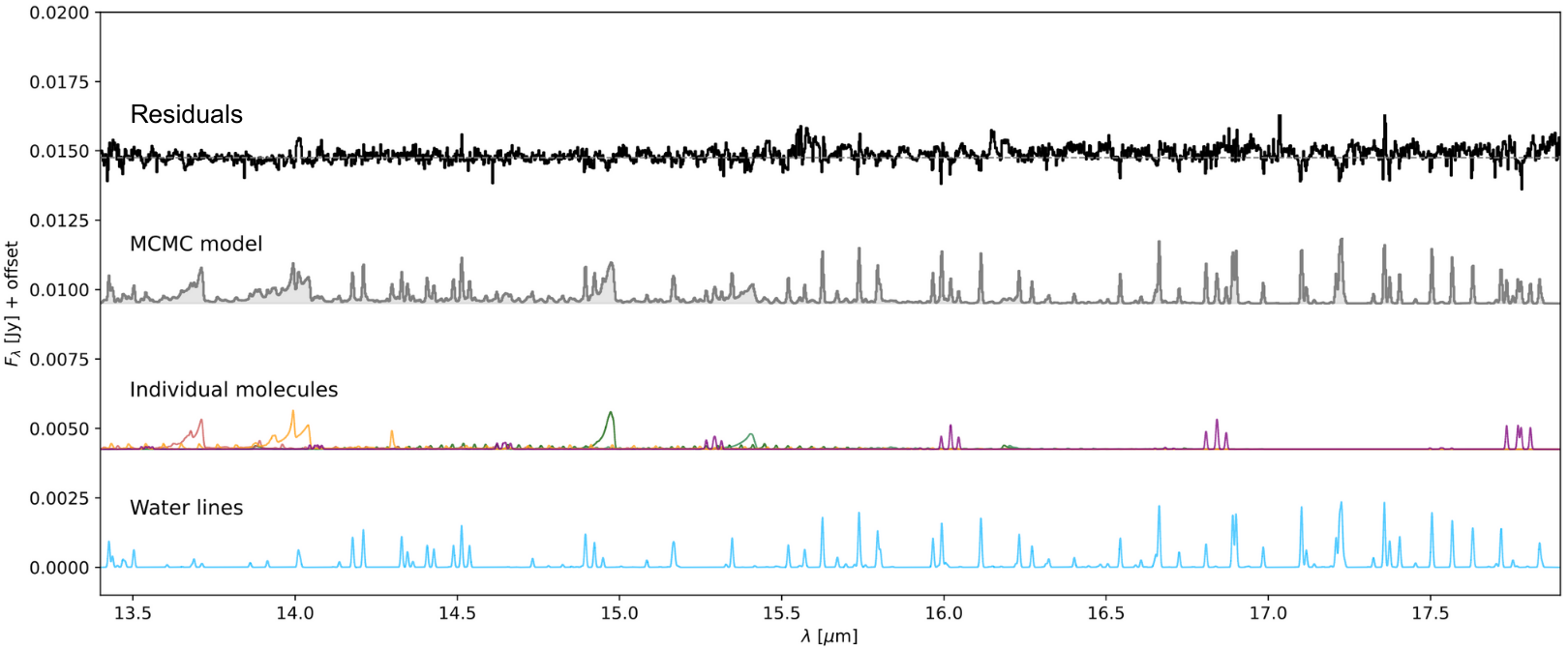}
\caption{\textbf{Residuals of the slab modelling (black).} Total MCMC model (grey), individual molecules (various colours), water lines (blue). We refer the reader to Figure~\ref{fig:slabs} for identification of the individual species.}
\label{fig:residuals}
\end{figure}

\begin{figure}[ht!]
\centering
\includegraphics[width=\hsize]{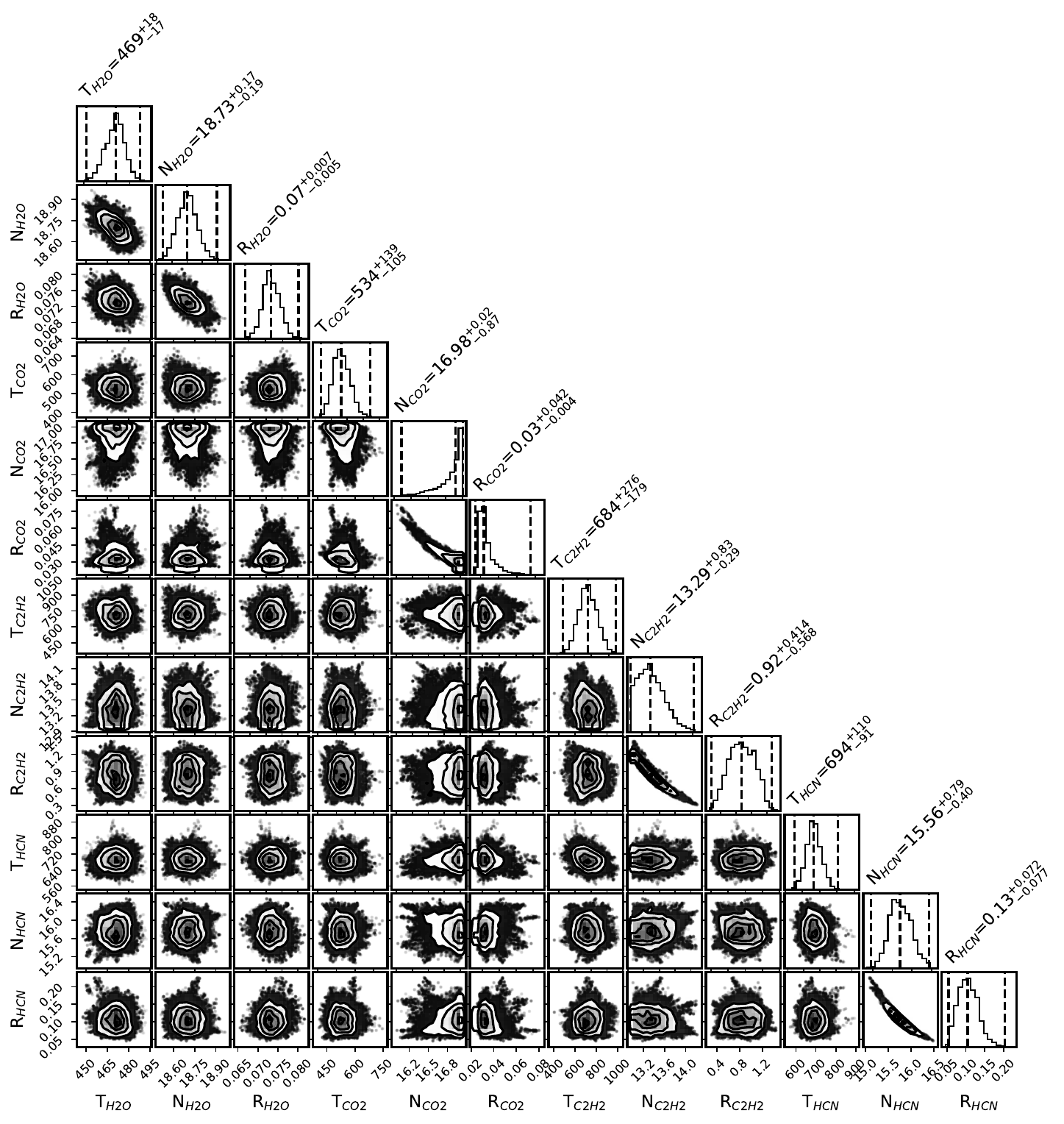}
\caption{\textbf{The posterior distributions of the MCMC slab modelling of J0438.} The column densities ($N$), temperature ($T$) and emitting radius ($R$) are reported with their $3\sigma$-uncertainties. The column densities are presented in log$_{10}$ space.}
\label{fig:correlations}
\end{figure}


\clearpage

\bibliography{references}{}

@ARTICLE{chauvin2005,
       author = {{Chauvin}, G. and {Lagrange}, A. -M. and {Dumas}, C. and {Zuckerman}, B. and {Mouillet}, D. and {Song}, I. and {Beuzit}, J. -L. and {Lowrance}, P.},
        title = "{Giant planet companion to 2MASSW J1207334-393254}",
      journal = {\aap},
         year = 2005,
        month = aug,
       volume = {438},
       number = {2},
        pages = {L25-L28}
}

@ARTICLE{Kumbhakar2023,
       author = {{Kumbhakar}, Rajib and {Mondal}, Soumen and {Ghosh}, Samrat and {Ram}, Diya},
        title = "{TESS Photometric Variability of Young Brown Dwarfs in the Taurus Star-forming Region}",
      journal = {\apj},
         year = 2023,
        month = sep,
       volume = {955},
       number = {1},
          eid = {18},
        pages = {18}
}

@ARTICLE{Xie2023,
       author = {{Xie}, Chengyan and {Pascucci}, Ilaria and {Long}, Feng and {Pontoppidan}, Klaus M. and {Banzatti}, Andrea and {Kalyaan}, Anusha and {Salyk}, Colette and {Liu}, Yao and {Najita}, Joan R. and {Pinilla}, Paola and {Arulanantham}, Nicole and {Herczeg}, Gregory J. and {Carr}, John and {Bergin}, Edwin A. and {Ballering}, Nicholas P. and {Krijt}, Sebastiaan and {Blake}, Geoffrey A. and {Zhang}, Ke and {{\"O}berg}, Karin I. and {Green}, Joel D. and {Jdiscs Collaboration}},
        title = "{Water-rich Disks around Late M Stars Unveiled: Exploring the Remarkable Case of Sz 114}",
      journal = {\apjl},
         year = 2023,
        month = dec,
       volume = {959},
       number = {2},
          eid = {L25},
        pages = {L25},
          doi = {10.3847/2041-8213/ad0ed9},
archivePrefix = {arXiv},
       eprint = {2310.13205},
 primaryClass = {astro-ph.EP},
       adsurl = {https://ui.adsabs.harvard.edu/abs/2023ApJ...959L..25X}
}

@ARTICLE{testi2016,
       author = {{Testi}, L. and {Natta}, A. and {Scholz}, A. and {Tazzari}, M. and {Ricci}, L. and {de Gregorio Monsalvo}, I.},
        title = "{Brown dwarf disks with ALMA: Evidence for truncated dust disks in Ophiuchus}",
      journal = {\aap},
         year = 2016,
        month = oct,
       volume = {593},
          eid = {A111},
        pages = {A111}
}

@ARTICLE{sanchis2020,
       author = {{Sanchis}, E. and {Testi}, L. and {Natta}, A. and {Manara}, C.~F. and {Ercolano}, B. and {Preibisch}, T. and {Henning}, T. and {Facchini}, S. and {Miotello}, A. and {de Gregorio-Monsalvo}, I. and {Lopez}, C. and {Mu{\v{z}}i{\'c}}, K. and {Pascucci}, I. and {Santamar{\'\i}a-Miranda}, A. and {Scholz}, A. and {Tazzari}, M. and {van Terwisga}, S. and {Williams}, J.~P.},
        title = "{Demographics of disks around young very low-mass stars and brown dwarfs in Lupus}",
      journal = {\aap},
         year = 2020,
        month = jan,
       volume = {633},
          eid = {A114},
        pages = {A114},
 primaryClass = {astro-ph.SR}
}

@ARTICLE{tabone2023,
       author = {{Tabone}, B. and {Bettoni}, G. and {van Dishoeck}, E.~F. and {Arabhavi}, A.~M. and {Grant}, S. and {Gasman}, D. and {Henning}, Th. and {Kamp}, I. and {G{\"u}del}, M. and {Lagage}, P.~O. and {Ray}, T. and {Vandenbussche}, B. and {Abergel}, A. and {Absil}, O. and {Argyriou}, I. and {Barrado}, D. and {Boccaletti}, A. and {Bouwman}, J. and {Caratti o Garatti}, A. and {Geers}, V. and {Glauser}, A.~M. and {Justannont}, K. and {Lahuis}, F. and {Mueller}, M. and {Nehm{\'e}}, C. and {Olofsson}, G. and {Pantin}, E. and {Scheithauer}, S. and {Waelkens}, C. and {Waters}, L.~B.~F.~M. and {Black}, J.~H. and {Christiaens}, V. and {Guadarrama}, R. and {Morales-Calder{\'o}n}, M. and {Jang}, H. and {Kanwar}, J. and {Pawellek}, N. and {Perotti}, G. and {Perrin}, A. and {Rodgers-Lee}, D. and {Samland}, M. and {Schreiber}, J. and {Schwarz}, K. and {Colina}, L. and {{\"O}stlin}, G. and {Wright}, G.},
        title = "{A rich hydrocarbon chemistry and high C to O ratio in the inner disk around a very low-mass star}",
      journal = {Nature Astronomy},
         year = 2023,
        month = jul,
       volume = {7},
        pages = {805-814},
 primaryClass = {astro-ph.EP}
}

@ARTICLE{Flaherty2010,
       author = {{Flaherty}, K.~M. and {Muzerolle}, J.},
        title = "{Modeling Mid-infrared Variability of Circumstellar Disks with Non-axisymmetric Structure}",
      journal = {\apj},
         year = 2010,
        month = aug,
       volume = {719},
       number = {2},
        pages = {1733-1749},
          doi = {10.1088/0004-637X/719/2/1733},
archivePrefix = {arXiv},
       eprint = {1007.1249},
 primaryClass = {astro-ph.GA},
       adsurl = {https://ui.adsabs.harvard.edu/abs/2010ApJ...719.1733F}
}

@ARTICLE{Pinilla2017,
       author = {{Pinilla}, P. and {Quiroga-Nu{\~n}ez}, L.~H. and {Benisty}, M. and {Natta}, A. and {Ricci}, L. and {Henning}, Th. and {van der Plas}, G. and {Birnstiel}, T. and {Testi}, L. and {Ward-Duong}, K.},
        title = "{Millimeter Spectral Indices and Dust Trapping By Planets in Brown Dwarf Disks}",
      journal = {\apj},
         year = 2017,
        month = sep,
       volume = {846},
       number = {1},
          eid = {70},
        pages = {70},
          doi = {10.3847/1538-4357/aa816f},
archivePrefix = {arXiv},
       eprint = {1709.00428},
 primaryClass = {astro-ph.EP},
       adsurl = {https://ui.adsabs.harvard.edu/abs/2017ApJ...846...70P}
}

@ARTICLE{Arabhavi2025a,
       author = {{Arabhavi}, Aditya M. and {Kamp}, Inga and {van Dishoeck}, Ewine F. and {Henning}, Thomas and {Jang}, Hyerin and {Christiaens}, Valentin and {Gasman}, Danny and {Pascucci}, Ilaria and {Perotti}, Giulia and {Grant}, Sierra L. and {Barrado}, David and {G{\"u}del}, Manuel and {Lagage}, Pierre-Olivier and {Caratti o Garatti}, Alessio and {Lahuis}, Fred and {Waters}, L.~B.~F.~M. and {Kaeufer}, Till and {Kanwar}, Jayatee and {Morales-Calder{\'o}n}, Maria and {Schwarz}, Kamber and {Sellek}, Andrew D. and {Tabone}, Beno{\^\i}t and {Temmink}, Milou and {Vlasblom}, Marissa},
        title = "{MINDS: The Very Low-mass Star and Brown Dwarf Sample Hidden Water in Carbon-dominated Protoplanetary Disks}",
      journal = {\apjl},
         year = 2025,
        month = may,
       volume = {984},
       number = {2},
          eid = {L62},
        pages = {L62},
          doi = {10.3847/2041-8213/adc692},
archivePrefix = {arXiv},
       eprint = {2504.11425},
 primaryClass = {astro-ph.EP},
       adsurl = {https://ui.adsabs.harvard.edu/abs/2025ApJ...984L..62A}
}

@ARTICLE{Arabhavi2025b,
       author = {{Arabhavi}, A.~M. and {Kamp}, I. and {Henning}, Th. and {van Dishoeck}, E.~F. and {Jang}, H. and {Waters}, L.~B.~F.~M. and {Christiaens}, V. and {Gasman}, D. and {Pascucci}, I. and {Perotti}, G. and {Grant}, S.~L. and {G{\"u}del}, M. and {Lagage}, P.-O. and {Barrado}, D. and {Caratti o Garatti}, A. and {Lahuis}, F. and {Kaeufer}, T. and {Kanwar}, J. and {Morales-Calder{\'o}n}, M. and {Schwarz}, K. and {Sellek}, A.~D. and {Tabone}, B. and {Temmink}, M. and {Vlasblom}, M. and {Patapis}, P.},
        title = "{MINDS: The very low-mass star and brown dwarf sample: Detections and trends in the inner disk gas}",
      journal = {\aap},
         year = 2025,
        month = jul,
       volume = {699},
          eid = {A194},
        pages = {A194},
          doi = {10.1051/0004-6361/202554109},
archivePrefix = {arXiv},
       eprint = {2506.02748},
 primaryClass = {astro-ph.EP},
       adsurl = {https://ui.adsabs.harvard.edu/abs/2025A&A...699A.194A}
}

@ARTICLE{Morales-Calderon2025,
       author = {{Morales-Calder{\'o}n}, Mar{\'\i}a and {Jang}, Hyerin and {Arabhavi}, Aditya M. and {Christiaens}, Valentin and {Barrado}, David and {Kamp}, Inga and {van Dishoeck}, Ewine F. and {Henning}, Thomas and {Waters}, L.~B.~F.~M. and {Temmink}, Milou and {G{\"u}del}, Manuel and {Lagage}, Pierre-Olivier and {Caratti o Garatti}, Alessio and {Glauser}, Adrian M. and {Ray}, Tom P. and {Franceschi}, Riccardo and {Gasman}, Danny and {Grant}, Sierra L. and {Kaeufer}, Till and {Kanwar}, Jayatee and {Perotti}, Giulia and {Samland}, Matthias and {Schwarz}, Kamber and {Vlasblom}, Marissa and {Colina}, Luis and {{\"O}stlin}, G{\"o}ran},
        title = "{MINDS: Cha H{\ensuremath{\alpha}} 1, a brown dwarf with a hydrocarbon-rich disk}",
      journal = {\aap},
         year = 2025,
        month = nov,
       volume = {703},
          eid = {A18},
        pages = {A18},
          doi = {10.1051/0004-6361/202555621},
archivePrefix = {arXiv},
       eprint = {2508.05155},
 primaryClass = {astro-ph.SR},
       adsurl = {https://ui.adsabs.harvard.edu/abs/2025A&A...703A..18M}
}

@ARTICLE{pinilla2013,
       author = {{Pinilla}, P. and {Birnstiel}, T. and {Benisty}, M. and {Ricci}, L. and {Natta}, A. and {Dullemond}, C.~P. and {Dominik}, C. and {Testi}, L.},
        title = "{Explaining millimeter-sized particles in brown dwarf disks}",
      journal = {\aap},
         year = 2013,
        month = jun,
       volume = {554},
          eid = {A95},
        pages = {A95}
}

@ARTICLE{banzatti2020,
       author = {{Banzatti}, Andrea and {Pascucci}, Ilaria and {Bosman}, Arthur D. and {Pinilla}, Paola and {Salyk}, Colette and {Herczeg}, Gregory J. and {Pontoppidan}, Klaus M. and {Vazquez}, Ivan and {Watkins}, Andrew and {Krijt}, Sebastiaan and {Hendler}, Nathan and {Long}, Feng},
        title = "{Hints for Icy Pebble Migration Feeding an Oxygen-rich Chemistry in the Inner Planet-forming Region of Disks}",
      journal = {\apj},
         year = 2020,
        month = nov,
       volume = {903},
       number = {2},
          eid = {124},
        pages = {124}
}

@ARTICLE{Kanwar2025,
       author = {{Kanwar}, Jayatee and {Kamp}, Inga and {Woitke}, Peter and {van Dishoeck}, Ewine F. and {Henning}, Thomas and {Liu}, Yao and {Kaeufer}, Till and {Tabone}, Beno{\^\i}t and {G{\"u}del}, Manuel and {Barrado}, David and {Arabhavi}, Aditya M. and {Franceschi}, Riccardo and {Vlasblom}, Marissa},
        title = "{MINDS. Strong oxygen depletion in the inner regions of a very low-mass star disk?}",
      journal = {arXiv e-prints},
         year = 2025,
        month = aug,
          eid = {arXiv:2508.11761},
        pages = {arXiv:2508.11761},
          doi = {10.48550/arXiv.2508.11761},
archivePrefix = {arXiv},
       eprint = {2508.11761},
 primaryClass = {astro-ph.EP},
       adsurl = {https://ui.adsabs.harvard.edu/abs/2025arXiv250811761K}
}

@ARTICLE{Salyk2025,
       author = {{Salyk}, Colette and {Pontoppidan}, Klaus M. and {Banzatti}, Andrea and {Bergin}, Edwin and {Arulanantham}, Nicole and {Najita}, Joan and {Blake}, Geoffrey A. and {Carr}, John and {Zhang}, Ke and {Xie}, Chengyan},
        title = "{Emission from Multiple Molecular Isotopologues in a High-inclination Protoplanetary Disk}",
      journal = {\aj},
         year = 2025,
        month = mar,
       volume = {169},
       number = {3},
          eid = {184},
        pages = {184},
          doi = {10.3847/1538-3881/adb397},
archivePrefix = {arXiv},
       eprint = {2502.05061},
 primaryClass = {astro-ph.SR},
       adsurl = {https://ui.adsabs.harvard.edu/abs/2025AJ....169..184S}
}

@ARTICLE{Kuruwita2020,
       author = {{Kuruwita}, Rajika L. and {Federrath}, Christoph and {Haugb{\o}lle}, Troels},
        title = "{The dependence of episodic accretion on eccentricity during the formation of binary stars}",
      journal = {\aap},
         year = 2020,
        month = sep,
       volume = {641},
          eid = {A59},
        pages = {A59},
          doi = {10.1051/0004-6361/202038181},
archivePrefix = {arXiv},
       eprint = {2004.07523},
 primaryClass = {astro-ph.SR},
       adsurl = {https://ui.adsabs.harvard.edu/abs/2020A&A...641A..59K}
}

@ARTICLE{Luhman2007,
       author = {{Luhman}, K.~L. and {Adame}, Luc{\'\i}a and {D'Alessio}, Paola and {Calvet}, Nuria and {McLeod}, Kim K. and {Bohac}, C.~J. and {Forrest}, William J. and {Hartmann}, Lee and {Sargent}, B. and {Watson}, Dan M.},
        title = "{Hubble and Spitzer Observations of an Edge-on Circumstellar Disk around a Brown Dwarf}",
      journal = {\apj},
         year = 2007,
        month = sep,
       volume = {666},
       number = {2},
        pages = {1219-1225}
}

@ARTICLE{Scholz2006,
       author = {{Scholz}, Alexander and {Jayawardhana}, Ray and {Wood}, Kenneth},
        title = "{Exploring Brown Dwarf Disks: A 1.3 mm Survey in Taurus}",
      journal = {\apj},
         year = 2006,
        month = jul,
       volume = {645},
       number = {2},
        pages = {1498-1508}
}

@ARTICLE{Gomez2017,
       author = {{G{\'o}mez}, Jos{\'e} F. and {Palau}, Aina and {Uscanga}, Lucero and {Manjarrez}, Guillermo and {Barrado}, David},
        title = "{A Search for Water Maser Emission from Brown Dwarfs and Low-luminosity Young Stellar Objects}",
      journal = {\aj},
         year = 2017,
        month = may,
       volume = {153},
       number = {5},
          eid = {221},
        pages = {221},
          doi = {10.3847/1538-3881/aa6622},
archivePrefix = {arXiv},
       eprint = {1703.04323},
 primaryClass = {astro-ph.SR},
       adsurl = {https://ui.adsabs.harvard.edu/abs/2017AJ....153..221G}
}

@ARTICLE{Rilinger2021,
       author = {{Rilinger}, Anneliese M. and {Espaillat}, Catherine C.},
        title = "{Disk Masses and Dust Evolution of Protoplanetary Disks around Brown Dwarfs}",
      journal = {\apj},
         year = 2021,
        month = nov,
       volume = {921},
       number = {2},
          eid = {182},
        pages = {182}
}

@ARTICLE{kurtovic2021,
       author = {{Kurtovic}, N.~T. and {Pinilla}, P. and {Long}, F. and {Benisty}, M. and {Manara}, C.~F. and {Natta}, A. and {Pascucci}, I. and {Ricci}, L. and {Scholz}, A. and {Testi}, L.},
        title = "{Size and structures of disks around very low mass stars in the Taurus star-forming region}",
      journal = {\aap},
         year = 2021,
        month = jan,
       volume = {645},
          eid = {A139},
        pages = {A139}
}

@ARTICLE{Mah2023,
       author = {{Mah}, Jingyi and {Bitsch}, Bertram and {Pascucci}, Ilaria and {Henning}, Thomas},
        title = "{Close-in ice lines and the super-stellar C/O ratio in discs around very low-mass stars}",
      journal = {\aap},
         year = 2023,
        month = sep,
       volume = {677},
          eid = {L7},
        pages = {L7},
}

@ARTICLE{Luhman2004,
       author = {{Luhman}, K.~L.},
        title = "{New Brown Dwarfs and an Updated Initial Mass Function in Taurus}",
      journal = {\apj},
         year = 2004,
        month = dec,
       volume = {617},
       number = {2},
        pages = {1216-1232}
}

@ARTICLE{gaia2021,
       author = {{Gaia Collaboration} and {Brown}, A.~G.~A. and {Vallenari}, A. and {Prusti}, T. and {de Bruijne}, J.~H.~J. and {Babusiaux}, C. and {Biermann}, M. and {Creevey}, O.~L. and {Evans}, D.~W. and {Eyer}, L. and {Hutton}, A. and {Jansen}, F. and {Jordi}, C. and {Klioner}, S.~A. and {Lammers}, U. and {Lindegren}, L. and {Luri}, X. and {Mignard}, F. and {Panem}, C. and {Pourbaix}, D. and {Randich}, S. and {Sartoretti}, P. and {Soubiran}, C. and {Walton}, N.~A. and {Arenou}, F. and {Bailer-Jones}, C.~A.~L. and {Bastian}, U. and {Cropper}, M. and {Drimmel}, R. and {Katz}, D. and {Lattanzi}, M.~G. and {van Leeuwen}, F. and {Bakker}, J. and {Cacciari}, C. and {Casta{\~n}eda}, J. and {De Angeli}, F. and {Ducourant}, C. and {Fabricius}, C. and {Fouesneau}, M. and {Fr{\'e}mat}, Y. and {Guerra}, R. and {Guerrier}, A. and {Guiraud}, J. and {Jean-Antoine Piccolo}, A. and {Masana}, E. and {Messineo}, R. and {Mowlavi}, N. and {Nicolas}, C. and {Nienartowicz}, K. and {Pailler}, F. and {Panuzzo}, P. and {Riclet}, F. and {Roux}, W. and {Seabroke}, G.~M. and {Sordo}, R. and {Tanga}, P. and {Th{\'e}venin}, F. and {Gracia-Abril}, G. and {Portell}, J. and {Teyssier}, D. and {Altmann}, M. and {Andrae}, R. and {Bellas-Velidis}, I. and {Benson}, K. and {Berthier}, J. and {Blomme}, R. and {Brugaletta}, E. and {Burgess}, P.~W. and {Busso}, G. and {Carry}, B. and {Cellino}, A. and {Cheek}, N. and {Clementini}, G. and {Damerdji}, Y. and {Davidson}, M. and {Delchambre}, L. and {Dell'Oro}, A. and {Fern{\'a}ndez-Hern{\'a}ndez}, J. and {Galluccio}, L. and {Garc{\'\i}a-Lario}, P. and {Garcia-Reinaldos}, M. and {Gonz{\'a}lez-N{\'u}{\~n}ez}, J. and {Gosset}, E. and {Haigron}, R. and {Halbwachs}, J. -L. and {Hambly}, N.~C. and {Harrison}, D.~L. and {Hatzidimitriou}, D. and {Heiter}, U. and {Hern{\'a}ndez}, J. and {Hestroffer}, D. and {Hodgkin}, S.~T. and {Holl}, B. and {Jan{\ss}en}, K. and {Jevardat de Fombelle}, G. and {Jordan}, S. and {Krone-Martins}, A. and {Lanzafame}, A.~C. and {L{\"o}ffler}, W. and {Lorca}, A. and {Manteiga}, M. and {Marchal}, O. and {Marrese}, P.~M. and {Moitinho}, A. and {Mora}, A. and {Muinonen}, K. and {Osborne}, P. and {Pancino}, E. and {Pauwels}, T. and {Petit}, J. -M. and {Recio-Blanco}, A. and {Richards}, P.~J. and {Riello}, M. and {Rimoldini}, L. and {Robin}, A.~C. and {Roegiers}, T. and {Rybizki}, J. and {Sarro}, L.~M. and {Siopis}, C. and {Smith}, M. and {Sozzetti}, A. and {Ulla}, A. and {Utrilla}, E. and {van Leeuwen}, M. and {van Reeven}, W. and {Abbas}, U. and {Abreu Aramburu}, A. and {Accart}, S. and {Aerts}, C. and {Aguado}, J.~J. and {Ajaj}, M. and {Altavilla}, G. and {{\'A}lvarez}, M.~A. and {{\'A}lvarez Cid-Fuentes}, J. and {Alves}, J. and {Anderson}, R.~I. and {Anglada Varela}, E. and {Antoja}, T. and {Audard}, M. and {Baines}, D. and {Baker}, S.~G. and {Balaguer-N{\'u}{\~n}ez}, L. and {Balbinot}, E. and {Balog}, Z. and {Barache}, C. and {Barbato}, D. and {Barros}, M. and {Barstow}, M.~A. and {Bartolom{\'e}}, S. and {Bassilana}, J. -L. and {Bauchet}, N. and {Baudesson-Stella}, A. and {Becciani}, U. and {Bellazzini}, M. and {Bernet}, M. and {Bertone}, S. and {Bianchi}, L. and {Blanco-Cuaresma}, S. and {Boch}, T. and {Bombrun}, A. and {Bossini}, D. and {Bouquillon}, S. and {Bragaglia}, A. and {Bramante}, L. and {Breedt}, E. and {Bressan}, A. and {Brouillet}, N. and {Bucciarelli}, B. and {Burlacu}, A. and {Busonero}, D. and {Butkevich}, A.~G. and {Buzzi}, R. and {Caffau}, E. and {Cancelliere}, R. and {C{\'a}novas}, H. and {Cantat-Gaudin}, T. and {Carballo}, R. and {Carlucci}, T. and {Carnerero}, M.~I. and {Carrasco}, J.~M. and {Casamiquela}, L. and {Castellani}, M. and {Castro-Ginard}, A. and {Castro Sampol}, P. and {Chaoul}, L. and {Charlot}, P. and {Chemin}, L. and {Chiavassa}, A. and {Cioni}, M. -R.~L. and {Comoretto}, G. and {Cooper}, W.~J. and {Cornez}, T. and {Cowell}, S. and {Crifo}, F. and {Crosta}, M. and {Crowley}, C. and {Dafonte}, C. and {Dapergolas}, A. and {David}, M. and {David}, P. and {de Laverny}, P. and {De Luise}, F. and {De March}, R. and {De Ridder}, J. and {de Souza}, R. and {de Teodoro}, P. and {de Torres}, A. and {del Peloso}, E.~F. and {del Pozo}, E. and {Delbo}, M. and {Delgado}, A. and {Delgado}, H.~E. and {Delisle}, J. -B. and {Di Matteo}, P. and {Diakite}, S. and {Diener}, C. and {Distefano}, E. and {Dolding}, C. and {Eappachen}, D. and {Edvardsson}, B. and {Enke}, H. and {Esquej}, P. and {Fabre}, C. and {Fabrizio}, M. and {Faigler}, S. and {Fedorets}, G. and {Fernique}, P. and {Fienga}, A. and {Figueras}, F. and {Fouron}, C. and {Fragkoudi}, F. and {Fraile}, E. and {Franke}, F. and {Gai}, M. and {Garabato}, D. and {Garcia-Gutierrez}, A. and {Garc{\'\i}a-Torres}, M. and {Garofalo}, A. and {Gavras}, P. and {Gerlach}, E. and {Geyer}, R. and {Giacobbe}, P. and {Gilmore}, G. and {Girona}, S. and {Giuffrida}, G. and {Gomel}, R. and {Gomez}, A. and {Gonzalez-Santamaria}, I. and {Gonz{\'a}lez-Vidal}, J.~J. and {Granvik}, M. and {Guti{\'e}rrez-S{\'a}nchez}, R. and {Guy}, L.~P. and {Hauser}, M. and {Haywood}, M. and {Helmi}, A. and {Hidalgo}, S.~L. and {Hilger}, T. and {H{\l}adczuk}, N. and {Hobbs}, D. and {Holland}, G. and {Huckle}, H.~E. and {Jasniewicz}, G. and {Jonker}, P.~G. and {Juaristi Campillo}, J. and {Julbe}, F. and {Karbevska}, L. and {Kervella}, P. and {Khanna}, S. and {Kochoska}, A. and {Kontizas}, M. and {Kordopatis}, G. and {Korn}, A.~J. and {Kostrzewa-Rutkowska}, Z. and {Kruszy{\'n}ska}, K. and {Lambert}, S. and {Lanza}, A.~F. and {Lasne}, Y. and {Le Campion}, J. -F. and {Le Fustec}, Y. and {Lebreton}, Y. and {Lebzelter}, T. and {Leccia}, S. and {Leclerc}, N. and {Lecoeur-Taibi}, I. and {Liao}, S. and {Licata}, E. and {Lindstr{\o}m}, E.~P. and {Lister}, T.~A. and {Livanou}, E. and {Lobel}, A. and {Madrero Pardo}, P. and {Managau}, S. and {Mann}, R.~G. and {Marchant}, J.~M. and {Marconi}, M. and {Marcos Santos}, M.~M.~S. and {Marinoni}, S. and {Marocco}, F. and {Marshall}, D.~J. and {Martin Polo}, L. and {Mart{\'\i}n-Fleitas}, J.~M. and {Masip}, A. and {Massari}, D. and {Mastrobuono-Battisti}, A. and {Mazeh}, T. and {McMillan}, P.~J. and {Messina}, S. and {Michalik}, D. and {Millar}, N.~R. and {Mints}, A. and {Molina}, D. and {Molinaro}, R. and {Moln{\'a}r}, L. and {Montegriffo}, P. and {Mor}, R. and {Morbidelli}, R. and {Morel}, T. and {Morris}, D. and {Mulone}, A.~F. and {Munoz}, D. and {Muraveva}, T. and {Murphy}, C.~P. and {Musella}, I. and {Noval}, L. and {Ord{\'e}novic}, C. and {Orr{\`u}}, G. and {Osinde}, J. and {Pagani}, C. and {Pagano}, I. and {Palaversa}, L. and {Palicio}, P.~A. and {Panahi}, A. and {Pawlak}, M. and {Pe{\~n}alosa Esteller}, X. and {Penttil{\"a}}, A. and {Piersimoni}, A.~M. and {Pineau}, F. -X. and {Plachy}, E. and {Plum}, G. and {Poggio}, E. and {Poretti}, E. and {Poujoulet}, E. and {Pr{\v{s}}a}, A. and {Pulone}, L. and {Racero}, E. and {Ragaini}, S. and {Rainer}, M. and {Raiteri}, C.~M. and {Rambaux}, N. and {Ramos}, P. and {Ramos-Lerate}, M. and {Re Fiorentin}, P. and {Regibo}, S. and {Reyl{\'e}}, C. and {Ripepi}, V. and {Riva}, A. and {Rixon}, G. and {Robichon}, N. and {Robin}, C. and {Roelens}, M. and {Rohrbasser}, L. and {Romero-G{\'o}mez}, M. and {Rowell}, N. and {Royer}, F. and {Rybicki}, K.~A. and {Sadowski}, G. and {Sagrist{\`a} Sell{\'e}s}, A. and {Sahlmann}, J. and {Salgado}, J. and {Salguero}, E. and {Samaras}, N. and {Sanchez Gimenez}, V. and {Sanna}, N. and {Santove{\~n}a}, R. and {Sarasso}, M. and {Schultheis}, M. and {Sciacca}, E. and {Segol}, M. and {Segovia}, J.~C. and {S{\'e}gransan}, D. and {Semeux}, D. and {Shahaf}, S. and {Siddiqui}, H.~I. and {Siebert}, A. and {Siltala}, L. and {Slezak}, E. and {Smart}, R.~L. and {Solano}, E. and {Solitro}, F. and {Souami}, D. and {Souchay}, J. and {Spagna}, A. and {Spoto}, F. and {Steele}, I.~A. and {Steidelm{\"u}ller}, H. and {Stephenson}, C.~A. and {S{\"u}veges}, M. and {Szabados}, L. and {Szegedi-Elek}, E. and {Taris}, F. and {Tauran}, G. and {Taylor}, M.~B. and {Teixeira}, R. and {Thuillot}, W. and {Tonello}, N. and {Torra}, F. and {Torra}, J. and {Turon}, C. and {Unger}, N. and {Vaillant}, M. and {van Dillen}, E. and {Vanel}, O. and {Vecchiato}, A. and {Viala}, Y. and {Vicente}, D. and {Voutsinas}, S. and {Weiler}, M. and {Wevers}, T. and {Wyrzykowski}, {\L}. and {Yoldas}, A. and {Yvard}, P. and {Zhao}, H. and {Zorec}, J. and {Zucker}, S. and {Zurbach}, C. and {Zwitter}, T.},
        title = "{Gaia Early Data Release 3. Summary of the contents and survey properties}",
      journal = {\aap},
         year = 2021,
        month = may,
       volume = {649},
          eid = {A1},
        pages = {A1}
}

@ARTICLE{Muzerolle2005,
       author = {{Muzerolle}, James and {Luhman}, Kevin L. and {Brice{\~n}o}, C{\'e}sar and {Hartmann}, Lee and {Calvet}, Nuria},
        title = "{Measuring Accretion in Young Substellar Objects: Approaching the Planetary Mass Regime}",
      journal = {\apj},
         year = 2005,
        month = jun,
       volume = {625},
       number = {2},
        pages = {906-912},
          doi = {10.1086/429483},
archivePrefix = {arXiv},
       eprint = {astro-ph/0502023},
 primaryClass = {astro-ph},
       adsurl = {https://ui.adsabs.harvard.edu/abs/2005ApJ...625..906M}
}

@ARTICLE{Monin2010,
       author = {{Monin}, J. -L. and {Guieu}, S. and {Pinte}, C. and {Rebull}, L. and {Goldsmith}, P. and {Fukagawa}, M. and {M{\'e}nard}, F. and {Padgett}, D. and {Stappelfeld}, K. and {McCabe}, C. and {Carey}, S. and {Noriega-Crespo}, A. and {Brooke}, T. and {Huard}, T. and {Terebey}, S. and {Hillenbrand}, L. and {Guedel}, M.},
        title = "{The large-scale disk fraction of brown dwarfs in the Taurus cloud as measured with Spitzer}",
      journal = {\aap},
         year = 2010,
        month = jun,
       volume = {515},
          eid = {A91},
        pages = {A91},
          doi = {10.1051/0004-6361/200912338},
archivePrefix = {arXiv},
       eprint = {1004.2541},
 primaryClass = {astro-ph.SR},
       adsurl = {https://ui.adsabs.harvard.edu/abs/2010A&A...515A..91M}
}

@ARTICLE{Pascucci2013,
       author = {{Pascucci}, I. and {Herczeg}, G. and {Carr}, J.~S. and {Bruderer}, S.},
        title = "{The Atomic and Molecular Content of Disks around Very Low-mass Stars and Brown Dwarfs}",
      journal = {\apj},
         year = 2013,
        month = dec,
       volume = {779},
       number = {2},
          eid = {178},
        pages = {178},
          doi = {10.1088/0004-637X/779/2/178},
archivePrefix = {arXiv},
       eprint = {1311.1228},
 primaryClass = {astro-ph.EP},
       adsurl = {https://ui.adsabs.harvard.edu/abs/2013ApJ...779..178P}
}

@ARTICLE{Cody2022,
       author = {{Cody}, Ann Marie and {Hillenbrand}, Lynne A. and {Rebull}, Luisa M.},
        title = "{The Many-faceted Light Curves of Young Disk-bearing Stars in Taurus as Seen by K2}",
      journal = {\aj},
         year = 2022,
        month = may,
       volume = {163},
       number = {5},
          eid = {212},
        pages = {212},
          doi = {10.3847/1538-3881/ac5b73},
archivePrefix = {arXiv},
       eprint = {2204.06646},
 primaryClass = {astro-ph.SR},
       adsurl = {https://ui.adsabs.harvard.edu/abs/2022AJ....163..212C}
}

@ARTICLE{Phan-Bao2014,
       author = {{Phan-Bao}, Ngoc and {Lee}, Chin-Fei and {Ho}, Paul T.~P. and {Dang-Duc}, Cuong and {Li}, Di},
        title = "{Characterization of Molecular Outflows in the Substellar Domain}",
      journal = {\apj},
         year = 2014,
        month = nov,
       volume = {795},
       number = {1},
          eid = {70},
        pages = {70},
          doi = {10.1088/0004-637X/795/1/70},
archivePrefix = {arXiv},
       eprint = {1408.4506},
 primaryClass = {astro-ph.SR},
       adsurl = {https://ui.adsabs.harvard.edu/abs/2014ApJ...795...70P}
}

@ARTICLE{Pascucci2009a,
       author = {{Pascucci}, I. and {Apai}, D. and {Luhman}, K. and {Henning}, Th. and {Bouwman}, J. and {Meyer}, M.~R. and {Lahuis}, F. and {Natta}, A.},
        title = "{The Different Evolution of Gas and Dust in Disks around Sun-Like and Cool Stars}",
      journal = {\apj},
         year = 2009,
        month = may,
       volume = {696},
       number = {1},
        pages = {143-159},
          doi = {10.1088/0004-637X/696/1/143},
archivePrefix = {arXiv},
       eprint = {0810.2552},
 primaryClass = {astro-ph},
       adsurl = {https://ui.adsabs.harvard.edu/abs/2009ApJ...696..143P}
}

@ARTICLE{Ward-Duong2018,
       author = {{Ward-Duong}, K. and {Patience}, J. and {Bulger}, J. and {van der Plas}, G. and {M{\'e}nard}, F. and {Pinte}, C. and {Jackson}, A.~P. and {Bryden}, G. and {Turner}, N.~J. and {Harvey}, P. and {Hales}, A. and {De Rosa}, R.~J.},
        title = "{The Taurus Boundary of Stellar/Substellar (TBOSS) Survey. II. Disk Masses from ALMA Continuum Observations}",
      journal = {\aj},
         year = 2018,
        month = feb,
       volume = {155},
       number = {2},
          eid = {54},
        pages = {54},
          doi = {10.3847/1538-3881/aaa128},
archivePrefix = {arXiv},
       eprint = {1712.07669},
 primaryClass = {astro-ph.EP},
       adsurl = {https://ui.adsabs.harvard.edu/abs/2018AJ....155...54W}
}

@ARTICLE{Hendler2017,
       author = {{Hendler}, Nathanial P. and {Mulders}, Gijs D. and {Pascucci}, Ilaria and {Greenwood}, Aaron and {Kamp}, Inga and {Henning}, Thomas and {M{\'e}nard}, Fran{\c{c}}ois and {Dent}, William R.~F. and {Evans}, Neal J., II},
        title = "{Hints for Small Disks around Very Low Mass Stars and Brown Dwarfs}",
      journal = {\apj},
         year = 2017,
        month = jun,
       volume = {841},
       number = {2},
          eid = {116},
        pages = {116},
          doi = {10.3847/1538-4357/aa71b8},
archivePrefix = {arXiv},
       eprint = {1705.01952},
 primaryClass = {astro-ph.EP},
       adsurl = {https://ui.adsabs.harvard.edu/abs/2017ApJ...841..116H}
}

@ARTICLE{Bulger2014,
       author = {{Bulger}, J. and {Patience}, J. and {Ward-Duong}, K. and {Pinte}, C. and {Bouy}, H. and {M{\'e}nard}, F. and {Monin}, J. -L.},
        title = "{The Taurus Boundary of Stellar/Substellar (TBOSS) Survey. I. Far-IR disk emission measured with Herschel}",
      journal = {\aap},
         year = 2014,
        month = oct,
       volume = {570},
          eid = {A29},
        pages = {A29},
          doi = {10.1051/0004-6361/201323088},
archivePrefix = {arXiv},
       eprint = {1407.4802},
 primaryClass = {astro-ph.SR},
       adsurl = {https://ui.adsabs.harvard.edu/abs/2014A&A...570A..29B}
}

@ARTICLE{Henning2024,
       author = {{Henning}, Thomas and {Kamp}, Inga and {Samland}, Matthias and {Arabhavi}, Aditya M. and {Kanwar}, Jayatee and {van Dishoeck}, Ewine F. and {G{\"u}del}, Manuel and {Lagage}, Pierre-Olivier and {Waelkens}, Christoffel and {Abergel}, Alain and {Absil}, Olivier and {Barrado}, David and {Boccaletti}, Anthony and {Bouwman}, Jeroen and {Caratti o Garatti}, Alessio and {Geers}, Vincent and {Glauser}, Adrian M. and {Lahuis}, Fred and {Mueller}, Michael and {Nehm{\'e}}, Cyrine and {Olofsson}, G{\"o}ran and {Pantin}, Eric and {Ray}, Tom P. and {Scheithauer}, Silvia and {Vandenbussche}, Bart and {Waters}, L.~B.~F.~M. and {Wright}, Gillian and {Argyriou}, Ioannis and {Christiaens}, Valentin and {Franceschi}, Riccardo and {Gasman}, Danny and {Grant}, Sierra L. and {Guadarrama}, Rodrigo and {Jang}, Hyerin and {Morales-Calder{\'o}n}, Maria and {Pawellek}, Nicole and {Perotti}, Giulia and {Rodgers-Lee}, Donna and {Schreiber}, J{\"u}rgen and {Schwarz}, Kamber and {Tabone}, Beno{\^\i}t and {Temmink}, Milou and {Vlasblom}, Marissa and {Colina}, Luis and {Greve}, Thomas R. and {{\"O}stlin}, G{\"o}ran},
        title = "{MINDS: The JWST MIRI Mid-INfrared Disk Survey}",
      journal = {\pasp},
         year = 2024,
        month = may,
       volume = {136},
       number = {5},
          eid = {054302},
        pages = {054302},
          doi = {10.1088/1538-3873/ad3455},
archivePrefix = {arXiv},
       eprint = {2403.09210},
 primaryClass = {astro-ph.EP},
       adsurl = {https://ui.adsabs.harvard.edu/abs/2024PASP..136e4302H}
}

@ARTICLE{Berne2023,
       author = {{Bern{\'e}}, Olivier and {Martin-Drumel}, Marie-Aline and {Schroetter}, Ilane and {Goicoechea}, Javier R. and {Jacovella}, Ugo and {Gans}, B{\'e}renger and {Dartois}, Emmanuel and {Coudert}, Laurent H. and {Bergin}, Edwin and {Alarcon}, Felipe and {Cami}, Jan and {Roueff}, Evelyne and {Black}, John H. and {Asvany}, Oskar and {Habart}, Emilie and {Peeters}, Els and {Canin}, Amelie and {Trahin}, Boris and {Joblin}, Christine and {Schlemmer}, Stephan and {Thorwirth}, Sven and {Cernicharo}, Jose and {Gerin}, Maryvonne and {Tielens}, Alexander and {Zannese}, Marion and {Abergel}, Alain and {Bernard-Salas}, Jeronimo and {Boersma}, Christiaan and {Bron}, Emeric and {Chown}, Ryan and {Cuadrado}, Sara and {Dicken}, Daniel and {Elyajouri}, Meriem and {Fuente}, Asunci{\'o}n and {Gordon}, Karl D. and {Issa}, Lina and {Kannavou}, Olga and {Khan}, Baria and {Lacinbala}, Ozan and {Languignon}, David and {Le Gal}, Romane and {Maragkoudakis}, Alexandros and {Meshaka}, Raphael and {Okada}, Yoko and {Onaka}, Takashi and {Pasquini}, Sofia and {Pound}, Marc W. and {Robberto}, Massimo and {R{\"o}llig}, Markus and {Schefter}, Bethany and {Schirmer}, Thi{\'e}baut and {Sidhu}, Ameek and {Tabone}, Benoit and {Van De Putte}, Dries and {Vicente}, S{\'\i}lvia and {Wolfire}, Mark G.},
        title = "{Formation of the methyl cation by photochemistry in a protoplanetary disk}",
      journal = {\nat},
         year = 2023,
        month = sep,
       volume = {621},
       number = {7977},
        pages = {56-59},
          doi = {10.1038/s41586-023-06307-x},
archivePrefix = {arXiv},
       eprint = {2401.03296},
 primaryClass = {astro-ph.GA},
       adsurl = {https://ui.adsabs.harvard.edu/abs/2023Natur.621...56B}
}

@ARTICLE{Luhman2010,
       author = {{Luhman}, K.~L. and {Allen}, P.~R. and {Espaillat}, C. and {Hartmann}, L. and {Calvet}, N.},
        title = "{The Disk Population of the Taurus Star-Forming Region}",
      journal = {\apjs},
         year = 2010,
        month = jan,
       volume = {186},
       number = {1},
        pages = {111-174},
          doi = {10.1088/0067-0049/186/1/111},
archivePrefix = {arXiv},
       eprint = {0911.5457},
 primaryClass = {astro-ph.GA},
       adsurl = {https://ui.adsabs.harvard.edu/abs/2010ApJS..186..111L}
}

@ARTICLE{Schwarz2024,
       author = {{Schwarz}, Kamber R. and {Henning}, Thomas and {Christiaens}, Valentin and {Gasman}, Danny and {Samland}, Matthias and {Perotti}, Giulia and {Jang}, Hyerin and {Grant}, Sierra L. and {Tabone}, Beno{\^\i}t and {Morales-Calder{\'o}n}, Maria and {Kamp}, Inga and {van Dishoeck}, Ewine F. and {G{\"u}del}, Manuel and {Lagage}, Pierre-Olivier and {Barrado}, David and {Caratti o Garatti}, Alessio and {Glauser}, Adrian M. and {Ray}, Tom P. and {Vandenbussche}, Bart and {Waters}, L.~B.~F.~M. and {Arabhavi}, Aditya M. and {Kanwar}, Jayatee and {Olofsson}, G{\"o}ran and {Rodgers-Lee}, Donna and {Schreiber}, J{\"u}rgen and {Temmink}, Milou},
        title = "{MINDS. JWST/MIRI Reveals a Dynamic Gas-rich Inner Disk inside the Cavity of SY Cha}",
      journal = {\apj},
         year = 2024,
        month = feb,
       volume = {962},
       number = {1},
          eid = {8},
        pages = {8},
          doi = {10.3847/1538-4357/ad1393},
archivePrefix = {arXiv},
       eprint = {2312.07135},
 primaryClass = {astro-ph.EP},
       adsurl = {https://ui.adsabs.harvard.edu/abs/2024ApJ...962....8S}
}

@ARTICLE{Bajaj2024,
       author = {{Bajaj}, Naman S. and {Pascucci}, Ilaria and {Gorti}, Uma and {Alexander}, Richard and {Sellek}, Andrew and {Morrison}, Jane and {Gaspar}, Andras and {Clarke}, Cathie and {Xie}, Chengyan and {Ballabio}, Giulia and {Deng}, Dingshan},
        title = "{JWST MIRI MRS Observations of T Cha: Discovery of a Spatially Resolved Disk Wind}",
      journal = {\aj},
         year = 2024,
        month = mar,
       volume = {167},
       number = {3},
          eid = {127},
        pages = {127},
          doi = {10.3847/1538-3881/ad22e1},
archivePrefix = {arXiv},
       eprint = {2403.01060},
 primaryClass = {astro-ph.EP},
       adsurl = {https://ui.adsabs.harvard.edu/abs/2024AJ....167..127B}
}

@ARTICLE{Labiano2021,
       author = {{Labiano}, A. and {Argyriou}, I. and {{\'A}lvarez-M{\'a}rquez}, J. and {Glasse}, A. and {Glauser}, A. and {Patapis}, P. and {Law}, D. and {Brandl}, B.~R. and {Justtanont}, K. and {Lahuis}, F. and {Mart{\'\i}nez-Galarza}, J.~R. and {Mueller}, M. and {Noriega-Crespo}, A. and {Royer}, P. and {Shaughnessy}, B. and {Vandenbussche}, B.},
        title = "{Wavelength calibration and resolving power of the JWST MIRI Medium Resolution Spectrometer}",
      journal = {\aap},
         year = 2021,
        month = dec,
       volume = {656},
          eid = {A57},
        pages = {A57},
          doi = {10.1051/0004-6361/202140614},
archivePrefix = {arXiv},
       eprint = {2109.04254},
 primaryClass = {astro-ph.IM},
       adsurl = {https://ui.adsabs.harvard.edu/abs/2021A&A...656A..57L}
}

@ARTICLE{Ramirez2023,
       author = {{Ram{\'\i}rez-Tannus}, Mar{\'\i}a Claudia and {Bik}, Arjan and {Cuijpers}, Lars and {Waters}, Rens and {G{\"o}ppl}, Christiane and {Henning}, Thomas and {Kamp}, Inga and {Preibisch}, Thomas and {Getman}, Konstantin V. and {Chaparro}, Germ{\'a}n and {Cuartas-Restrepo}, Pablo and {de Koter}, Alex and {Feigelson}, Eric D. and {Grant}, Sierra L. and {Haworth}, Thomas J. and {Hern{\'a}ndez}, Sebasti{\'a}n and {Kuhn}, Michael A. and {Perotti}, Giulia and {Povich}, Matthew S. and {Reiter}, Megan and {Roccatagliata}, Veronica and {Sabbi}, Elena and {Tabone}, Beno{\^\i}t and {Winter}, Andrew J. and {McLeod}, Anna F. and {van Boekel}, Roy and {van Terwisga}, Sierk E.},
        title = "{XUE: Molecular Inventory in the Inner Region of an Extremely Irradiated Protoplanetary Disk}",
      journal = {\apjl},
         year = 2023,
        month = dec,
       volume = {958},
       number = {2},
          eid = {L30},
        pages = {L30},
          doi = {10.3847/2041-8213/ad03f8},
archivePrefix = {arXiv},
       eprint = {2310.11074},
 primaryClass = {astro-ph.SR},
       adsurl = {https://ui.adsabs.harvard.edu/abs/2023ApJ...958L..30R}
}

@ARTICLE{Espaillat2023,
       author = {{Espaillat}, C.~C. and {Thanathibodee}, T. and {Pittman}, C.~V. and {Sturm}, J.~A. and {McClure}, M.~K. and {Calvet}, N. and {Walter}, F.~M. and {Franco-Hern{\'a}ndez}, R. and {Muzerolle Page}, J.},
        title = "{JWST Detects Neon Line Variability in a Protoplanetary Disk}",
      journal = {\apjl},
         year = 2023,
        month = nov,
       volume = {958},
       number = {1},
          eid = {L4},
        pages = {L4},
          doi = {10.3847/2041-8213/ad023d},
archivePrefix = {arXiv},
       eprint = {2311.07739},
 primaryClass = {astro-ph.SR},
       adsurl = {https://ui.adsabs.harvard.edu/abs/2023ApJ...958L...4E}
}

@ARTICLE{Sturm2024,
       author = {{Sturm}, J.~A. and {McClure}, M.~K. and {Harsono}, D. and {Bergner}, J.~B. and {Dartois}, E. and {Boogert}, A.~C.~A. and {Cordiner}, M.~A. and {Drozdovskaya}, M.~N. and {Ioppolo}, S. and {Law}, C.~J. and {Lis}, D.~C. and {McGuire}, B.~A. and {Melnick}, G.~J. and {Noble}, J.~A. and {{\"O}berg}, K.~I. and {Palumbo}, M.~E. and {Pendleton}, Y.~J. and {Perotti}, G. and {Rocha}, W.~R.~M. and {Urso}, R.~G. and {van Dishoeck}, E.~F.},
        title = "{A JWST/MIRI analysis of the ice distribution and polycyclic aromatic hydrocarbon emission in the protoplanetary disk HH 48 NE}",
      journal = {\aap},
         year = 2024,
        month = sep,
       volume = {689},
          eid = {A92},
        pages = {A92},
          doi = {10.1051/0004-6361/202450865},
archivePrefix = {arXiv},
       eprint = {2407.09627},
 primaryClass = {astro-ph.EP},
       adsurl = {https://ui.adsabs.harvard.edu/abs/2024A&A...689A..92S}
}

@article{rieke2015,
  title={The mid-infrared instrument for the james webb space telescope, i: Introduction},
  author={Rieke, George H and Wright, GS and B{\"o}ker, T and Bouwman, J and Colina, L and Glasse, Alistair and Gordon, KD and Greene, TP and G{\"u}del, Manuel and Henning, Th and others},
  journal={Publ. Astron. Soc. Pac.},
  volume={127},
  number={953},
  pages={584},
  year={2015},
  publisher={IOP Publishing}
}

@ARTICLE{Wright2015,
       author = {{Wright}, G.~S. and {Wright}, David and {Goodson}, G.~B. and {Rieke}, G.~H. and {Aitink-Kroes}, Gabby and {Amiaux}, J. and {Aricha-Yanguas}, Ana and {Azzollini}, Ruym{\'a}n and {Banks}, Kimberly and {Barrado-Navascues}, D. and {Belenguer-Davila}, T. and {Bloemmart}, J.~A.~D.~L. and {Bouchet}, Patrice and {Brandl}, B.~R. and {Colina}, L. and {Detre}, {\"O}rs and {Diaz-Catala}, Eva and {Eccleston}, Paul and {Friedman}, Scott D. and {Garc{\'\i}a-Mar{\'\i}n}, Macarena and {G{\"u}del}, Manuel and {Glasse}, Alistair and {Glauser}, Adrian M. and {Greene}, T.~P. and {Groezinger}, Uli and {Grundy}, Tim and {Hastings}, Peter and {Henning}, Th. and {Hofferbert}, Ralph and {Hunter}, Faye and {Jessen}, N.~C. and {Justtanont}, K. and {Karnik}, Avinash R. and {Khorrami}, Mori A. and {Krause}, Oliver and {Labiano}, Alvaro and {Lagage}, P. -O. and {Langer}, Ulrich and {Lemke}, Dietrich and {Lim}, Tanya and {Lorenzo-Alvarez}, Jose and {Mazy}, Emmanuel and {McGowan}, Norman and {Meixner}, M.~E. and {Morris}, Nigel and {Morrison}, Jane E. and {M{\"u}ller}, Friedrich and {rgaard-Nielson}, H. -U. N{\o} and {Olofsson}, G{\"o}ran and {O'Sullivan}, Brian and {Pel}, J. -W. and {Penanen}, Konstantin and {Petach}, M.~B. and {Pye}, J.~P. and {Ray}, T.~P. and {Renotte}, Etienne and {Renouf}, Ian and {Ressler}, M.~E. and {Samara-Ratna}, Piyal and {Scheithauer}, Silvia and {Schneider}, Analyn and {Shaughnessy}, Bryan and {Stevenson}, Tim and {Sukhatme}, Kalyani and {Swinyard}, Bruce and {Sykes}, Jon and {Thatcher}, John and {Tikkanen}, Tuomo and {van Dishoeck}, E.~F. and {Waelkens}, C. and {Walker}, Helen and {Wells}, Martyn and {Zhender}, Alex},
        title = "{The Mid-Infrared Instrument for the James Webb Space Telescope, II: Design and Build}",
      journal = {\pasp},
         year = 2015,
        month = jul,
       volume = {127},
       number = {953},
        pages = {595},
          doi = {10.1086/682253},
archivePrefix = {arXiv},
       eprint = {1508.02333},
 primaryClass = {astro-ph.IM},
       adsurl = {https://ui.adsabs.harvard.edu/abs/2015PASP..127..595W}
}

@ARTICLE{Wells2015,
       author = {{Wells}, Martyn and {Pel}, J. -W. and {Glasse}, Alistair and {Wright}, G.~S. and {Aitink-Kroes}, Gabby and {Azzollini}, Ruym{\'a}n and {Beard}, Steven and {Brandl}, B.~R. and {Gallie}, Angus and {Geers}, V.~C. and {Glauser}, A.~M. and {Hastings}, Peter and {Henning}, Th. and {Jager}, Rieks and {Justtanont}, K. and {Kruizinga}, Bob and {Lahuis}, Fred and {Lee}, David and {Martinez-Delgado}, I. and {Mart{\'\i}nez-Galarza}, J.~R. and {Meijers}, M. and {Morrison}, Jane E. and {M{\"u}ller}, Friedrich and {Nakos}, Thodori and {O'Sullivan}, Brian and {Oudenhuysen}, Ad and {Parr-Burman}, P. and {Pauwels}, Evert and {Rohloff}, R. -R. and {Schmalzl}, Eva and {Sykes}, Jon and {Thelen}, M.~P. and {van Dishoeck}, E.~F. and {Vandenbussche}, Bart and {Venema}, Lars B. and {Visser}, Huib and {Waters}, L.~B.~F.~M. and {Wright}, David},
        title = "{The Mid-Infrared Instrument for the James Webb Space Telescope, VI: The Medium Resolution Spectrometer}",
      journal = {\pasp},
         year = 2015,
        month = jul,
       volume = {127},
       number = {953},
        pages = {646},
          doi = {10.1086/682281},
archivePrefix = {arXiv},
       eprint = {1508.03070},
 primaryClass = {astro-ph.IM},
       adsurl = {https://ui.adsabs.harvard.edu/abs/2015PASP..127..646W}
}

@ARTICLE{Grant2023,
       author = {{Grant}, Sierra L. and {van Dishoeck}, Ewine F. and {Tabone}, Beno{\^\i}t and {Gasman}, Danny and {Henning}, Thomas and {Kamp}, Inga and {G{\"u}del}, Manuel and {Lagage}, Pierre-Olivier and {Bettoni}, Giulio and {Perotti}, Giulia and {Christiaens}, Valentin and {Samland}, Matthias and {Arabhavi}, Aditya M. and {Argyriou}, Ioannis and {Abergel}, Alain and {Absil}, Olivier and {Barrado}, David and {Boccaletti}, Anthony and {Bouwman}, Jeroen and {o Garatti}, Alessio Caratti and {Geers}, Vincent and {Glauser}, Adrian M. and {Guadarrama}, Rodrigo and {Jang}, Hyerin and {Kanwar}, Jayatee and {Lahuis}, Fred and {Morales-Calder{\'o}n}, Maria and {Mueller}, Michael and {Nehm{\'e}}, Cyrine and {Olofsson}, G{\"o}ran and {Pantin}, Eric and {Pawellek}, Nicole and {Ray}, Tom P. and {Rodgers-Lee}, Donna and {Scheithauer}, Silvia and {Schreiber}, J{\"u}rgen and {Schwarz}, Kamber and {Temmink}, Milou and {Vandenbussche}, Bart and {Vlasblom}, Marissa and {Waters}, L.~B.~F.~M. and {Wright}, Gillian and {Colina}, Luis and {Greve}, Thomas R. and {Justannont}, Kay and {{\"O}stlin}, G{\"o}ran},
        title = "{MINDS. The Detection of $^{13}$CO$_{2}$ with JWST-MIRI Indicates Abundant CO$_{2}$ in a Protoplanetary Disk}",
      journal = {\apjl},
         year = 2023,
        month = apr,
       volume = {947},
       number = {1},
          eid = {L6},
        pages = {L6},
          doi = {10.3847/2041-8213/acc44b},
archivePrefix = {arXiv},
       eprint = {2212.08047},
 primaryClass = {astro-ph.SR},
       adsurl = {https://ui.adsabs.harvard.edu/abs/2023ApJ...947L...6G}
}

@software{Bushouse2024,
       author = {{Bushouse}, Howard and {Eisenhamer}, Jonathan and {Dencheva}, Nadia and {Davies}, James and {Greenfield}, Perry and {Morrison}, Jane and {Hodge}, Phil and {Simon}, Bernie and {Grumm}, David and {Droettboom}, Michael and {Slavich}, Edward and {Sosey}, Megan and {Pauly}, Tyler and {Miller}, Todd and {Jedrzejewski}, Robert and {Hack}, Warren and {Davis}, David and {Crawford}, Steven and {Law}, David and {Gordon}, Karl and {Regan}, Michael and {Cara}, Mihai and {MacDonald}, Ken and {Bradley}, Larry and {Shanahan}, Clare and {Jamieson}, William and {Teodoro}, Mairan and {Williams}, Thomas and {Pena-Guerrero}, Maria},
        title = "{JWST Calibration Pipeline}",
         year = 2024,
        month = mar,
          eid = {10.5281/zenodo.10870758},
          doi = {10.5281/zenodo.10870758},
      version = {1.14.0},
    publisher = {Zenodo},
       adsurl = {https://ui.adsabs.harvard.edu/abs/2024zndo..10870758B}
}

@ARTICLE{Perotti2023,
       author = {{Perotti}, G. and {Christiaens}, V. and {Henning}, Th. and {Tabone}, B. and {Waters}, L.~B.~F.~M. and {Kamp}, I. and {Olofsson}, G. and {Grant}, S.~L. and {Gasman}, D. and {Bouwman}, J. and {Samland}, M. and {Franceschi}, R. and {van Dishoeck}, E.~F. and {Schwarz}, K. and {G{\"u}del}, M. and {Lagage}, P. -O. and {Ray}, T.~P. and {Vandenbussche}, B. and {Abergel}, A. and {Absil}, O. and {Arabhavi}, A.~M. and {Argyriou}, I. and {Barrado}, D. and {Boccaletti}, A. and {Caratti o Garatti}, A. and {Geers}, V. and {Glauser}, A.~M. and {Justannont}, K. and {Lahuis}, F. and {Mueller}, M. and {Nehm{\'e}}, C. and {Pantin}, E. and {Scheithauer}, S. and {Waelkens}, C. and {Guadarrama}, R. and {Jang}, H. and {Kanwar}, J. and {Morales-Calder{\'o}n}, M. and {Pawellek}, N. and {Rodgers-Lee}, D. and {Schreiber}, J. and {Colina}, L. and {Greve}, T.~R. and {{\"O}stlin}, G. and {Wright}, G.},
        title = "{Water in the terrestrial planet-forming zone of the PDS 70 disk}",
      journal = {\nat},
         year = 2023,
        month = aug,
       volume = {620},
       number = {7974},
        pages = {516-520},
          doi = {10.1038/s41586-023-06317-9},
archivePrefix = {arXiv},
       eprint = {2307.12040},
 primaryClass = {astro-ph.EP},
       adsurl = {https://ui.adsabs.harvard.edu/abs/2023Natur.620..516P}
}

@article{GomezGonzalez2017,
	author = {{Gomez Gonzalez}, C.~A. and {Wertz}, O. and {Absil}, O. and {Christiaens}, V. and {Defr{\`e}re}, D. and {Mawet}, D. and {Milli}, J. and {Absil}, P.-A. and {Van Droogenbroeck}, M. and {Cantalloube}, F. and {Hinz}, P.~M. and {Skemer}, A.~J. and {Karlsson}, M. and {Surdej}, J.},
	journal = {\aj},
	month = jul,
	pages = {7},
	title = {{VIP: Vortex Image Processing Package for High-contrast Direct Imaging}},
	volume = {154},
	year = {2017}}

@ARTICLE{Hunter2007,
  author={Hunter, John D.},
  journal={Computing in Science \& Engineering}, 
  title={Matplotlib: A 2D Graphics Environment}, 
  year={2007},
  volume={9},
  number={3},
  pages={90-95},
  doi={10.1109/MCSE.2007.55}}

@misc{carnall2017,
      title={SpectRes: A Fast Spectral Resampling Tool in Python}, 
      author={A. C. Carnall},
      year={2017},
      eprint={1705.05165},
      archivePrefix={arXiv},
      primaryClass={astro-ph.IM}
}

@ARTICLE{Ballering2021,
       author = {{Ballering}, Nicholas P. and {Cleeves}, L. Ilsedore and {Anderson}, Dana E.},
        title = "{Simulating Observations of Ices in Protoplanetary Disks}",
      journal = {\apj},
         year = 2021,
        month = oct,
       volume = {920},
       number = {2},
          eid = {115},
        pages = {115},
          doi = {10.3847/1538-4357/ac17ed},
archivePrefix = {arXiv},
       eprint = {2105.12169},
 primaryClass = {astro-ph.EP},
       adsurl = {https://ui.adsabs.harvard.edu/abs/2021ApJ...920..115B}
}

@ARTICLE{Arabhavi2022,
       author = {{Arabhavi}, Aditya M. and {Woitke}, Peter and {Cazaux}, St{\'e}phanie M. and {Kamp}, Inga and {Rab}, Christian and {Thi}, Wing-Fai},
        title = "{Ices in planet-forming disks: Self-consistent ice opacities in disk models}",
      journal = {\aap},
         year = 2022,
        month = oct,
       volume = {666},
          eid = {A139},
        pages = {A139},
          doi = {10.1051/0004-6361/202141825},
archivePrefix = {arXiv},
       eprint = {2208.12739},
 primaryClass = {astro-ph.EP},
       adsurl = {https://ui.adsabs.harvard.edu/abs/2022A&A...666A.139A}
}

@ARTICLE{Husser2013,
       author = {{Husser}, T. -O. and {Wende-von Berg}, S. and {Dreizler}, S. and {Homeier}, D. and {Reiners}, A. and {Barman}, T. and {Hauschildt}, P.~H.},
        title = "{A new extensive library of PHOENIX stellar atmospheres and synthetic spectra}",
      journal = {\aap},
         year = 2013,
        month = may,
       volume = {553},
          eid = {A6},
        pages = {A6},
          doi = {10.1051/0004-6361/201219058},
archivePrefix = {arXiv},
       eprint = {1303.5632},
 primaryClass = {astro-ph.SR},
       adsurl = {https://ui.adsabs.harvard.edu/abs/2013A&A...553A...6H}
}

@ARTICLE{Temmink2024,
       author = {{Temmink}, Milou and {van Dishoeck}, Ewine F. and {Grant}, Sierra L. and {Tabone}, Beno{\^\i}t and {Gasman}, Danny and {Christiaens}, Valentin and {Samland}, Matthias and {Argyriou}, Ioannis and {Perotti}, Giulia and {G{\"u}del}, Manuel and {Henning}, Thomas and {Lagage}, Pierre-Olivier and {Abergel}, Alain and {Absil}, Olivier and {Barrado}, David and {Caratti o Garatti}, Alessio and {Glauser}, Adrian M. and {Kamp}, Inga and {Lahuis}, Fred and {Olofsson}, G{\"o}ran and {Ray}, Tom P. and {Scheithauer}, Silvia and {Vandenbussche}, Bart and {Waters}, L.~B.~F.~M. and {Arabhavi}, Aditya M. and {Jang}, Hyerin and {Kanwar}, Jayatee and {Morales-Calder{\'o}n}, Maria and {Rodgers-Lee}, Donna and {Schreiber}, J{\"u}rgen and {Schwarz}, Kamber and {Colina}, Luis},
        title = "{MINDS: The DR Tau disk. I. Combining JWST-MIRI data with high-resolution CO spectra to characterise the hot gas}",
      journal = {\aap},
         year = 2024,
        month = jun,
       volume = {686},
          eid = {A117},
        pages = {A117},
          doi = {10.1051/0004-6361/202348911},
       adsurl = {https://ui.adsabs.harvard.edu/abs/2024A&A...686A.117T}
}

@software{Christiaens2024,
       author = {{Christiaens}, Valentin and {Samland}, Matthias and {Gasman}, Danny and {Temmink}, Milou and {Perotti}, Giulia},
        title = "{MINDS: Hybrid pipeline for the reduction of JWST/MIRI-MRS data}",
 howpublished = {Astrophysics Source Code Library, record ascl:2403.007},
         year = 2024,
        month = mar,
          eid = {ascl:2403.007},
       adsurl = {https://ui.adsabs.harvard.edu/abs/2024ascl.soft03007C}
}

@ARTICLE{Lebouteiller2011,
       author = {{Lebouteiller}, V. and {Barry}, D.~J. and {Spoon}, H.~W.~W. and {Bernard-Salas}, J. and {Sloan}, G.~C. and {Houck}, J.~R. and {Weedman}, D.~W.},
        title = "{CASSIS: The Cornell Atlas of Spitzer/Infrared Spectrograph Sources}",
      journal = {\apjs},
         year = 2011,
        month = sep,
       volume = {196},
       number = {1},
          eid = {8},
        pages = {8},
          doi = {10.1088/0067-0049/196/1/8},
archivePrefix = {arXiv},
       eprint = {1108.3507},
 primaryClass = {astro-ph.IM},
       adsurl = {https://ui.adsabs.harvard.edu/abs/2011ApJS..196....8L}
}

@ARTICLE{Lebouteiller2015,
       author = {{Lebouteiller}, V. and {Barry}, D.~J. and {Goes}, C. and {Sloan}, G.~C. and {Spoon}, H.~W.~W. and {Weedman}, D.~W. and {Bernard-Salas}, J. and {Houck}, J.~R.},
        title = "{CASSIS: The Cornell Atlas of Spitzer/Infrared Spectrograph Sources. II. High-resolution Observations}",
      journal = {\apjs},
         year = 2015,
        month = jun,
       volume = {218},
       number = {2},
          eid = {21},
        pages = {21},
          doi = {10.1088/0067-0049/218/2/21},
archivePrefix = {arXiv},
       eprint = {1506.07610},
 primaryClass = {astro-ph.IM},
       adsurl = {https://ui.adsabs.harvard.edu/abs/2015ApJS..218...21L}
}

@ARTICLE{Andrews2013,
       author = {{Andrews}, Sean M. and {Rosenfeld}, Katherine A. and {Kraus}, Adam L. and {Wilner}, David J.},
        title = "{The Mass Dependence between Protoplanetary Disks and their Stellar Hosts}",
      journal = {\apj},
         year = 2013,
        month = jul,
       volume = {771},
       number = {2},
          eid = {129},
        pages = {129},
          doi = {10.1088/0004-637X/771/2/129},
archivePrefix = {arXiv},
       eprint = {1305.5262},
 primaryClass = {astro-ph.SR},
       adsurl = {https://ui.adsabs.harvard.edu/abs/2013ApJ...771..129A}
}

@ARTICLE{Furlan2006,
       author = {{Furlan}, E. and {Hartmann}, L. and {Calvet}, N. and {D'Alessio}, P. and {Franco-Hern{\'a}ndez}, R. and {Forrest}, W.~J. and {Watson}, D.~M. and {Uchida}, K.~I. and {Sargent}, B. and {Green}, J.~D. and {Keller}, L.~D. and {Herter}, T.~L.},
        title = "{A Survey and Analysis of Spitzer Infrared Spectrograph Spectra of T Tauri Stars in Taurus}",
      journal = {\apjs},
         year = 2006,
        month = aug,
       volume = {165},
       number = {2},
        pages = {568-605},
          doi = {10.1086/505468},
archivePrefix = {arXiv},
       eprint = {astro-ph/0608038},
 primaryClass = {astro-ph},
       adsurl = {https://ui.adsabs.harvard.edu/abs/2006ApJS..165..568F}
}

@ARTICLE{Watson2009,
       author = {{Watson}, Dan M. and {Leisenring}, Jarron M. and {Furlan}, Elise and {Bohac}, C.~J. and {Sargent}, B. and {Forrest}, W.~J. and {Calvet}, Nuria and {Hartmann}, Lee and {Nordhaus}, Jason T. and {Green}, Joel D. and {Kim}, K.~H. and {Sloan}, G.~C. and {Chen}, C.~H. and {Keller}, L.~D. and {d'Alessio}, Paola and {Najita}, J. and {Uchida}, Keven I. and {Houck}, J.~R.},
        title = "{Crystalline Silicates and Dust Processing in the Protoplanetary Disks of the Taurus Young Cluster}",
      journal = {\apjs},
         year = 2009,
        month = jan,
       volume = {180},
       number = {1},
        pages = {84-101},
          doi = {10.1088/0067-0049/180/1/84},
archivePrefix = {arXiv},
       eprint = {0704.1518},
 primaryClass = {astro-ph},
       adsurl = {https://ui.adsabs.harvard.edu/abs/2009ApJS..180...84W}
}

@ARTICLE{Espaillat2011,
       author = {{Espaillat}, C. and {Furlan}, E. and {D'Alessio}, P. and {Sargent}, B. and {Nagel}, E. and {Calvet}, N. and {Watson}, Dan M. and {Muzerolle}, J.},
        title = "{A Spitzer IRS Study of Infrared Variability in Transitional and Pre-transitional Disks Around T Tauri Stars}",
      journal = {\apj},
         year = 2011,
        month = feb,
       volume = {728},
       number = {1},
          eid = {49},
        pages = {49},
          doi = {10.1088/0004-637X/728/1/49},
archivePrefix = {arXiv},
       eprint = {1012.3500},
 primaryClass = {astro-ph.SR},
       adsurl = {https://ui.adsabs.harvard.edu/abs/2011ApJ...728...49E}
}

@ARTICLE{Muzerolle2009,
       author = {{Muzerolle}, James and {Flaherty}, Kevin and {Balog}, Zoltan and {Furlan}, Elise and {Smith}, Paul S. and {Allen}, Lori and {Calvet}, Nuria and {D'Alessio}, Paola and {Megeath}, S. Thomas and {Muench}, August and {Rieke}, George H. and {Sherry}, William H.},
        title = "{Evidence for Dynamical Changes in a Transitional Protoplanetary Disk with Mid-Infrared Variability}",
      journal = {\apjl},
         year = 2009,
        month = oct,
       volume = {704},
       number = {1},
        pages = {L15-L19},
          doi = {10.1088/0004-637X/704/1/L15},
archivePrefix = {arXiv},
       eprint = {0909.5201},
 primaryClass = {astro-ph.SR},
       adsurl = {https://ui.adsabs.harvard.edu/abs/2009ApJ...704L..15M}
}

@INPROCEEDINGS{Soderblom2014,
       author = {{Soderblom}, D.~R. and {Hillenbrand}, L.~A. and {Jeffries}, R.~D. and {Mamajek}, E.~E. and {Naylor}, T.},
        title = "{Ages of Young Stars}",
    booktitle = {Protostars and Planets VI},
         year = 2014,
       editor = {{Beuther}, Henrik and {Klessen}, Ralf S. and {Dullemond}, Cornelis P. and {Henning}, Thomas},
        month = jan,
        pages = {219-241},
          doi = {10.2458/azu_uapress_9780816531240-ch010},
archivePrefix = {arXiv},
       eprint = {1311.7024},
 primaryClass = {astro-ph.SR},
       adsurl = {https://ui.adsabs.harvard.edu/abs/2014prpl.conf..219S}
}

@INPROCEEDINGS{Manara2023,
       author = {{Manara}, C.~F. and {Ansdell}, M. and {Rosotti}, G.~P. and {Hughes}, A.~M. and {Armitage}, P.~J. and {Lodato}, G. and {Williams}, J.~P.},
        title = "{Demographics of Young Stars and their Protoplanetary Disks: Lessons Learned on Disk Evolution and its Connection to Planet Formation}",
    booktitle = {Protostars and Planets VII},
         year = 2023,
       editor = {{Inutsuka}, S. and {Aikawa}, Y. and {Muto}, T. and {Tomida}, K. and {Tamura}, M.},
       series = {Astronomical Society of the Pacific Conference Series},
       volume = {534},
        month = jul,
        pages = {539},
          doi = {10.48550/arXiv.2203.09930},
archivePrefix = {arXiv},
       eprint = {2203.09930},
 primaryClass = {astro-ph.SR},
       adsurl = {https://ui.adsabs.harvard.edu/abs/2023ASPC..534..539M}
}

@software{Erb_2022,
  author       = {Erb, Donald},
  title        = {{pybaselines: A Python library of algorithms for 
                   the baseline correction of experimental data}},
  month        = oct,
  year         = 2022,
  publisher    = {Zenodo},
  version      = {1.0.0},
  doi          = {10.5281/zenodo.7255880},
  url          = {https://doi.org/10.5281/zenodo.7255880}
}

@ARTICLE{Gordon2022,
       author = {{Gordon}, I.~E. and {Rothman}, L.~S. and {Hargreaves}, R.~J. and {Hashemi}, R. and {Karlovets}, E.~V. and {Skinner}, F.~M. and {Conway}, E.~K. and {Hill}, C. and {Kochanov}, R.~V. and {Tan}, Y. and {Wcis{\l}o}, P. and {Finenko}, A.~A. and {Nelson}, K. and {Bernath}, P.~F. and {Birk}, M. and {Boudon}, V. and {Campargue}, A. and {Chance}, K.~V. and {Coustenis}, A. and {Drouin}, B.~J. and {Flaud}, J. -M. and {Gamache}, R.~R. and {Hodges}, J.~T. and {Jacquemart}, D. and {Mlawer}, E.~J. and {Nikitin}, A.~V. and {Perevalov}, V.~I. and {Rotger}, M. and {Tennyson}, J. and {Toon}, G.~C. and {Tran}, H. and {Tyuterev}, V.~G. and {Adkins}, E.~M. and {Baker}, A. and {Barbe}, A. and {Can{\`e}}, E. and {Cs{\'a}sz{\'a}r}, A.~G. and {Dudaryonok}, A. and {Egorov}, O. and {Fleisher}, A.~J. and {Fleurbaey}, H. and {Foltynowicz}, A. and {Furtenbacher}, T. and {Harrison}, J.~J. and {Hartmann}, J. -M. and {Horneman}, V. -M. and {Huang}, X. and {Karman}, T. and {Karns}, J. and {Kassi}, S. and {Kleiner}, I. and {Kofman}, V. and {Kwabia-Tchana}, F. and {Lavrentieva}, N.~N. and {Lee}, T.~J. and {Long}, D.~A. and {Lukashevskaya}, A.~A. and {Lyulin}, O.~M. and {Makhnev}, V. Yu. and {Matt}, W. and {Massie}, S.~T. and {Melosso}, M. and {Mikhailenko}, S.~N. and {Mondelain}, D. and {M{\"u}ller}, H.~S.~P. and {Naumenko}, O.~V. and {Perrin}, A. and {Polyansky}, O.~L. and {Raddaoui}, E. and {Raston}, P.~L. and {Reed}, Z.~D. and {Rey}, M. and {Richard}, C. and {T{\'o}bi{\'a}s}, R. and {Sadiek}, I. and {Schwenke}, D.~W. and {Starikova}, E. and {Sung}, K. and {Tamassia}, F. and {Tashkun}, S.~A. and {Vander Auwera}, J. and {Vasilenko}, I.~A. and {Vigasin}, A.~A. and {Villanueva}, G.~L. and {Vispoel}, B. and {Wagner}, G. and {Yachmenev}, A. and {Yurchenko}, S.~N.},
        title = "{The HITRAN2020 molecular spectroscopic database}",
      journal = {\jqsrt},
         year = 2022,
        month = jan,
       volume = {277},
          eid = {107949},
        pages = {107949},
          doi = {10.1016/j.jqsrt.2021.107949},
       adsurl = {https://ui.adsabs.harvard.edu/abs/2022JQSRT.27707949G}
}

@ARTICLE{Gasman2023,
       author = {{Gasman}, Danny and {van Dishoeck}, Ewine F. and {Grant}, Sierra L. and {Temmink}, Milou and {Tabone}, Beno{\^\i}t and {Henning}, Thomas and {Kamp}, Inga and {G{\"u}del}, Manuel and {Lagage}, Pierre-Olivier and {Perotti}, Giulia and {Christiaens}, Valentin and {Samland}, Matthias and {Arabhavi}, Aditya M. and {Argyriou}, Ioannis and {Abergel}, Alain and {Absil}, Olivier and {Barrado}, David and {Boccaletti}, Anthony and {Bouwman}, Jeroen and {Caratti o Garatti}, Alessio and {Geers}, Vincent and {Glauser}, Adrian M. and {Guadarrama}, Rodrigo and {Jang}, Hyerin and {Kanwar}, Jayatee and {Lahuis}, Fred and {Morales-Calder{\'o}n}, Maria and {Mueller}, Michael and {Nehm{\'e}}, Cyrine and {Olofsson}, G{\"o}ran and {Pantin}, {\'E}ric and {Pawellek}, Nicole and {Ray}, Tom P. and {Rodgers-Lee}, Donna and {Scheithauer}, Silvia and {Schreiber}, J{\"u}rgen and {Schwarz}, Kamber and {Vandenbussche}, Bart and {Vlasblom}, Marissa and {Waters}, Rens L.~B.~F.~M. and {Wright}, Gillian and {Colina}, Luis and {Greve}, Thomas R. and {{\"O}stlin}, G{\"o}ran},
        title = "{MINDS. Abundant water and varying C/O across the disk of Sz 98 as seen by JWST/MIRI}",
      journal = {\aap},
         year = 2023,
        month = nov,
       volume = {679},
          eid = {A117},
        pages = {A117},
          doi = {10.1051/0004-6361/202347005},
archivePrefix = {arXiv},
       eprint = {2307.09301},
 primaryClass = {astro-ph.EP},
       adsurl = {https://ui.adsabs.harvard.edu/abs/2023A&A...679A.117G}
}

@ARTICLE{Salyk2011b,
       author = {{Salyk}, C. and {Pontoppidan}, K.~M. and {Blake}, G.~A. and {Najita}, J.~R. and {Carr}, J.~S.},
        title = "{A Spitzer Survey of Mid-infrared Molecular Emission from Protoplanetary Disks. II. Correlations and Local Thermal Equilibrium Models}",
      journal = {\apj},
         year = 2011,
        month = apr,
       volume = {731},
       number = {2},
          eid = {130},
        pages = {130},
          doi = {10.1088/0004-637X/731/2/130},
archivePrefix = {arXiv},
       eprint = {1104.0948},
 primaryClass = {astro-ph.GA},
       adsurl = {https://ui.adsabs.harvard.edu/abs/2011ApJ...731..130S}
}

@article{Jones2023,
    author = {Jones, O C and Álvarez-Márquez, J and Sloan, G C and Kavanagh, P J and Argyriou, I and Law, D R and Labiano, A and Patapis, P and Mueller, Michael and Larson, Kirsten L and Bright, Stacey N and Klaassen, P D and Fox, O D and Gasman, Danny and Geers, V C and Glauser, Adrian M and Guillard, Pierre and Nayak, Omnarayani and Noriega-Crespo, A and Ressler, Michael E and Sargent, B and Temim, T and Vandenbussche, B and García Marín, Macarena},
    title = "{Observations of the planetary nebula SMP LMC 058 with the JWST MIRI medium resolution spectrometer}",
    journal = {Monthly Notices of the Royal Astronomical Society},
    volume = {523},
    number = {2},
    pages = {2519-2529},
    year = {2023},
    month = {05},
    issn = {0035-8711},
    doi = {10.1093/mnras/stad1609},
    url = {https://doi.org/10.1093/mnras/stad1609},
    eprint = {https://academic.oup.com/mnras/article-pdf/523/2/2519/50519730/stad1609.pdf},
}

@ARTICLE{Li2024,
       author = {{Li}, Jialu and {Boogert}, Adwin and {Tielens}, Alexander G.~G.~M.},
        title = "{On the Interpretation of Mid-infrared Absorption Lines of Gas-phase H$_{2}$O as Observed by JWST/MIRI}",
      journal = {\apjs},
         year = 2024,
        month = aug,
       volume = {273},
       number = {2},
          eid = {32},
        pages = {32},
          doi = {10.3847/1538-4365/ad571a},
archivePrefix = {arXiv},
       eprint = {2406.05198},
 primaryClass = {astro-ph.IM},
       adsurl = {https://ui.adsabs.harvard.edu/abs/2024ApJS..273...32L}
}

@ARTICLE{Grant2024,
       author = {{Grant}, Sierra L. and {Kurtovic}, Nicolas T. and {van Dishoeck}, Ewine F. and {Henning}, Thomas and {Kamp}, Inga and {Nowacki}, Hugo and {Perraut}, Karine and {Banzatti}, Andrea and {Temmink}, Milou and {Christiaens}, Valentin and {Samland}, Matthias and {Gasman}, Danny and {Tabone}, Beno{\^\i}t and {G{\"u}del}, Manuel and {Lagage}, Pierre-Olivier and {Arabhavi}, Aditya M. and {Barrado}, David and {Caratti o Garatti}, Alessio and {Glauser}, Adrian M. and {Jang}, Hyerin and {Kanwar}, Jayatee and {Lahuis}, Fred and {Morales-Calder{\'o}n}, Maria and {Olofsson}, G{\"o}ran and {Perotti}, Giulia and {Schwarz}, Kamber and {Vlasblom}, Marissa and {Garcia Lopez}, Rebeca and {Long}, Feng},
        title = "{MINDS: A multi-instrument investigation into the molecule-rich JWST-MIRI spectrum of the DF Tau binary system}",
      journal = {\aap},
         year = 2024,
        month = sep,
       volume = {689},
          eid = {A85},
        pages = {A85},
          doi = {10.1051/0004-6361/202450768},
archivePrefix = {arXiv},
       eprint = {2406.10217},
 primaryClass = {astro-ph.EP},
       adsurl = {https://ui.adsabs.harvard.edu/abs/2024A&A...689A..85G}
}

@article{Arabhavi2024,
author = {A. M. Arabhavi  and I. Kamp  and Th. Henning  and E. F. van Dishoeck  and V. Christiaens  and D. Gasman  and A. Perrin  and M. Güdel  and B. Tabone  and J. Kanwar  and L. B. F. M. Waters  and I. Pascucci  and M. Samland  and G. Perotti  and G. Bettoni  and S. L. Grant  and P. O. Lagage  and T. P. Ray  and B. Vandenbussche  and O. Absil  and I. Argyriou  and D. Barrado  and A. Boccaletti  and J. Bouwman  and A. Caratti o Garatti  and A. M. Glauser  and F. Lahuis  and M. Mueller  and G. Olofsson  and E. Pantin  and S. Scheithauer  and M. Morales-Calderón  and R. Franceschi  and H. Jang  and N. Pawellek  and D. Rodgers-Lee  and J. Schreiber  and K. Schwarz  and M. Temmink  and M. Vlasblom  and G. Wright  and L. Colina  and G. Östlin },
title = {Abundant hydrocarbons in the disk around a very-low-mass star},
journal = {Science},
volume = {384},
number = {6700},
pages = {1086-1090},
year = {2024},
doi = {10.1126/science.adi8147}}

@ARTICLE{Foreman-Mackey2013,
       author = {{Foreman-Mackey}, Daniel and {Hogg}, David W. and {Lang}, Dustin and {Goodman}, Jonathan},
        title = "{emcee: The MCMC Hammer}",
      journal = {\pasp},
         year = 2013,
        month = mar,
       volume = {125},
       number = {925},
        pages = {306},
          doi = {10.1086/670067},
archivePrefix = {arXiv},
       eprint = {1202.3665},
 primaryClass = {astro-ph.IM},
       adsurl = {https://ui.adsabs.harvard.edu/abs/2013PASP..125..306F}
}

@ARTICLE{Christiaens2023,
       author = {{Christiaens}, Valentin and {Gonzalez}, Carlos and {Farkas}, Ralf and {Dahlqvist}, Carl-Henrik and {Nasedkin}, Evert and {Milli}, Julien and {Absil}, Olivier and {Ngo}, Henry and {Cantero}, Carles and {Rainot}, Alan and {Hammond}, Iain and {Bonse}, Markus and {Cantalloube}, Faustine and {Vigan}, Arthur and {Kompella}, Vijay and {Hancock}, Paul},
        title = "{VIP: A Python package for high-contrast imaging}",
      journal = {The Journal of Open Source Software},
         year = 2023,
        month = jan,
       volume = {8},
       number = {81},
          eid = {4774},
        pages = {4774},
          doi = {10.21105/joss.04774},
       adsurl = {https://ui.adsabs.harvard.edu/abs/2023JOSS....8.4774C}
}

@ARTICLE{vanderMarel2023,
       author = {{van der Marel}, Nienke and {Pinilla}, Paola},
        title = "{Dust evolution in protoplanetary disks}",
      journal = {arXiv e-prints},
         year = 2023,
        month = oct,
          eid = {arXiv:2310.09077},
        pages = {arXiv:2310.09077},
          doi = {10.48550/arXiv.2310.09077},
archivePrefix = {arXiv},
       eprint = {2310.09077},
 primaryClass = {astro-ph.EP},
       adsurl = {https://ui.adsabs.harvard.edu/abs/2023arXiv231009077V}
}

@ARTICLE{Mah2024,
       author = {{Mah}, Jingyi and {Savvidou}, Sofia and {Bitsch}, Bertram},
        title = "{Mind the gap: Distinguishing disc substructures and their impact on the inner disc composition}",
      journal = {\aap},
         year = 2024,
        month = jun,
       volume = {686},
          eid = {L17},
        pages = {L17},
          doi = {10.1051/0004-6361/202450322},
archivePrefix = {arXiv},
       eprint = {2406.06219},
 primaryClass = {astro-ph.EP},
       adsurl = {https://ui.adsabs.harvard.edu/abs/2024A&A...686L..17M}
}

@ARTICLE{Luhman2023,
       author = {{Luhman}, K.~L.},
        title = "{A Census of the Taurus Star-forming Region and Neighboring Associations with Gaia}",
      journal = {\aj},
         year = 2023,
        month = feb,
       volume = {165},
       number = {2},
          eid = {37},
        pages = {37},
          doi = {10.3847/1538-3881/ac9da3},
archivePrefix = {arXiv},
       eprint = {2211.09785},
 primaryClass = {astro-ph.GA},
       adsurl = {https://ui.adsabs.harvard.edu/abs/2023AJ....165...37L}
}

@ARTICLE{Galli2021,
       author = {{Galli}, P.~A.~B. and {Bouy}, H. and {Olivares}, J. and {Miret-Roig}, N. and {Sarro}, L.~M. and {Barrado}, D. and {Berihuete}, A. and {Bertin}, E. and {Cuillandre}, J. -C.},
        title = "{Chamaeleon DANCe. Revisiting the stellar populations of Chamaeleon I and Chamaeleon II with Gaia-DR2 data}",
      journal = {\aap},
         year = 2021,
        month = feb,
       volume = {646},
          eid = {A46},
        pages = {A46},
          doi = {10.1051/0004-6361/202039395},
archivePrefix = {arXiv},
       eprint = {2012.00329},
 primaryClass = {astro-ph.GA},
       adsurl = {https://ui.adsabs.harvard.edu/abs/2021A&A...646A..46G}
}

@ARTICLE{Tripathi2017,
       author = {{Tripathi}, Anjali and {Andrews}, Sean M. and {Birnstiel}, Tilman and {Wilner}, David J.},
        title = "{A millimeter Continuum Size-Luminosity Relationship for Protoplanetary Disks}",
      journal = {\apj},
         year = 2017,
        month = aug,
       volume = {845},
       number = {1},
          eid = {44},
        pages = {44},
          doi = {10.3847/1538-4357/aa7c62},
archivePrefix = {arXiv},
       eprint = {1706.08977},
 primaryClass = {astro-ph.EP},
       adsurl = {https://ui.adsabs.harvard.edu/abs/2017ApJ...845...44T}
}

@ARTICLE{Hendler2020,
       author = {{Hendler}, Nathanial and {Pascucci}, Ilaria and {Pinilla}, Paola and {Tazzari}, Marco and {Carpenter}, John and {Malhotra}, Renu and {Testi}, Leonardo},
        title = "{The Evolution of Dust Disk Sizes from a Homogeneous Analysis of 1-10 Myr old Stars}",
      journal = {\apj},
         year = 2020,
        month = jun,
       volume = {895},
       number = {2},
          eid = {126},
        pages = {126},
          doi = {10.3847/1538-4357/ab70ba},
archivePrefix = {arXiv},
       eprint = {2001.02666},
 primaryClass = {astro-ph.EP},
       adsurl = {https://ui.adsabs.harvard.edu/abs/2020ApJ...895..126H}
}

@ARTICLE{Schneider2021,
       author = {{Schneider}, Aaron David and {Bitsch}, Bertram},
        title = "{How drifting and evaporating pebbles shape giant planets. I. Heavy element content and atmospheric C/O}",
      journal = {\aap},
         year = 2021,
        month = oct,
       volume = {654},
          eid = {A71},
        pages = {A71},
          doi = {10.1051/0004-6361/202039640},
archivePrefix = {arXiv},
       eprint = {2105.13267},
 primaryClass = {astro-ph.EP},
       adsurl = {https://ui.adsabs.harvard.edu/abs/2021A&A...654A..71S}
}

@ARTICLE{Mulders2015,
       author = {{Mulders}, Gijs D. and {Ciesla}, Fred J. and {Min}, Michiel and {Pascucci}, Ilaria},
        title = "{The Snow Line in Viscous Disks around Low-mass Stars: Implications for Water Delivery to Terrestrial Planets in the Habitable Zone}",
      journal = {\apj},
         year = 2015,
        month = jul,
       volume = {807},
       number = {1},
          eid = {9},
        pages = {9},
          doi = {10.1088/0004-637X/807/1/9},
archivePrefix = {arXiv},
       eprint = {1505.03516},
 primaryClass = {astro-ph.EP},
       adsurl = {https://ui.adsabs.harvard.edu/abs/2015ApJ...807....9M}
}

@ARTICLE{Kalyaan2023,
       author = {{Kalyaan}, Anusha and {Pinilla}, Paola and {Krijt}, Sebastiaan and {Banzatti}, Andrea and {Rosotti}, Giovanni and {Mulders}, Gijs D. and {Lambrechts}, Michiel and {Long}, Feng and {Herczeg}, Gregory J.},
        title = "{The Effect of Dust Evolution and Traps on Inner Disk Water Enrichment}",
      journal = {\apj},
         year = 2023,
        month = sep,
       volume = {954},
       number = {1},
          eid = {66},
        pages = {66},
          doi = {10.3847/1538-4357/ace535},
archivePrefix = {arXiv},
       eprint = {2307.01789},
 primaryClass = {astro-ph.EP},
       adsurl = {https://ui.adsabs.harvard.edu/abs/2023ApJ...954...66K}
}

@ARTICLE{Pinilla2024,
       author = {{Pinilla}, Paola and {Benisty}, Myriam and {Waters}, Rens and {Bae}, Jaehan and {Facchini}, Stefano},
        title = "{Survival of the long-lived inner disk of PDS70}",
      journal = {\aap},
         year = 2024,
        month = jun,
       volume = {686},
          eid = {A135},
        pages = {A135},
          doi = {10.1051/0004-6361/202348707},
archivePrefix = {arXiv},
       eprint = {2403.07057},
 primaryClass = {astro-ph.EP},
       adsurl = {https://ui.adsabs.harvard.edu/abs/2024A&A...686A.135P}
}

@ARTICLE{Ratzenbock2023,
       author = {{Ratzenb{\"o}ck}, Sebastian and {Gro{\ss}schedl}, Josefa E. and {Alves}, Jo{\~a}o and {Miret-Roig}, N{\'u}ria and {Bomze}, Immanuel and {Forbes}, John and {Goodman}, Alyssa and {Hacar}, {\'A}lvaro and {Lin}, Doug and {Meingast}, Stefan and {M{\"o}ller}, Torsten and {Piecka}, Martin and {Posch}, Laura and {Rottensteiner}, Alena and {Swiggum}, Cameren and {Zucker}, Catherine},
        title = "{The star formation history of the Sco-Cen association. Coherent star formation patterns in space and time}",
      journal = {\aap},
         year = 2023,
        month = oct,
       volume = {678},
          eid = {A71},
        pages = {A71},
          doi = {10.1051/0004-6361/202346901},
archivePrefix = {arXiv},
       eprint = {2302.07853},
 primaryClass = {astro-ph.SR},
       adsurl = {https://ui.adsabs.harvard.edu/abs/2023A&A...678A..71R}
}

@INPROCEEDINGS{Zucker2023,
       author = {{Zucker}, C. and {Alves}, J. and {Goodman}, A. and {Meingast}, S. and {Galli}, P.},
        title = "{The Solar Neighborhood in the Age of Gaia}",
    booktitle = {Protostars and Planets VII},
         year = 2023,
       editor = {{Inutsuka}, S. and {Aikawa}, Y. and {Muto}, T. and {Tomida}, K. and {Tamura}, M.},
       series = {Astronomical Society of the Pacific Conference Series},
       volume = {534},
        month = jul,
        pages = {43},
          doi = {10.48550/arXiv.2212.00067},
archivePrefix = {arXiv},
       eprint = {2212.00067},
 primaryClass = {astro-ph.GA},
       adsurl = {https://ui.adsabs.harvard.edu/abs/2023ASPC..534...43Z}
}

@ARTICLE{Kospal2023,
       author = {{K{\'o}sp{\'a}l}, {\'A}gnes and {{\'A}brah{\'a}m}, P{\'e}ter and {Diehl}, Lindsey and {Banzatti}, Andrea and {Bouwman}, Jeroen and {Chen}, Lei and {Cruz-S{\'a}enz de Miera}, Fernando and {Green}, Joel D. and {Henning}, Thomas and {Rab}, Christian},
        title = "{JWST/MIRI Spectroscopy of the Disk of the Young Eruptive Star EX Lup in Quiescence}",
      journal = {\apjl},
         year = 2023,
        month = mar,
       volume = {945},
       number = {1},
          eid = {L7},
        pages = {L7},
          doi = {10.3847/2041-8213/acb58a},
archivePrefix = {arXiv},
       eprint = {2301.08770},
 primaryClass = {astro-ph.SR},
       adsurl = {https://ui.adsabs.harvard.edu/abs/2023ApJ...945L...7K}
}

@ARTICLE{Ansdell2018,
       author = {{Ansdell}, M. and {Williams}, J.~P. and {Trapman}, L. and {van Terwisga}, S.~E. and {Facchini}, S. and {Manara}, C.~F. and {van der Marel}, N. and {Miotello}, A. and {Tazzari}, M. and {Hogerheijde}, M. and {Guidi}, G. and {Testi}, L. and {van Dishoeck}, E.~F.},
        title = "{ALMA Survey of Lupus Protoplanetary Disks. II. Gas Disk Radii}",
      journal = {\apj},
         year = 2018,
        month = may,
       volume = {859},
       number = {1},
          eid = {21},
        pages = {21},
          doi = {10.3847/1538-4357/aab890},
archivePrefix = {arXiv},
       eprint = {1803.05923},
 primaryClass = {astro-ph.EP},
       adsurl = {https://ui.adsabs.harvard.edu/abs/2018ApJ...859...21A}
}

@ARTICLE{Jennings2022,
       author = {{Jennings}, Jeff and {Booth}, Richard A. and {Tazzari}, Marco and {Clarke}, Cathie J. and {Rosotti}, Giovanni P.},
        title = "{A super-resolution analysis of the DSHARP survey: substructure is common in the inner 30 au}",
      journal = {\mnras},
         year = 2022,
        month = jan,
       volume = {509},
       number = {2},
        pages = {2780-2799},
          doi = {10.1093/mnras/stab3185},
archivePrefix = {arXiv},
       eprint = {2103.02392},
 primaryClass = {astro-ph.EP},
       adsurl = {https://ui.adsabs.harvard.edu/abs/2022MNRAS.509.2780J}
}

@ARTICLE{Huang2018,
       author = {{Huang}, Jane and {Andrews}, Sean M. and {Dullemond}, Cornelis P. and {Isella}, Andrea and {P{\'e}rez}, Laura M. and {Guzm{\'a}n}, Viviana V. and {{\"O}berg}, Karin I. and {Zhu}, Zhaohuan and {Zhang}, Shangjia and {Bai}, Xue-Ning and {Benisty}, Myriam and {Birnstiel}, Tilman and {Carpenter}, John M. and {Hughes}, A. Meredith and {Ricci}, Luca and {Weaver}, Erik and {Wilner}, David J.},
        title = "{The Disk Substructures at High Angular Resolution Project (DSHARP). II. Characteristics of Annular Substructures}",
      journal = {\apjl},
         year = 2018,
        month = dec,
       volume = {869},
       number = {2},
          eid = {L42},
        pages = {L42},
          doi = {10.3847/2041-8213/aaf740},
archivePrefix = {arXiv},
       eprint = {1812.04041},
 primaryClass = {astro-ph.EP},
       adsurl = {https://ui.adsabs.harvard.edu/abs/2018ApJ...869L..42H}
}

@ARTICLE{Haugbolle2019,
       author = {{Haugb{\o}lle}, Troels and {Weber}, Philipp and {Wielandt}, Daniel P. and {Ben{\'\i}tez-Llambay}, Pablo and {Bizzarro}, Martin and {Gressel}, Oliver and {Pessah}, Martin E.},
        title = "{Probing the Protosolar Disk Using Dust Filtering at Gaps in the Early Solar System}",
      journal = {\aj},
         year = 2019,
        month = aug,
       volume = {158},
       number = {2},
          eid = {55},
        pages = {55},
          doi = {10.3847/1538-3881/ab1591},
archivePrefix = {arXiv},
       eprint = {1903.12274},
 primaryClass = {astro-ph.EP},
       adsurl = {https://ui.adsabs.harvard.edu/abs/2019AJ....158...55H}
}
\bibliographystyle{aasjournal}


\end{document}